\documentclass[reqno]{amsart}
\usepackage{tikz}
\usepackage{mathptmx}
\usepackage{xspace}
\usepackage{amsmath,amsfonts,amsthm,eucal}
\usepackage{graphicx}
\usepackage{epstopdf}
\newtheorem{Thm}{Theorem}[section]
\newtheorem{Prop}[Thm]{Proposition}
\newtheorem{Lem}[Thm]{Lemma}
\newtheorem{Coro}[Thm]{Corollary}
\newtheorem{Def}{Definition}
\newcommand{\epais}[0]{very thick}
\newcommand{\epaisdfn}[0]{very thick}
\newcommand{\vac}[1]{\draw (#1) circle (1/5);
 	\fill[white] (#1) circle (1/5)}
\newcommand{\defect}[1]{\draw (#1) --++ (3,0)}
\newcommand{\bord}[2]{\draw[\epais] (#1,-1/2) --++ (0,#2)}
\newcommand{\link}[4]{\draw (#1) .. controls (#2) and (#3) .. (#4)}
\newcommand{\defects}[1]{\draw (#1) --++ (3/2,0)}
\newcommand{\wave}[2]{\draw[dashed] (#1,#2) --++ (3,0);}
\newcommand{\tilea}[2]{\begin{tikzpicture}[baseline={(current bounding box.center)},scale=1/3]
\draw (#1,#2) --++ (2,0) --++ (0,2) --++ (-2,0) --++ (0,-2);
\draw (#1,#2+1) .. controls (#1+1/2,#2+1) and (#1+1,#2+3/2).. (#1+1,#2+2);
\draw (#1+1,#2) .. controls (#1+1,#2+1/2) and (#1+3/2,#2+1).. (#1+2,#2+1);
\end{tikzpicture} }
\newcommand{\tileb}[2]{\begin{tikzpicture}[baseline={(current bounding box.center)},scale=1/3]
\draw (#1,#2) --++ (2,0) --++ (0,2) --++ (-2,0) --++ (0,-2);
\draw (#1,#2+1) .. controls (#1+1/2,#2+1) and (#1+1,#2+1/2).. (#1+1,#2);
\draw (#1+1,#2+2) .. controls (#1+1,#2+3/2) and (#1+3/2,#2+1).. (#1+2,#2+1); 
\end{tikzpicture}}
\newcommand{\tilec}[2]{\begin{tikzpicture}[baseline={(current bounding box.center)},scale=1/3]
\draw (#1,#2) --++ (2,0) --++ (0,2) --++ (-2,0) --++ (0,-2);
\vac{#1,#2+1};
\vac{#1+2,#2+1};
\draw (#1+1,#2) --  (#1+1,#2+2); 
\end{tikzpicture}}
\newcommand{\tilecMini}[2]{\begin{tikzpicture}[scale=1/8]
\draw (#1,#2) --++ (2,0) --++ (0,2) --++ (-2,0) --++ (0,-2);
\vac{#1,#2+1};
\vac{#1+2,#2+1};
\draw (#1+1,#2) --  (#1+1,#2+2); 
\end{tikzpicture}}
\newcommand{\dtile}[2]{\begin{tikzpicture}[baseline={(current bounding box.center)},scale=1/3]
\draw (#1,#2) --++ (0,2) --++ (2,0) --++ (0,-2) --++ (-2,0); 
\draw (#1 +1,#2) --++ (0,2);
\draw (#1,#2 +1) --++ (3/4,0);
\draw (#1 +2,#2 +1) --++ (-3/4,0);
\end{tikzpicture}}
\newcommand{\dtiler}[2]{\begin{tikzpicture}[baseline={(current bounding box.center)},scale=1/3]
\draw (#1,#2) --++ (0,2) --++ (2,0) --++ (0,-2) --++ (-2,0); 
\draw (#1,#2 +1) --++ (2,0);
\draw (#1 + 1,#2) --++ (0,3/4);
\draw (#1 + 1 ,#2 + 2) --++ (0,-3/4);
\end{tikzpicture}}
\newcommand{\cuteBubble}{\begin{tikzpicture}[scale=1/8]
\bord{0}{2};
\link{0,0}{1,0}{1,1}{0,1};; 
\end{tikzpicture}}
\newcommand{\dtl}[1]{\mathsf{dTL}_{#1}}
\newcommand{\odtl}[1]{\mathsf{odTL}_{#1}}
\newcommand{\edtl}[1]{\mathsf{edTL}_{#1}}
\newcommand{\tl}[1]{\mathsf{TL}_{#1}}
\newcommand{\idtl}{\mathsf{id}}
\newcommand{\eid}{\mathsf{eid}}
\newcommand{\oid}{\mathsf{oid}}

\newcommand{\LM}[1]{\mathsf A_{#1}}
\newcommand{\TM}[1]{\mathsf H_{#1}}
\newcommand{\U}[1]{\mathsf S_{#1}}
\newcommand{\X}[1]{X_{#1}}
\newcommand{\Y}[1]{Y_{#1}}
\newcommand{\irred}[1]{I_{#1}}
\newcommand{\DF}[1]{\mathsf F_{#1}}
\newcommand{\df}[1]{\delta_{#1}}
\newcommand{\vs}[1]{V_{#1}}
\newcommand{\dr}[1]{\mathsf R_{#1}}
\newcommand{\R}[1]{R_{#1}}
\newcommand{\dl}[1]{\mathsf I_{#1}}
\newcommand{\dG}[1]{dG_{#1}}
\newcommand{\G}[1]{G_{#1}}
\newcommand{\F}[1]{F_{#1}}
\newcommand{\A}[2]{\mathsf P^{#1}_{#2}}
\newcommand{\dP}[1]{\mathsf P{}\hspace{-0.14em}_{#1}}

\newcommand{\rl}[1]{I_{#1}} 
\newcommand{\dU}[1]{s_{#1}} 
\newcommand{\ddl}[1]{i_{#1}} 
\newcommand{\ddr}[1]{r_{#1}}
\newcommand{\drl}[1]{\tilde{i}_{#1}} 
\newcommand{\kc}{k_c}
\newcommand{\Res}[1]{#1{}\hspace{-0.18em} \downarrow{}\hskip -0.18em}
\newcommand{\Ind}[1]{#1{}\hspace{-0.18em} \uparrow{}\hskip -0.18em}
\DeclareMathOperator{\id}{id}
\DeclareMathOperator{\im}{im}

\DeclareMathOperator{\Hom}{Hom}
\definecolor{rougePompier}{rgb}{0.93,0.11,0.14}
\definecolor{vertForet}{rgb}{0.04,0.75,0.07}

\newcommand{\invisible}[1]{} 
\title[The dilute Temperley-Lieb algebra]{The principal indecomposable modules\\ of the dilute Temperley-Lieb algebra}
\author[J Bellet\^ete]{Jonathan Bellet\^ete}
\address[Jonathan Bellet\^ete]{
CRM and D\'epartement de physique\\ 
Universit\'e de Montr\'eal\\
Montr\'eal, QC, Canada, H3C 3J7}
\email{\tt jonathan.belletete@umontreal.ca}
\author[Y Saint-Aubin]{Yvan Saint-Aubin}
\address[Yvan Saint-Aubin]{
D\'{e}partement de math\'{e}matiques et de statistique\\
Universit\'{e} de Montr\'{e}al\\
Montr\'eal, QC, Canada, H3C 3J7.}
\email{yvan.saint-aubin@umontreal.ca}
\date{\today}
\begin{document}
  
\begin{abstract}The Temperley-Lieb algebra $\tl n(\beta)$ can be defined as the set of rectangular diagrams with $n$ points on each of their vertical sides, with all points joined pairwise by non-intersecting strings. The multiplication is then the concatenation of diagrams. The dilute Temperley-Lieb algebra $\dtl n(\beta)$ has a similar diagrammatic definition where, now, points on the sides may remain free of strings. Like $\tl n$, the dilute $\dtl n$ depends on a parameter $\beta\in\mathbb C$, often given as $\beta=q+q^{-1}$ for some $q\in\mathbb C^\times$. In statistical physics, the algebra plays a central role in the study of dilute loop models. The paper is devoted to the construction of its principal indecomposable modules.
  
Basic definitions and properties are first given: the dimension of $\dtl n$, its break up into even and odd subalgebras and its filtration through $n+1$ ideals. 
The standard modules $\U{n,k}$ are then introduced and their behaviour under restriction and induction is described. A bilinear form, the Gram product, is used to identify their (unique) maximal submodule $\dr{n,k}$ which is then shown to be irreducible or trivial.
It is then noted that $\dtl n$ is a cellular algebra. This fact allows for the identification of complete sets of non-isomorphic irreducible modules and projective indecomposable ones. The structure of $\dtl n$ as a left module over itself is then given for all values of the parameter $q$, that is, for both $q$ generic and a root of unity.
  
\medskip
  
\end{abstract}
  
\maketitle  
  
%
\begin{section}{Introduction}\label{sec:intro}
%

Since its introduction in the 1970s \cite{TemperleyLieb}, the Temperley-Lieb algebra has played a central role in several domains of mathematical physics, mainly in the statistical physics description of lattice models and in conformal field theory. But, since its ``rediscovery'' by mathematicians --- Jones' seminal paper \cite{Jones} comes here to mind, --- algebraists have contributed significantly to its understanding. Its representation theory was first described independently by Goodman and Wenzl \cite{GoodmanWenzl} and by Martin \cite{Martin} and is now widely used. 
  
Several generalizations have been introduced, many suggested by physical problems:  the periodic (affine) Temperley-Lieb algebra \cite{GrahamLehrer, GreenFan, Greenseul, ErdmannGreen, MartinSaleur}, polychromatic algebras \cite{GrimmPearce}, the Birman-Wenzl-Murakami algebra \cite{BirmanWenzl, Murakami} and the dilute Temperley-Lieb algebra \cite{Grimm}. Their role in mathematical physics has developed over the years, particularly since their intimate relationship with infinite-dimensional Lie algebras appearing in the description of continuum limits of lattice models have been recognized. The fact that some hamiltonians or transfer matrices could be seen as representatives, within given modules, of an abstract element of the Temperley-Lieb algebra was already in Temperley and Lieb's work. But the following fact is Pasquier and Saleur's crucial observation \cite{PasquierSaleur}: the representation theory of the Temperley-Lieb algebra can be used to understand the Virasoro representations appearing in the limit, when the mesh goes to zero, of the finite-size lattice models.
  
The origin of the dilute Temperley-Lieb algebra $\dtl n$ can be tied to Nienhuis' work \cite{Nienhuis}. It was known since early works by Yang and Baxter that some algebraic conditions on Boltzmann weights of statistical lattice models assure some form of integrability. Trying to find integrable $O(N)$ models, Nienhuis introduced a family of such weights satisfying these conditions. He noticed soon after that these weights, labeled by two parameters $\lambda$ and $u$, were part of a larger family defined by Izergin and Korepin \cite{IzerginKorepin}. With Bl\"ote he explored the large $n$ limit through numerical simulations \cite{BloteNienhuis}. (Note that we use small $n\geq 1$ for the size of the lattice. This integer $n$ is independent of the $N$ appearing in the usual name of the $O(N)$ model.) Under the hypothesis that such lattice models would go to conformal field theories in the limit $n\rightarrow \infty$, they found a simple relation between the parameter $\lambda$ and the central charge of these continuum theories. Nienhuis' weights are attached to the tiles forming the lattice. The various states of the tiles of these models are described by non-intersecting links joining their edges pairwise, exactly as in the Temperley-Lieb description of (fully-packed) loop models. But contrarily to the Temperley-Lieb case, some of the edges of the tiles may be left free of links in dilute models. Generalisations of these dilute models \cite{Roche, WarnaarNienhuisSeaton, ZhouPearceGrimm} and sets of integrable boundary conditions \cite{YungBatchelor,BehrendPearce} to match the (bulk) Boltzmann weights were found in the years that followed.
  
Even though the representation theory of the (original) Temperley-Lieb algebra \cite{GoodmanWenzl, Martin} and that of the periodic version \cite{GrahamLehrer, MartinSaleur} are well-established, that of the dilute Temperley-Lieb lags behind. A few years ago the dichromatic Temperley-Lieb algebra has been studied \cite{GrimmMartin} and one might be able to retrieve, at least partially, some properties of the dilute $\dtl n$ from some quotient of the dichromatic one. But the dilute Temperley-Lieb algebra $\dtl n(\beta), \beta\in\mathbb C$, has now become such an important tool in mathematical physics that a direct and systematic description of its properties is necessary. The structure uncovered and tools developed should be powerful enough to study questions like, for example, the computation of the fusion ring of its standard and projective modules, the possible existence of a Schur-Weyl duality with some other (quantum) algebra, or the identification of modules in which transfer matrices have non-trivial Jordan structure. The present paper is a first step toward this goal. It gives an explicit construction of all its principal indecomposable modules, for both cases when the algebra $\dtl n(\beta)$ is semisimple and non-semisimple.
  
Several approaches surrounding the families of Temperley-Lieb algebras are based on diagrammatic techniques. Several rigorous mathematical works resort to them and they are used to define many lattice models. So it is not surprising that the early construction of the principal indecomposable modules of the (original) Temperley-Lieb algebra $\tl n$ by Martin has been reformulated through methods based on link diagrams \cite{Westbury, RiSY}. It is this approach that we choose to follow here. Both the elements and the multiplication of the dilute Temperley-Lieb algebra $\dtl n(\beta)$ are defined through diagrams in section \ref{sec:dtln}. (Another parameter, $q\in\mathbb C^\times$, is also used. It is related to the first by $\beta=q+q^{-1}$.) These definitions lead to the identification of a natural subalgebra $S_n\subset \dtl n$ and several copies of the usual Temperley-Lieb algebras $\tl{n'}, n'\leq n$, the computation of its dimension and its decomposition into even and odd parts. 
A natural filtration of $\dtl n$ by ideals is also introduced here.
Section \ref{sec:modules} is devoted to the construction of standard modules. Their basic characteristics are there established: they are cyclic and indecomposable and their dimensions are expressed in terms of those of the standard modules of $\tl n$. Restriction and induction are used to probe their inner structure. Section \ref{sec:Gram} introduces another classical tool of representation theory. A bilinear form, called the Gram product, is defined on the standard modules. The radical of this form, that is the subspace of vectors with vanishing Gram coupling with all others, is shown to be the unique maximal submodule of the standard module. The determinant of the Gram matrix representing the bilinear form in some basis is easily computed. Its zeroes occur when the parameter $q$ is a root of unity and, consequently, the algebra $\dtl n$ is semisimple when $q$ is generic, that is when it is not a root of unity. Finally the radical, when it is non-trivial, is shown to be irreducible and isomorphic to the irreducible quotient of another standard module.
   
In section \ref{sec:AtRootUnity} the definition of cellular algebras \cite{GrahamL, Mathas} is recalled and results from previous sections show that $\dtl n$ is indeed cellular. The fundamental properties of cellular algebras then provide a complete list of non-isomorphic irreducible modules and of projective indecomposable ones. Some information about the structure of the standard and principal modules can be retrieved as well as the structure of $\dtl n$ as a left module over itself when $q$ is a root of unity. The induction of the principal modules, from $\dtl{n-1}$ to $\dtl n$, is finally expressed in terms of the principal ones of $\dtl n$. This gives an explicit way to construct bases for these modules. 
  
The conclusion reviews the main results and discusses possible extensions. Some results of this paper are based on the analogous ones for the algebra $\tl n$. These are reviewed in appendix \ref{app.tln}. Appendices \ref{app:Fn} and \ref{app.dlnk} contain some technical computations and proofs. Finally appendix \ref{sec:algebre} reviews the algebraic tools that are used throughout the paper, but particularly in section \ref{sec:AtRootUnity}.

\end{section}
%
%
\begin{section}{ Basic properties of the dilute algebra $\dtl n$}\label{sec:dtln}
%
%
This section introduces the dilute Temperley-Lieb algebra whose elements and product are defined diagrammatically. It is shown to split naturally into a direct sum of two ideals, its even and odd parts, and to be filtered by ideals labeled by an integer running from $0$ to $n$. Another subalgebra $S_n\subset \dtl n$ will play a role in the subsequent section and it is also defined. The section ends with the computation of the dimension of $\dtl n$. Several techniques used here are borrowed from previous studies of the (original) Temperley-Lieb algebra. Appendix \ref{app.tln} gathers some basic results for this algebra. Reading this appendix in parallel will ease the understanding of this section and of the next one. Some results of this section and the two next ones will be crucial to recognize $\dtl n$ as a cellular algebra. This will be done in subsection \ref{sec:cellalgebra}.

Techniques and results are borrowed from previous works. First are diagrammatic methods. These were introduced early on in topology (see, e.g.~\cite{Jones2,KauffmanLins} for landmarks of their use). Martin's book \cite{Martin} contains $n$-diagrams and $n$-links (see his chapter 9), but they do not play a major role in the classification of indecomposable projective modules. But both play a crucial one in Martin and Saleur's definition of the Temperley-Lieb algebra of type $B$, also known in the physics literature as the ``blob algebra'' \cite{MartinSaleur2}. 
In topology, representation theory and physics, these diagrammatic methods have shown their power. Second the filtration \eqref{eq:filtrationByIdeals} of $\dtl n$ is a crucial observation. Again it appears in \cite{MartinSaleur2} (see their Proposition 1). But it is in Graham and Lehrer's work \cite{GrahamL} that the deep consequences of this filtration are recognized. Our lemma \ref{lem.hom.unk} (their proposition 2.6 for cellular algebras) follows from it and is a key step in our study. The filtration's final role will be played in subsection \ref{sec:cellalgebra}, thanks to Graham and Lehrer's results. A third tool is the Gram bilinear form. It is harder to put a date on its first use in the representation theory of the Temperley-Lieb families, but it was already in use in the early nineties, e.g.~\cite{MartinSaleur, MartinSaleur2}.

\begin{subsection}{Definition of $\dtl n(\beta)$}\label{sec:definition}
  
The basic objects, $n$-diagrams, are first introduced. Draw two vertical lines, each with $n$ points on it, $n$ being a positive integer. 
Choose first $2m$ points, $0\leq m \leq n$ an integer, and put a $\circ$ on each of them. A point with a $\circ$ will be called a \emph{vacancy}. Now connect the remaining points, pairwise, with non-intersecting strings. The resulting object is called a {\em  dilute $n$-diagram}.
  
On the set of formal linear combinations of all dilute $n$-diagrams a product is defined by extending linearly the product of two $n$-diagrams obtained as follows. The two diagrams are put side by side, the inner borders and the points on them are identified, then removed. A string which no longer ties two points is called a \emph{floating string}. A floating string that closes on itself is called a \emph{closed loop}. If all floating strings are closed loops, the result of the product of the two dilute $n$-diagrams is then the diagram obtained by reading the vacancies on the left and right vertical lines and the strings between them multiplied by a factor of $\beta$ for each closed loop. Otherwise, the product is the zero element of the algebra. 
  
The three following products give examples of these definitions. The second contains two floating strings that are not closed and the product is therefore zero, and the third has one closed floating string leading to the factor $\beta$: \newline
  
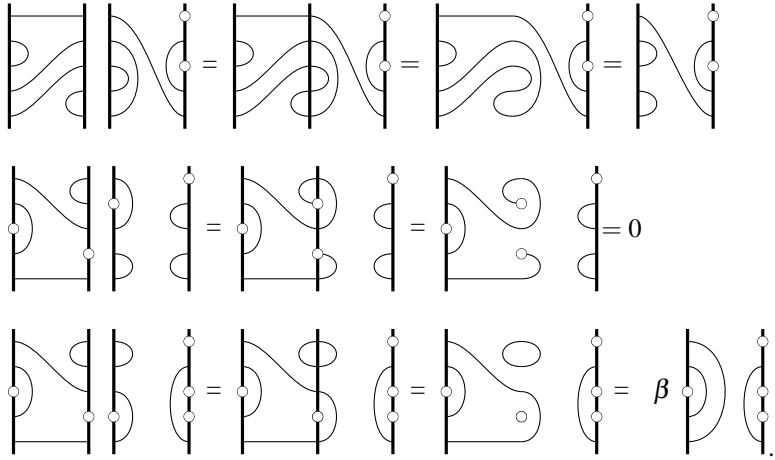
\begin{tikzpicture}[scale = 1/3]
	\bord{0}{5};
	\bord{3}{5};
	\bord{4}{5};
	\bord{7}{5};
	\link{0,0}{1,0}{2,2}{3,2};
	\link{0,1}{1,1}{2,3}{3,3};
	\link{0,4}{1,4}{2,4}{3,4};
	\link{0,2}{1,2}{1,3}{0,3};
	\link{3,0}{2,0}{2,1}{3,1};
	\link{4,0}{11/2,0}{11/2,3}{4,3};
	\link{4,1}{5,1}{5,2}{4,2};
	\link{4,4}{11/2,4}{6,0}{7,0};
	\link{7,1}{6,1}{6,3}{7,3};
	\vac{7,2};
	\vac{7,4};
	\node at (8,2) { = };
\end{tikzpicture}
\begin{tikzpicture}[scale = 1/3]
	\bord{0}{5};
	\bord{3}{5};
	\bord{6}{5};
	\link{0,0}{1,0}{2,2}{3,2};
	\link{0,1}{1,1}{2,3}{3,3};
	\link{0,4}{1,4}{2,4}{3,4};
	\link{0,2}{1,2}{1,3}{0,3};
	\link{3,0}{2,0}{2,1}{3,1};
	\link{3,0}{9/2,0}{9/2,3}{3,3};
	\link{3,1}{4,1}{4,2}{3,2};
	\link{3,4}{9/2,4}{5,0}{6,0};
	\link{6,1}{5,1}{5,3}{6,3};
	\vac{6,2};
	\vac{6,4};
	\node at (7,2) { $=$ };
\end{tikzpicture}
\begin{tikzpicture}[scale = 1/3]
	\bord{0}{5};
	\bord{6}{5};
	\link{0,0}{1,0}{2,2}{3,2};
	\link{0,1}{1,1}{2,3}{3,3};
	\link{0,4}{1,4}{2,4}{3,4};
	\link{0,2}{1,2}{1,3}{0,3};
	\link{3,0}{2,0}{2,1}{3,1};
	\link{3,0}{9/2,0}{9/2,3}{3,3};
	\link{3,1}{4,1}{4,2}{3,2};
	\link{3,4}{9/2,4}{5,0}{6,0};
	\link{6,1}{5,1}{5,3}{6,3};
	\vac{6,2};
	\vac{6,4};
	\node at (7,2) {$=$};
	\bord{8}{5};
	\bord{11}{5};
	\link{8,0}{9,0}{9,1}{8,1};
	\link{8,2}{9,2}{9,3}{8,3};
	\link{8,4}{9,4}{10,0}{11,0};
	\link{11,1}{10,1}{10,3}{11,3};
	\vac{11,2};
	\vac{11,4};
\end{tikzpicture} \newline
  
\begin{tikzpicture}[scale = 1/3]
	\bord{0}{5};
	\bord{3}{5};
	\bord{4}{5};
	\bord{7}{5};
	\defect{0,0};
	\link{0,1}{1,1}{1,3}{0,3};
	\vac{0,2};
	\link{0,4}{1,4}{2,2}{3,2};
	\vac{3,1};
	\link{3,3}{2,3}{2,4}{3,4};
	\link{4,0}{5,0}{5,1}{4,1};
	\link{4,2}{5,2}{5,4}{4,4};
	\vac{4,3};
	\vac{7,4};
	\link{7,0}{6,0}{6,1}{7,1};
	\link{7,2}{6,2}{6,3}{7,3};
	\node at (8,2) { = };
\end{tikzpicture}
\begin{tikzpicture}[scale = 1/3]
	\bord{0}{5};
	\bord{3}{5};
	\bord{6}{5};
	\defect{0,0};
	\link{0,1}{1,1}{1,3}{0,3};
	\vac{0,2};
	\link{0,4}{1,4}{2,2}{3,2};
	\link{3,3}{2,3}{2,4}{3,4};
	\link{3,0}{4,0}{4,1}{3,1};
	\vac{3,1};
	\link{3,2}{4,2}{4,4}{3,4};
	\vac{3,3};
	\vac{6,4};
	\link{6,0}{5,0}{5,1}{6,1};
	\link{6,2}{5,2}{5,3}{6,3};
	\node at (7,2) { = };
\end{tikzpicture}
\begin{tikzpicture}[scale = 1/3]
	\bord{0}{5};
	\bord{6}{5};
	\defect{0,0};
	\link{0,1}{1,1}{1,3}{0,3};
	\vac{0,2};
	\link{0,4}{1,4}{2,2}{3,2};
	\link{3,3}{2,3}{2,4}{3,4};
	\link{3,0}{4,0}{4,1}{3,1};
	\vac{3,1};
	\link{3,2}{4,2}{4,4}{3,4};
	\vac{3,3};
	\vac{6,4};
	\link{6,0}{5,0}{5,1}{6,1};
	\link{6,2}{5,2}{5,3}{6,3};
	\node at (7,2) { $=$ $0$ };
\end{tikzpicture}\newline

\begin{tikzpicture}[scale = 1/3]
	\bord{0}{5};
	\bord{3}{5};
	\bord{4}{5};
	\bord{7}{5};
	\defect{0,0};
	\link{0,1}{1,1}{1,3}{0,3};
	\vac{0,2};
	\link{0,4}{1,4}{2,2}{3,2};
	\vac{3,1};
	\link{3,3}{2,3}{2,4}{3,4};
	\link{4,3}{5,3}{5,4}{4,4};
	\link{4,0}{5,0}{5,2}{4,2};
	\vac{4,1};
	\link{7,0}{6,0}{6,3}{7,3};
	\vac{7,1};
	\vac{7,2};
	\vac{7,4};
	\node at (8,2) { = };
\end{tikzpicture}
\begin{tikzpicture}[scale = 1/3]
	\bord{0}{5};
	\bord{3}{5};
	\bord{6}{5};
	\defect{0,0};
	\link{0,1}{1,1}{1,3}{0,3};
	\vac{0,2};
	\link{0,4}{1,4}{2,2}{3,2};
	\vac{3,1};
	\link{3,3}{2,3}{2,4}{3,4};
	\link{3,3}{4,3}{4,4}{3,4};
	\link{3,0}{4,0}{4,2}{3,2};
	\vac{3,1};
	\link{6,0}{5,0}{5,3}{6,3};
	\vac{6,1};
	\vac{6,2};
	\vac{6,4};
	\node at (7,2) { = };
\end{tikzpicture}
\begin{tikzpicture}[scale = 1/3]
	\bord{0}{5};
	\bord{6}{5};
	\defect{0,0};
	\link{0,1}{1,1}{1,3}{0,3};
	\vac{0,2};
	\link{0,4}{1,4}{2,2}{3,2};
	\vac{3,1};
	\link{3,3}{2,3}{2,4}{3,4};
	\link{3,3}{4,3}{4,4}{3,4};
	\link{3,0}{4,0}{4,2}{3,2};
	\vac{3,1};
	\link{6,0}{5,0}{5,3}{6,3};
	\vac{6,1};
	\vac{6,2};
	\vac{6,4};
	\node at (7,2) { = };
\end{tikzpicture}
\begin{tikzpicture}[scale = 1/3]
	\bord{1}{5};
	\bord{4}{5};
	\node at (0,2) {$\beta$};
	
	\link{1,0}{3,0}{3,4}{1,4};
	\link{1,1}{2,1}{2,3}{1,3};
	\vac{1,2};
	\link{4,0}{3,0}{3,3}{4,3};
	\vac{4,1};
	\vac{4,2};
	\vac{4,4};
\end{tikzpicture}.\newline
A dashed string represents the formal sum of two diagrams: one where the points are linked by a regular string, and one where the points are both vacancies. For example,
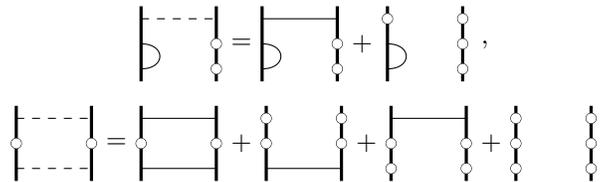
\begin{equation*}
\begin{tikzpicture}[baseline={(current bounding box.center)},scale=1/3]
	\bord{0}{3};
	\bord{3}{3};
	\link{0,0}{1,0}{1,1}{0,1};
	\wave{0}{2}
	\vac{3,0};
	\vac{3,1};
	\node at (4,1) {$=$};
\end{tikzpicture}
\begin{tikzpicture}[baseline={(current bounding box.center)},scale=1/3]
	\bord{0}{3};
	\bord{3}{3};
	\link{0,0}{1,0}{1,1}{0,1};
	\defect{0,2};
	\vac{3,0};
	\vac{3,1};
	\node at (4,1) {$+$};
	\bord{5}{3};
	\bord{8}{3};
	\link{5,0}{6,0}{6,1}{5,1};
	\vac{5,2};
	\vac{8,2};
	\vac{8,0};
	\vac{8,1};
\end{tikzpicture}\ \ ,
\end{equation*}
\begin{equation*}
\begin{tikzpicture}[baseline={(current bounding box.center)},scale=1/3]
	\bord{0}{3};
	\bord{3}{3};
	\wave{0}{0}
	\wave{0}{2}
	\vac{0,1};
	\vac{3,1};
	\node at (4,1) {$=$};
\end{tikzpicture}
\begin{tikzpicture}[baseline={(current bounding box.center)},scale=1/3]
	\bord{0}{3};
	\bord{3}{3};
	\defect{0,0};
	\defect{0,2};
	\vac{0,1};
	\vac{3,1};
	\node at (4,1) {$+$};
\end{tikzpicture}
\begin{tikzpicture}[baseline={(current bounding box.center)},scale=1/3]
	\bord{0}{3};
	\bord{3}{3};
	\defect{0,0};
	\vac{0,2};
	\vac{3,2};
	\vac{0,1};
	\vac{3,1};
	\node at (4,1) {$+$};
\end{tikzpicture}
\begin{tikzpicture}[baseline={(current bounding box.center)},scale=1/3]
	\bord{0}{3};
	\bord{3}{3};
	\vac{0,0};
	\vac{3,0};
	\defect{0,2};
	\vac{0,1};
	\vac{3,1};
	\node at (4,1) {$+$};
\end{tikzpicture}
\begin{tikzpicture}[baseline={(current bounding box.center)},scale=1/3]
	\bord{0}{3};
	\bord{3}{3};
	\vac{0,0};
	\vac{3,0};
	\vac{0,2};
	\vac{3,2};
	\vac{0,1};
	\vac{3,1};
\end{tikzpicture}\ \ .
\end{equation*}
Note that the diagram where each point is linked by a dashed line to the corresponding point on the opposite side acts as the identity on all dilute $n$-diagrams. It is a sum of $2^n$ diagrams. For example, when $n=3$
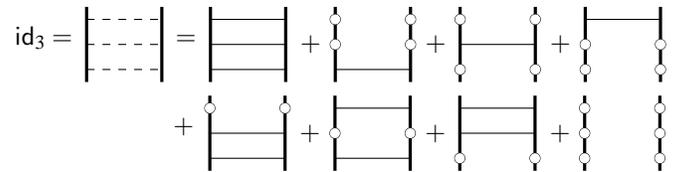
\begin{align*}
\idtl_{3}=\ \begin{tikzpicture}[baseline={(current bounding box.center)},scale=1/3]	
	\bord{0}{3};
	\bord{3}{3};
	\wave{0}{0}
	\wave{0}{1}
	\wave{0}{2}
\end{tikzpicture}
&\ =\  
\begin{tikzpicture}[baseline={(current bounding box.center)},scale=1/3]	
	\bord{0}{3};
	\bord{3}{3};
	\defect{0,0};
	\defect{0,1};
	\defect{0,2};
	\node at (4,1) {$+$};	
\end{tikzpicture}
\begin{tikzpicture}[baseline={(current bounding box.center)},scale=1/3]	
	\bord{0}{3};
	\bord{3}{3};
	\defect{0,0};
	\vac{0,1};
	\vac{3,1};
	\vac{0,2};
	\vac{3,2};
	\node at (4,1) {$+$};	
\end{tikzpicture}
\begin{tikzpicture}[baseline={(current bounding box.center)},scale=1/3]	
	\bord{0}{3};
	\bord{3}{3};
	\defect{0,1};
	\vac{0,0};
	\vac{3,0};
	\vac{0,2};
	\vac{3,2};
	\node at (4,1) {$+$};
\end{tikzpicture}
\begin{tikzpicture}[baseline={(current bounding box.center)},scale=1/3]	
	\bord{0}{3};
	\bord{3}{3};
	\defect{0,2};
	\vac{0,1};
	\vac{3,1};
	\vac{0,0};
	\vac{3,0};
\end{tikzpicture}\\
&\ + \ 
\begin{tikzpicture}[baseline={(current bounding box.center)},scale=1/3]	
	\bord{0}{3};
	\bord{3}{3};
	\defect{0,0};
	\defect{0,1};
	\vac{0,2};
	\vac{3,2};
	\node at (4,1) {$+$};	
\end{tikzpicture}
\begin{tikzpicture}[baseline={(current bounding box.center)},scale=1/3]	
	\bord{0}{3};
	\bord{3}{3};
	\defect{0,0};
	\defect{0,2};
	\vac{0,1};
	\vac{3,1};
	\node at (4,1) {$+$};	
\end{tikzpicture}
\begin{tikzpicture}[baseline={(current bounding box.center)},scale=1/3]	
	\bord{0}{3};
	\bord{3}{3};
	\defect{0,2};
	\defect{0,1};
	\vac{0,0};
	\vac{3,0};
	\node at (4,1) {$+$};	
\end{tikzpicture}
\begin{tikzpicture}[baseline={(current bounding box.center)},scale=1/3]	
	\bord{0}{3};
	\bord{3}{3};
	\vac{0,0};
	\vac{3,0};
	\vac{0,1};
	\vac{3,1};
	\vac{0,2};
	\vac{3,2};
	\node at (4,1) {.};
\end{tikzpicture}
\end{align*}
Note finally that the product is clearly associative: the reading of how the left and right sides are connected in a product of three diagrams is blind to the order of glueing, and so is the number of closed loops. The set of $n$-diagrams with the formal sum and product just introduced is the dilute Temperley-Lieb algebra $\dtl{n}=\dtl{n}(\beta)$. We also define $\dtl 0=\mathbb C$. When the parameter $\beta$ is chosen to be a formal one, then the algebra is over $\mathbb C[\beta]$. We shall be interested mostly in the case $\beta\in\mathbb C$ for which the algebra is over $\mathbb C$.
  
Several generating sets for $\dtl n$ can be found. For instance, the set
$\{ a_{i}, a^{t}_{i}, b_{i}, b^{t}_{i}, e_{j}, x_{j},\allowbreak i \in [1,n-1], j\in [1,n]\}$ where\newline
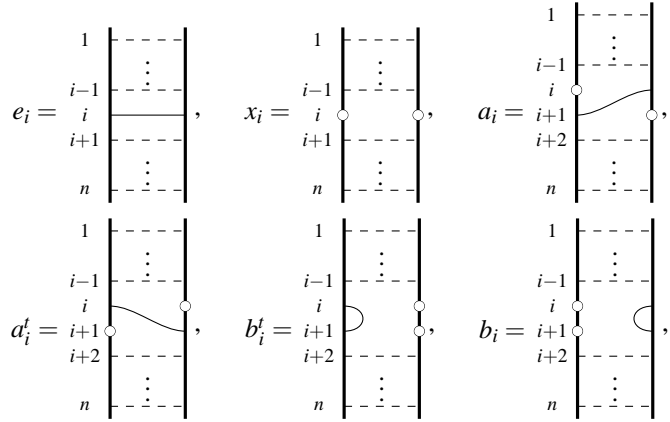
\begin{equation*}
\begin{tikzpicture}[scale=1/3]
	\bord{0}{7};
	\bord{3}{7};
	\defect{0,3};
	\wave{0}{4}	
	\wave{0}{2}
	\node at (3/2,5) {$\vdots$};
	\node at (3/2,1) {$\vdots$};
	\wave{0}{0}
	\wave{0}{6}
	\node at (-1,3) {$\scriptstyle{i}$};
	\node at (-1,6) {$\scriptstyle{1}$};
	\node at (-1,4) {$\scriptstyle{i-1}$};
	\node at (-1,2) {$\scriptstyle{i+1}$};
	\node at (-1,0) {$\scriptstyle{n}$};
	\node at (-3,3) {$e_{i} =$};
	\node at (3.5,3) {,};
\end{tikzpicture} \quad 
\begin{tikzpicture}[scale=1/3]
	\bord{0}{7};
	\bord{3}{7};
	\vac{0,3};
	\vac{3,3};
	\wave{0}{4}	
	\wave{0}{2}
	\node at (3/2,5) {$\vdots$};
	\node at (3/2,1) {$\vdots$};
	\wave{0}{0}
	\wave{0}{6}
	\node at (-1,3) {$\scriptstyle{i}$};
	\node at (-1,6) {$\scriptstyle{1}$};
	\node at (-1,4) {$\scriptstyle{i-1}$};
	\node at (-1,2) {$\scriptstyle{i+1}$};
	\node at (-1,0) {$\scriptstyle{n}$};
	\node at (-3,3) {$x_{i} =$};
	\node at (3.5,3) {,};
\end{tikzpicture} \quad 
\begin{tikzpicture}[scale=1/3]
	\bord{0}{8};
	\bord{3}{8};
	\link{0,3}{1,3}{2,4}{3,4};
	\vac{0,4};
	\vac{3,3};
	\wave{0}{5}	
	\wave{0}{2}
	\node at (3/2,6) {$\vdots$};
	\node at (3/2,1) {$\vdots$};
	\wave{0}{0}
	\wave{0}{7}
	\node at (-1,4) {$\scriptstyle{i}$};
	\node at (-1,7) {$\scriptstyle{1}$};
	\node at (-1,5) {$\scriptstyle{i-1}$};
	\node at (-1,3) {$\scriptstyle{i+1}$};
	\node at (-1,2) {$\scriptstyle{i+2}$};
	\node at (-1,0) {$\scriptstyle{n}$};
	\node at (-3,3) {$a_{i} =$};
	\node at (3.5,3) {,};
\end{tikzpicture}
\end{equation*}
\begin{equation*}
\begin{tikzpicture}[scale=1/3]
	\bord{0}{8};
	\bord{3}{8};
	\link{0,4}{1,4}{2,3}{3,3};
	\vac{0,3};
	\vac{3,4};
	\wave{0}{5}	
	\wave{0}{2}
	\node at (3/2,6) {$\vdots$};
	\node at (3/2,1) {$\vdots$};
	\wave{0}{0}
	\wave{0}{7}
	\node at (-1,4) {$\scriptstyle{i}$};
	\node at (-1,7) {$\scriptstyle{1}$};
	\node at (-1,5) {$\scriptstyle{i-1}$};
	\node at (-1,3) {$\scriptstyle{i+1}$};
	\node at (-1,2) {$\scriptstyle{i+2}$};
	\node at (-1,0) {$\scriptstyle{n}$};
	\node at (-3,3) {$a^{t}_{i} =$};
	\node at (3.5,3) {,};
\end{tikzpicture}\quad
\begin{tikzpicture}[scale=1/3]
	\bord{0}{8};
	\bord{3}{8};
	\link{0,4}{1,4}{1,3}{0,3};
	\vac{3,3};
	\vac{3,4};
	\wave{0}{5}	
	\wave{0}{2}
	\node at (3/2,6) {$\vdots$};
	\node at (3/2,1) {$\vdots$};
	\wave{0}{0}
	\wave{0}{7}
	\node at (-1,4) {$\scriptstyle{i}$};
	\node at (-1,7) {$\scriptstyle{1}$};
	\node at (-1,5) {$\scriptstyle{i-1}$};
	\node at (-1,3) {$\scriptstyle{i+1}$};
	\node at (-1,2) {$\scriptstyle{i+2}$};
	\node at (-1,0) {$\scriptstyle{n}$};
	\node at (-3,3) {$b^{t}_{i} =$};
	\node at (3.5,3) {,};
\end{tikzpicture}\quad
\begin{tikzpicture}[scale=1/3]
	\bord{0}{8};
	\bord{3}{8};
	\link{3,4}{2,4}{2,3}{3,3};
	\vac{0,3};
	\vac{0,4};
	\wave{0}{5}	
	\wave{0}{2}
	\node at (3/2,6) {$\vdots$};
	\node at (3/2,1) {$\vdots$};
	\wave{0}{0}
	\wave{0}{7}
	\node at (-1,4) {$\scriptstyle{i}$};
	\node at (-1,7) {$\scriptstyle{1}$};
	\node at (-1,5) {$\scriptstyle{i-1}$};
	\node at (-1,3) {$\scriptstyle{i+1}$};
	\node at (-1,2) {$\scriptstyle{i+2}$};
	\node at (-1,0) {$\scriptstyle{n}$};
	\node at (-3,3) {$b_{i} =$};
	\node at (3.5,3) {,};
\end{tikzpicture}
\end{equation*}
generates the algebra. However, they do not form a minimal set, as for all $1\le i\leq n$, $e_i + x_i = \idtl_{n}$. Through the identification $u_i = b^{t}_{i}b_{i}$, the connection with the regular $n$-diagram algebra $\tl{n}$ should be clear. A set of relations was proposed by Grimm \cite{Grimm} to define $\dtl n$ through generators and relations. The equivalence between the diagrammatic definition, the one used here, and that with relations is stated there without proof.
  
The numbers of vacancies on either side of a dilute $n$-diagram always have the same parity. If these numbers are even (odd), the diagram will be called an even (odd) diagram. The subset spanned by only even (odd) diagrams is closed under the product and this subalgebra will be called the even (odd) dilute Temperley-Lieb subalgebra, denoted by $\edtl{n}$ ($\odtl{n}$).  Clearly any dilute $n$-diagram is either even or odd. Since the product of two diagrams of distinct parities is zero, it is clear that the even and odd subalgebras are two-sided ideals of $\dtl{n}$ and
$$\dtl n=\edtl n\oplus \odtl n.$$
For example \newline
\begin{align}
& \begin{tikzpicture}[baseline={(current bounding box.center)},scale=1/3]
\node at (-2,1/2) {$\dtl{2} \simeq \textrm{span}\Big\{ $};
\end{tikzpicture}
\begin{tikzpicture}[baseline={(current bounding box.center)},scale=1/3]
	\bord{0}{2};
	\bord{3}{2};
	\defect{0,0};
	\defect{0,1};
	\node at (4,0) {, };
\end{tikzpicture}
\begin{tikzpicture}[baseline={(current bounding box.center)},scale=1/3]
	\bord{0}{2};
	\bord{3}{2};
	\link{0,0}{1,0}{1,1}{0,1};
	\link{3,0}{2,0}{2,1}{3,1};
	\node at (4,0) {, };
\end{tikzpicture}
\begin{tikzpicture}[baseline={(current bounding box.center)},scale=1/3]
	\bord{0}{2};
	\bord{3}{2};
	\link{0,0}{1,0}{1,1}{0,1};
	\vac{3,0};
	\vac{3,1};
	\node at (4,0) {, };
\end{tikzpicture}
\begin{tikzpicture}[baseline={(current bounding box.center)},scale=1/3]
	\bord{0}{2};
	\bord{3}{2};
	\vac{0,0};
	\vac{0,1};
	\link{3,0}{2,0}{2,1}{3,1};
	\node at (4,0) {, };
\end{tikzpicture}
\begin{tikzpicture}[baseline={(current bounding box.center)},scale=1/3]
	\bord{0}{2};
	\bord{3}{2};
	\vac{0,0};
	\vac{0,1};
	\vac{3,0};
	\vac{3,1};
\end{tikzpicture}
\begin{tikzpicture}[baseline={(current bounding box.center)},scale=1/3]
\node at (0,1/2) {$\Big\}$};
\end{tikzpicture}\notag \\ 
& \ \qquad\quad \begin{tikzpicture}[baseline={(current bounding box.center)},scale=1/3]
\node at (0,1/2) {$\oplus\ \ \textrm{span}\Big\{$};
\end{tikzpicture}
\begin{tikzpicture}[baseline={(current bounding box.center)},scale=1/3]
	\bord{0}{2};
	\bord{3}{2};
	\vac{0,1};
	\defect{0,0};
	\vac{3,1};
	\node at (4,0) {, };
\end{tikzpicture}
\begin{tikzpicture}[baseline={(current bounding box.center)},scale=1/3]
	\bord{0}{2};
	\bord{3}{2};
	\vac{0,1};
	\link{0,0}{1,0}{2,1}{3,1};
	\vac{3,0};
	\node at (4,0) {, };
\end{tikzpicture}
\begin{tikzpicture}[baseline={(current bounding box.center)},scale=1/3]
	\bord{0}{2};
	\bord{3}{2};
	\vac{0,0};
	\link{0,1}{1,1}{2,0}{3,0};
	\vac{3,1};
	\node at (4,0) {, };
\end{tikzpicture}
\begin{tikzpicture}[baseline={(current bounding box.center)},scale=1/3]
	\bord{0}{2};
	\bord{3}{2};
	\vac{0,0};
	\defect{0,1};
	\vac{3,0};
\end{tikzpicture}
\begin{tikzpicture}[baseline={(current bounding box.center)},scale=1/3]
\node at (0,1/2) {$\Big\}$};
\end{tikzpicture}\ .
\end{align}
\newline
The unit $\idtl\in \dtl{n}$ decomposes into $\idtl=\eid+\oid$ with $\eid\in\edtl n$ and $\oid\in\odtl n$. The odd and even units are orthogonal idempotents: $\eid^2=\eid$, $\oid^2=\oid$ and $\oid\cdot \eid= \eid\cdot\oid=0$. (On the previous example of $\idtl_3$, the four $3$-diagrams of the first line of the rhs form $\eid$ and the last line is $\oid$.) Let $M$ be a $\dtl n$-module and decompose it, as vector space, into $M=\eid\cdot M\oplus\oid\cdot M$. Clearly $\eid\cdot(\oid\cdot M)=0$ and $\oid\cdot(\eid\cdot M)=0$. But $a=a\cdot \eid$ for any $a\in\edtl n$ and therefore $\edtl n$ acts trivially on $\oid\cdot M$ and, similarly, so does $\odtl n$ on $\eid\cdot M$. The decomposition into a direct sum of subspaces is thus a direct sum of modules. The two summands $\oid\cdot M$ and $\eid\cdot M$ will be called the odd and even submodules of $M$. If the odd submodule of $M$ is trivial, $M$ will be said to be {\em even} and vice versa. An indecomposable module $M$ is either odd or even.

The dilute algebra $\dtl n$ can filtered by a family of ideals defined diagrammatically. Let a {\em crossing string} be a string in an $n$-diagram that ties a point on the left vertical line to one on the right and, for each $n$-diagram $a\in\dtl n$, define the integer $c=c(a)$ with $0\le c\le n$ to be its number of crossing strings. The diagrammatic definition of the multiplication in $\dtl n$ implies that $c(ab)\le \min(c(a),c(b))$ for all pairs of $n$-diagrams $a$ and $b$ for which $ab\neq0$.  
Therefore the linear span $I_k\subset \dtl n$ of all $n$ diagrams $a$ such that $c(a)\le k$ is an ideal of $\dtl n$ and
\begin{equation}\label{eq:filtrationByIdeals}
0\subset I_0\subset I_1\subset \dots\subset I_n=\dtl n.
\end{equation}
  
Consider now $S_{n}$, the subset of $\dtl{n}$ spanned by dilute $n$-diagrams having symmetric vacancies, that is, a position on one of their sides is a vacancy if and only if it is also on their other side. Multiplying two symmetric $n$-diagrams gives either zero if the vacancies do not match perfectly or is a symmetric diagram. The subset $S_n$ is therefore a subalgebra of $\dtl{n}$. Now, choose a subset $A\subset\{1,2,\dots, n\}$ of $\iota$ integers and define $\pi_A=\prod _{i\in A\text{, } j\notin A} x_j e_i$. Note that $\pi_A^2=\pi_A$ and thus $\pi_A(\dtl{n})\pi_A$ is a subalgebra of $S_n$. It is spanned by all $n$-diagrams with links starting and ending at positions labeled by $A$ and vacancies at all other positions. Therefore $\pi_A(\dtl{n})\pi_A$ is isomorphic to $\tl{\iota}$ and any $n$-diagram in $S_n$ belongs to precisely one of these subalgebras. For a given $\iota$, there are $\binom n\iota$ distinct such subalgebras in $S_n$, all isomorphic to $\tl{\iota}$.  Finally, since the product of two diagrams with different vacancies is always zero, it follows that $S_{n}$ is isomorphic to the direct sum of all subalgebras $\pi_A(\dtl{n})\pi_A$ obtained from subsets of $A\subset \{1,2,\dots, n\}$. We arrive at the following proposition.
\begin{Prop}\label{prop.symmetric.salg}The subalgebra $S_n\subset \dtl n$ is isomorphic to
\begin{equation} 
S_{n} \simeq \bigoplus_{0\leq \iota\leq n}\Big(\bigoplus_{1\leq p\leq\binom{n}{\iota} }\tl{\iota} \Big)
\end{equation}
where $\tl{0}=\mathbb C$.  
\end{Prop}
\end{subsection}
  
\begin{subsection}{The dimension of $\dtl{n}$}\label{sec:dimension}
  
The ressemblance with the Temperley-Lieb algebra $\tl n$ provides a fairly straightforward method to obtain the dimension of $\dtl{n}$. In fact, the same technique of ``slicing diagrams" can be used here. The procedure goes as follows: first, take a dilute $n$-diagram and rotate its right side so that it sits below its left side, stretching the strings so that the points remains connected. Second, connect the two-sides together. For example,
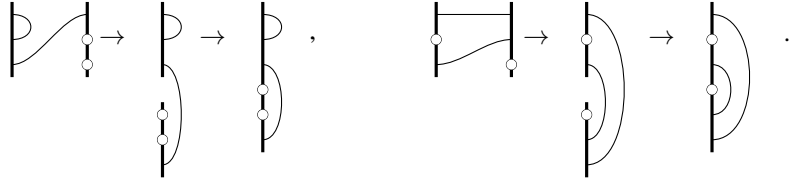
\begin{equation*}
\begin{tikzpicture}[scale = 1/3]
	\bord{0}{3};
	\bord{3}{3};
	\link{0,0}{1,0}{2,2}{3,2};
	\link{0,1}{1,1}{1,2}{0,2};
	\vac{3,0};
	\vac{3,1};
	\node at (4,1) {$\rightarrow$};
	\bord{6}{3};
	\draw[\epais] (6,-3/2) -- (6,-9/2);
	\link{6,1}{7,1}{7,2}{6,2};
	\link{6,0}{7,0}{7,-4}{6,-4};
	\vac{6,-2};
	\vac{6,-3};
	\node at (8,1) {$\rightarrow$};
	\draw[\epais] (10,5/2) -- (10,-7/2);
	\link{10,1}{11,1}{11,2}{10,2};
	\link{10,0}{11,0}{11,-3}{10,-3};
	\vac{10,-1};
	\vac{10,-2};
	\node at (12,1) {,};
\end{tikzpicture}
\qquad \qquad
\begin{tikzpicture}[scale=1/3]
	\bord{0}{3};
	\bord{3}{3};
	\link{0,0}{1,0}{2,1}{3,1};
	\link{0,2}{1,2}{2,2}{3,2};
	\vac{3,0};
	\vac{0,1};
	\node at (4,1) {$\rightarrow$};
	\bord{6}{3};
	\draw[\epais] (6,-3/2) -- (6,-9/2);
	\link{6,2}{8,2}{8,-4}{6,-4};
	\link{6,0}{7,0}{7,-3}{6,-3};
	\vac{6,-2};
	\vac{6,1};
	\node at (9,1) {$\rightarrow$};
	\draw[\epais] (11,5/2) -- (11,-7/2);
	\link{11,2}{13,2}{13,-3}{11,-3};
	\link{11,0}{12,0}{12,-2}{11,-2};
	\vac{11,1};
	\vac{11,-1};
	\node at (14,1) {.};
\end{tikzpicture}
\end{equation*}
Now, consider the dilute $n$-diagrams whose vacancies are all at the same places, apply the procedure, then remove the points where the vacancies are. For $n=3$ and two vacancies located as below, the result look like this:
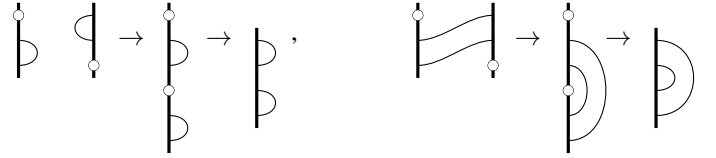
\begin{equation*}
\begin{tikzpicture}[scale = 1/3]
	\bord{0}{3};
	\bord{3}{3};
	\link{0,0}{1,0}{1,1}{0,1};
	\link{3,1}{2,1}{2,2}{3,2};
	\vac{0,2};
	\vac{3,0};
	\node at (4.5,1) { $\rightarrow$};
	\draw[\epais] (6,-7/2) -- (6,5/2);
	\link{6,0}{7,0}{7,1}{6,1};
	\link{6,-2}{7,-2}{7,-3}{6,-3};
	\vac{6,2};
	\vac{6,-1};
	\node at (8,1) {$\rightarrow$};
	\draw[\epais] (9.5,-5/2) -- (9.5,3/2);
	\link{9.5,-2}{10.5,-2}{10.5,-1}{9.5,-1};
	\link{9.5,1}{10.5,1}{10.5,0}{9.5,0};
	\node at (11,1) {, };
\end{tikzpicture}
\qquad \qquad
\begin{tikzpicture}[scale = 1/3]
	\bord{0}{3};
	\bord{3}{3};
	\link{0,0}{1,0}{2,1}{3,1};
	\link{0,1}{1,1}{2,2}{3,2};
	\vac{0,2};
	\vac{3,0};
	\node at (4.4,1) { $\rightarrow$};
	\draw[\epais] (6,-7/2) -- (6,5/2);
	\link{6,0}{7,0}{7,-2}{6,-2};
	\link{6,1}{8,1}{8,-3}{6,-3};
	\vac{6,2};
	\vac{6,-1};
	\node at (8,1) {$\rightarrow$};
	\draw[\epais] (9.5,-5/2) -- (9.5,3/2);
	\link{9.5,0}{10.5,0}{10.5,-1}{9.5,-1};
	\link{9.5,1}{11.5,1}{11.5,-2}{9.5,-2};
	\node at (12,1) {.};
\end{tikzpicture}
\end{equation*}
One should recognize in the results two elements of the link basis of the standard module $\vs{4,0}$ of $\tl{4}$ or, in general, of the $\tl{2m}$-module $\vs{2m,0}$ with no defects. (See section \ref{sec:modules} for a formal definition of links and standard modules for $\dtl n$ and also appendix \ref{app.tln} for their $\tl n$ analogues.)  By the reverse procedure just described, it was shown in \cite{RiSY} that $\dim \tl n=\dim \vs{2n,0}$. This leads to the following expression for the dimension of $\dtl n$.
\begin{Prop}The dimension of the associative algebra $\dtl n$ is
\begin{equation}
\dim  \dtl{n} = \sum _{k=0}^n \binom{2 n}{2 k} \dim \tl k = \sum _{k=0}^n \frac1{k+1}\binom{2k}k\binom{2 n}{2 k}.
\end{equation}
where $\tl 0=\mathbb C$.
\end{Prop}
\begin{proof}
Choose $2m\leq 2n$ positions and form the subset of dilute $n$-diagrams that have vacancies at (and only at) these fixed positions. The previous procedure applied to this subset will lead to the link basis of $\vs{2(n-m),0}$, irrespective of the chosen positions. Since there are $\binom{2 n}{2 m}$ different ways of choosing these positions, it follows that the space of dilute $n$-diagrams with $2m$ vacancies has dimension $\binom{2 n}{2 m} \dim \vs{2(n-m),0}$. The proof is completed by recalling that, for all $n$, $\dim \vs{2n,0}= \dim \tl{n}$.
\end{proof}
  
\noindent Motzkin numbers $M_n, n\geq 0,$ are defined as the number of ways of drawing any number of nonintersecting chords joining $n$ (labeled) points on a circle. The first Motzkin numbers are:
$$	1, 1, 2, 4, 9, 21, 51, 127, 323, 835, 2188, 5798, 15511, 41835, 113634, 310572, 853467, \dots $$
Clearly each $n$-diagram of $\dtl n$ with its $2n$ points leads to such a drawing of non-inter\-secting chords on a circle with $2n$ points and vice versa. The dimension of $\dtl n$ is thus the Motzkin number $M_{2n}$ and, for example, $\dim \dtl{8}=M_{16}=853467$.

\end{subsection}
\end{section}

%
\begin{section}{Left (and right) $\dtl n$-modules}\label{sec:modules}
%
  
This section introduces some of the basic modules over the dilute Temperley-Lieb algebra $\dtl n$: the link modules $\LM n$ and then the standard modules $\U{n,k}$. The latter will turn out to form a complete set of non-isomorphic irreducible modules when $q$ is not a root of unity. They will also be seen in subsection \ref{sec:cellalgebra} to be the cell modules that are naturally defined for cellular algebras. Some of their properties will be proved here. The modules $\U{n,k}$ are cyclic and indecomposable, their dimensions can be computed, and both their restriction to $\dtl {n-1}$ and induction to $\dtl {n+1}$ satisfy short exact sequences.
  
\begin{subsection}{The link modules $\LM n$ and $\TM{n,k}$}
A left (right) $n$-link diagram, with $n\geq 1$, is built in the following way. First, take a dilute $n$-diagram and remove its right (left) side as well as the points that were on it. An object, whether it is a string or a vacancy that no longer touches any point, is simply removed. The other floating strings are straightened out and called \emph{defects}. For example,
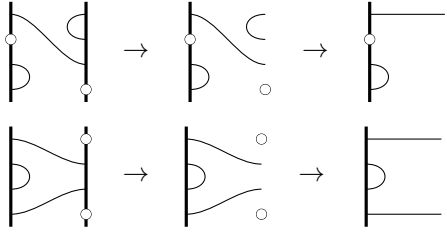
\begin{alignat*}{3}
\begin{tikzpicture}[scale = 1/3]
	\bord{0}{4};
	\bord{3}{4};	
	\link{0,3}{1,3}{2,1}{3,1};
	\link{0,0}{1,0}{1,1}{0,1};
	\link{3,2}{2,2}{2,3}{3,3};
	\vac{0,2};
	\vac{3,0};
	\node at (5,1.5) {$\rightarrow$     };
\end{tikzpicture}&\quad
\begin{tikzpicture}[scale=1/3]
	\bord{0}{4};
	\link{0,3}{1,3}{2,1}{3,1};
	\link{0,0}{1,0}{1,1}{0,1};
	\link{3,2}{2,2}{2,3}{3,3};
	\vac{0,2};
	\vac{3,0};
	\node at (5,1.5) {$\rightarrow$     };
\end{tikzpicture}& \quad
\begin{tikzpicture}[scale=1/3]
	\bord{0}{4};
	\defect{0,3};
	\link{0,0}{1,0}{1,1}{0,1};
	\vac{0,2};
	\node at (4,1.5) {\ };
\end{tikzpicture}\\
\begin{tikzpicture}[scale=1/3]
	\bord{0}{4};
	\bord{3}{4};
	\link{0,3}{1,3}{2,2}{3,2};
	\link{0,2}{1,2}{1,1}{0,1};
	\link{0,0}{1,0}{2,1}{3,1};
	\vac{3,3};
	\vac{3,0};
	\node at (5,1.5) {$\rightarrow$     };
\end{tikzpicture}&\quad
\begin{tikzpicture}[scale=1/3]
	\bord{0}{4};
	\link{0,3}{1,3}{2,2}{3,2};
	\link{0,2}{1,2}{1,1}{0,1};
	\link{0,0}{1,0}{2,1}{3,1};
	\vac{3,3};
	\vac{3,0};
	\node at (5,1.5) {$\rightarrow$     };
\end{tikzpicture}&\quad
\begin{tikzpicture}[scale=1/3]
	\bord{0}{4};
	\defect{0,3};
	\link{0,2}{1,2}{1,1}{0,1};
	\defect{0,0};
	\node at (4,1.5) {.};
\end{tikzpicture}
\end{alignat*}
The resulting diagram is called a {\em left $n$-link} ({\em right $n$-link}). It is seen that a dilute $n$-diagram induces a unique pair of one left and one right $n$-link diagrams and that, given such a pair, there can be at most one $n$-diagram, if any, that could have induced them. It will thus be useful to denote an $n$-diagram by its induced $n$-links, which we will denote by $b= |l r|$, where $l$ ($r$) is the left (right) link diagram induced from $b$.  This notation can also be used for linear combinations of $n$-diagrams as in $b=|(l+j)r| + |uv|$ where $l, j, u$ are left $n$-links and $r, v$ right ones. If $u$ is a left link, then $\bar u$ will denote its (right) mirror image.

A natural action can be defined of $\dtl n$ on left (and right) $n$-link diagrams. We start with the left action. Draw the $n$-diagram on the left side of the left $n$-link, identify the points on its right side with those on the link and remove them. Each floating string that is not connected to the remaining side is removed and yields a factor $\beta$ if it is closed and zero if it touches a vacancy. If a floating string starting on the remaining side is connected to a defect in the $n$-link diagram, it becomes a defect. Finally, if a floating string contains two distinct defects of the original diagram, it is simply removed, as any remaining vacancies. The remaining drawing is the resulting $n$-link diagram, weighted by factors of $\beta$, one for each closed floating strings. For example
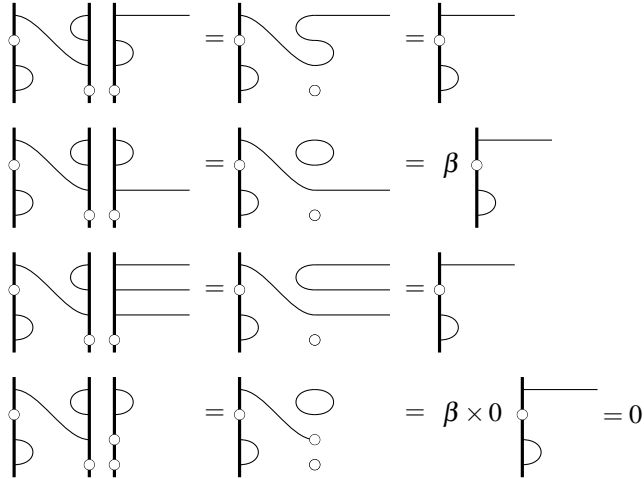
\begin{align*}
\begin{tikzpicture}[scale = 1/3]
	\bord{0}{4};
	\bord{3}{4};
	\link{0,3}{1,3}{2,1}{3,1};
	\link{0,0}{1,0}{1,1}{0,1};
	\link{3,2}{2,2}{2,3}{3,3};
	\vac{0,2};
	\vac{3,0};
	\bord{4}{4};
	\defect{4,3};
	\link{4,2}{5,2}{5,1}{4,1};
	\vac{4,0};
	\node at (8,2) {$=$ };
\end{tikzpicture}&
\begin{tikzpicture}[scale = 1/3]
	\bord{0}{4};
	\link{0,3}{1,3}{2,1}{3,1};
	\link{0,0}{1,0}{1,1}{0,1};
	\link{3,2}{2,2}{2,3}{3,3};
	\vac{0,2};
	\vac{3,0};
	\defect{3,3};
	\link{3,2}{4,2}{4,1}{3,1};
	\vac{3,0};
	\node at (7,2) {$=$ };
\end{tikzpicture}
\begin{tikzpicture}[scale = 1/3]
	\bord{0}{4};
	\defect{0,3};
	\link{0,0}{1,0}{1,1}{0,1};
	\vac{0,2};
\end{tikzpicture}\\
\begin{tikzpicture}[scale = 1/3]
	\bord{0}{4};
	\bord{3}{4};
	\link{0,3}{1,3}{2,1}{3,1};
	\link{0,0}{1,0}{1,1}{0,1};
	\link{3,2}{2,2}{2,3}{3,3};
	\vac{0,2};
	\vac{3,0};
	\bord{4}{4};
	\defect{4,1};
	\link{4,2}{5,2}{5,3}{4,3};
	\vac{4,0};
	\node at (8,2) {$=$ };
\end{tikzpicture}&
\begin{tikzpicture}[scale = 1/3]
	\bord{0}{4};
	\link{0,3}{1,3}{2,1}{3,1};
	\link{0,0}{1,0}{1,1}{0,1};
	\link{3,2}{2,2}{2,3}{3,3};
	\vac{0,2};
	\vac{3,0};
	\defect{3,1};
	\link{3,2}{4,2}{4,3}{3,3};
	\vac{3,0};
	\node at (7,2) {$=$   };
\end{tikzpicture}
\begin{tikzpicture}[scale = 1/3]
	\bord{0}{4};
	\node at (-1,2) {$\beta$};
	\defect{0,3};
	\link{0,0}{1,0}{1,1}{0,1};
	\vac{0,2};
\end{tikzpicture}\\
\begin{tikzpicture}[scale = 1/3]
	\bord{0}{4};
	\bord{3}{4};
	\link{0,3}{1,3}{2,1}{3,1};
	\link{0,0}{1,0}{1,1}{0,1};
	\link{3,2}{2,2}{2,3}{3,3};
	\vac{0,2};
	\vac{3,0};
	\bord{4}{4};
	\defect{4,3};
	\defect{4,2};
	\defect{4,1};
	\vac{4,0};
	\node at (8,2) {$=$ };
\end{tikzpicture}&
\begin{tikzpicture}[scale = 1/3]
	\bord{0}{4};
	\link{0,3}{1,3}{2,1}{3,1};
	\link{0,0}{1,0}{1,1}{0,1};
	\link{3,2}{2,2}{2,3}{3,3};
	\vac{0,2};
	\vac{3,0};
	\defect{3,3};
	\defect{3,2};
	\defect{3,1};
	\vac{3,0};
	\node at (7,2) {$=$ };
\end{tikzpicture}
\begin{tikzpicture}[scale = 1/3]
	\bord{0}{4};
	\defect{0,3};
	\link{0,0}{1,0}{1,1}{0,1};
	\vac{0,2};
\end{tikzpicture}\\
\begin{tikzpicture}[scale = 1/3]
	\bord{0}{4};
	\bord{3}{4};
	\link{0,3}{1,3}{2,1}{3,1};
	\link{0,0}{1,0}{1,1}{0,1};
	\link{3,2}{2,2}{2,3}{3,3};
	\vac{0,2};
	\vac{3,0};
	\bord{4}{4};
	\link{4,2}{5,2}{5,3}{4,3};
	\vac{4,1};
	\vac{4,0};
	\node at (8,2) {$=$ };
\end{tikzpicture}&
\begin{tikzpicture}[scale = 1/3]
	\bord{0}{4};
	\link{0,3}{1,3}{2,1}{3,1};
	\link{0,0}{1,0}{1,1}{0,1};
	\link{3,2}{2,2}{2,3}{3,3};
	\vac{0,2};
	\vac{3,0};
	\link{3,2}{4,2}{4,3}{3,3};
	\vac{3,1};
	\vac{3,0};
	\node at (7,2) {$=$   };
\end{tikzpicture}
\begin{tikzpicture}[scale = 1/3]
	\bord{0}{4};
	\node at (-2,2) {$\beta \times 0$};
	\defect{0,3};
	\link{0,0}{1,0}{1,1}{0,1};
	\vac{0,2};	
	\node at (4,2) {$= 0$};
\end{tikzpicture}
\end{align*}
This action can be extended linearly to any element of $\dtl n$. Let $\LM{n}$ be the vector space of all formal linear combinations of $n$-link diagrams. Again the above action can be extended linearly to any element of this space. This action is associative. (The connectivities of each floating string in $(ab)v$ and $a(bv)$, for $a,b\in\dtl n$ and $v\in \LM n$, are the same.) The vector space $\LM n$ is therefore a left $\dtl n$-module for this action. Right modules can be defined similarly by putting the elements of $\dtl n$ to the right of right $n$-links. A general element of $\LM n$ will be called a \emph{$n$-link state}. The modules $\LM n$ extends the link modules of the Temperley-Lieb algebra. One should note that, unlike for Temperley-Lieb link modules, the number of arcs in $\dtl n$-link modules can vary freely. However, as in Temperley-Lieb link modules, the action of an element of $\dtl n$ on a link diagram cannot increase its number of defects. The submodule of $\LM{n}$ spanned by $n$-link diagrams having at most $k$ defects is called $\TM{n,k}$, $0\le k\le n$, and these submodules $\TM{n,k}$ form a filtration of $\LM n$:
\begin{equation}\label{eq:filtrationOfLinks}\TM{n,0}\subset\TM{n,1}\subset \dots \subset \TM{n,n}=\LM n.
\end{equation}
The submodules $\TM{n,k}$ and the module $A_n$ will be called {\em link modules}.
\end{subsection}
  
  
\begin{subsection}{The standard modules $\U{n,k}$}\label{sec:standard}
  
The filtration \eqref{eq:filtrationOfLinks} leads to the definition of another family of left modules, obtained simply by the quotient of two consecutive link modules $\TM{n,k}$ and $\TM{n,k-1}$, namely:
$$\U{n,k} \simeq {\TM{n,k}}/{\TM{n,k-1}}, \textrm{\ for }1\le k\le n,\qquad\textrm{and}\qquad \U{n,0}=\TM{n,0}.$$
It will also be useful to set $\U{n,k}=\{0\}$ for integers $k\in\mathbb Z$ not in the set $\{0,1,\dots,n\}$. The left $\dtl n$-modules $\U{n,k}$ are called the {\em standard modules} and extend those of the Temperley-Lieb algebra. 
(In \cite{RiSY}, the standard modules of $\tl n$ were denoted by $\mathcal{V}_{n,p}$ where $p$ stands for the number of arcs. The number of defects is then $n-2p$, as there are no vacancies in the diagrammatic definition of $\tl n$. As noted before, the number of arcs is not constant in $\U{n,k}$. This explains the discrepancy in labelling between the present text and \cite{RiSY}. From now on, we shall use defects instead of arcs even for objects related to $\tl n$ and will translate results of \cite{RiSY} accordingly.)

By construction the number of defects is always conserved in $\U{n,k}$. More precisely, a basis of ${\TM{n,k}}/{\TM{n,k-1}}$ can be chosen to be the set of equivalence classes of $n$-links with precisely $k$ defects. If $v$ is such an $n$-link diagram, then the class $[v]\in {\TM{n,k}}/{\TM{n,k-1}}$ contains a unique $n$-link with $k$ defects and it is precisely $v$. For that reason we shall write $v$ for $[v]$.
  
As an example, the equivalence classes corresponding to the following $4$-links form a basis of $\U{4,2}$:
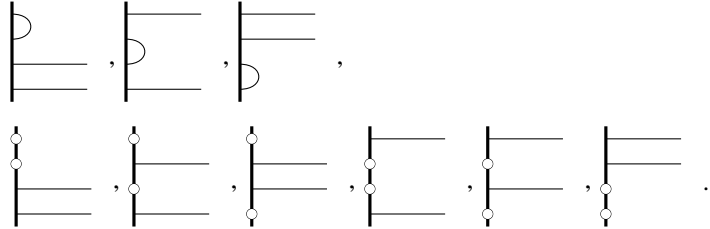
\begin{align*}
\ &\begin{tikzpicture}[scale=1/3]
	\bord{0}{4};
	\defect{0,0};
	\defect{0,1};
	\link{0,2}{1,2}{1,3}{0,3};
	\node at (4,1) {, };
\end{tikzpicture}
\begin{tikzpicture}[scale=1/3]
	\bord{0}{4};
	\defect{0,0};
	\defect{0,3};
	\link{0,1}{1,1}{1,2}{0,2};
	\node at (4,1) {, };
\end{tikzpicture}
\begin{tikzpicture}[scale=1/3]
	\bord{0}{4};
	\defect{0,2};
	\defect{0,3};
	\link{0,0}{1,0}{1,1}{0,1};
	\node at (4,1) {, };
\end{tikzpicture}
\\
\ &
\begin{tikzpicture}[scale=1/3]
	\bord{0}{4};
	\vac{0,3};
	\vac{0,2};
	\defect{0,1};
	\defect{0,0};
	\node at (4,1) {, };
\end{tikzpicture}
\begin{tikzpicture}[scale=1/3]
	\bord{0}{4};
	\vac{0,3};
	\vac{0,1};
	\defect{0,2};
	\defect{0,0};
	\node at (4,1) {, };
\end{tikzpicture}
\begin{tikzpicture}[scale=1/3]
	\bord{0}{4};
	\vac{0,3};
	\vac{0,0};
	\defect{0,1};
	\defect{0,2};
	\node at (4,1) {, };
\end{tikzpicture}
\begin{tikzpicture}[scale=1/3]
	\bord{0}{4};
	\vac{0,2};
	\vac{0,1};
	\defect{0,3};
	\defect{0,0};
	\node at (4,1) {, };
\end{tikzpicture}
\begin{tikzpicture}[scale=1/3]
	\bord{0}{4};
	\vac{0,2};
	\vac{0,0};
	\defect{0,3};
	\defect{0,1};
	\node at (4,1) {, };
\end{tikzpicture}
\begin{tikzpicture}[scale=1/3]
	\bord{0}{4};
	\vac{0,1};
	\vac{0,0};
	\defect{0,3};
	\defect{0,2};
	\node at (4,1) {. };
\end{tikzpicture}
\end{align*}
  
Note that if $n-k$ is even (odd), then only $\edtl{n}$ ($\odtl{n}$) can act non-trivially on it. That is, only an element of $\edtl{n}$ ($\odtl{n}$) may lead to a non-zero result. For that reason, the standard module $\U{n,k}$ has a given parity, that of the number $(n-k)$. Also, note that the number of vacancies on a link diagram restricts the elements of $\dtl{n}$ that can act non-trivially on it. For example, $n$-diagrams with more than $n-k$ vacancies on either of their sides act as zero on $\U{n,k}$.

Let $\Y{n,k}$ be the set of left $n$-links with precisely $k$ defects. By the previous discussion, it is clear that the set $\Y{n,k}$ is a basis of $\U{n,k}$ (or, more precisely, the set of equivalence classes $[y], y\in\Y{n,k}$, is). Moreover the set $\Y{n,k}, 0\le k\le n$, can be used to build a basis of $\dtl n$ itself. The glueing of left and right $n$-links described at the beginning of the section leads to an $n$-diagram if and only if their number of defects coincide. Note that the ideals appearing in the filtration \eqref{eq:filtrationByIdeals} are such that the quotient $I_k/I_{k-1}$ has, as a basis, the $n$-diagrams with precisely $k$ crossing strings. If $C:\sqcup_{0\le k\le n} \Y{n,k}\times\Y{n,k}\rightarrow \dtl n$ denotes the map that sends the pair $(x,y)$ of $n$-links with $k$ defects onto the $n$-diagram $|x\bar y|$, then the map $C$ is seen to be injective and its image is a basis of $\dtl n$.
  
It is also useful to define the subset $\X{n,k}\subset \Y{n,k}$ 
of $n$-links having precisely $k$ defects and $n-k$ vacancies. In general the subspace $\textrm{span\,}\X{n,k}$ {\em is not} a $\dtl{n}$-submodule, but it will be important for the analysis now to be carried.
  
Let $z\in \X{n,k}$, $u$ and $v$ be any left $n$-link diagrams in $\U{n,k}$. For the action in $\U{n,k}$, the element $|u \bar v|$ of $\dtl n$ acts as zero on $z$ unless $v$ and $z$ are equal. Similarly, if $|u\bar z| v$ is non-zero in $\U{n,k}$, then again $v$ and $z$ are equal.
(Note that this fails to be true if $v$ is a general link state and not a link diagram. We will see how this property generalizes to link states soon.) Note finally that, for all link states $u\in \U{n,k}$, $|u \bar z|z=u$. This property leads to the following result.
  
\begin{Prop}\label{prop:cyclic}
$\U{n,k}$ is cyclic, with any non-zero element of $\text{\em{span} }\X{n,k}$ being a generator.
\end{Prop}
\begin{proof}
The property just outlined means that any element $z$ in $\X{n,k}$ is a generator:  $\left(\dtl{n}\right)z = \U{n,k}$. Let $v$ be a non-zero element in $\textrm{span\,}\X{n,k}$. Since the elements of $\X{n,k}$ are linearly independent, $v\in \textrm{span }\X{n,k}$
has a non-zero component along some $n$-link $z$ and $|z\bar z|v$ is equal to $z$ up to a non-zero constant. Therefore $v$ is also a generator of $\U{n,k}$.
\end{proof}
  
This property is also used in the following propositions.
\begin{Prop}\label{prop:UnkIndecomposable}
$\U{n,k}$ is indecomposable.
\end{Prop}
\begin{proof}
Recall that, for any pair of $n$-link diagrams $u\in \U{n,k}$ and $z\in \X{n,k}$, $|z\bar z|u = z$ if $u=z$ and zero otherwise. So, suppose that $\U{n,k}\simeq A \oplus B$ for some submodules $A$ and $B$. Since $z$ generates the whole module, it cannot belong to either $A$ or $B$, unless one of them is trivial. There must be two non-zero link states $a\in A$ and $b\in B$ such that $z=a+b$, with $z=|z\bar z|z=|z\bar z|(a+b)=a'+b'$ with $a'=|z\bar z|a\in A$ and $b'=|z\bar z|b\in B$. If $a'$ is zero, then $b'=z\in B$ and $B=\U{n,k}$ and $A=\{0\}$. If $a'$ is not zero, it must have a non-zero component along $z$ in the basis of $n$-links. Therefore $a'=|z\bar z|a=\alpha z$ for some $\alpha\in\mathbb C^\times$. Again this implies that $A=\U{n,k}$ and $B=\{0\}$. So $\U{n,k}$ is indecomposable.
\end{proof}
\begin{Prop}\label{prop:nonIsomorphe}
$\U{n,k}\simeq \U{n,j}\Leftrightarrow k=j$.
\end{Prop}
\begin{proof} 
Only the statement ``$\Rightarrow$'' is non-trivial.
Choose $k\leq j$ and let $\theta: \U{n,k} \to \U{n,j} $ be a $\dtl n$-isomorphism. Choose $x\in\X{n,k}$ and a $\sigma=|u\bar x|\in\dtl n$, with a non-zero $u\in \U{n,k}$. Then $\sigma x$ is non-zero and, since $\theta$ is an isomorphism, so is $\theta (\sigma x) = \sigma \theta(x)$. This means that $\theta(x)$ is a linear combination of states, one of which must have precisely $n-k$ vacancies, all of them coinciding with those of $x$. Since $j \ge k$, all other positions of this state must be occupied by defects, and $j$ and $k$ must actually be equal.
\end{proof}
\noindent Proposition \ref{prop:cyclic} has shown that any vector in $\textrm{span\,}\X{n,k}$ generates the standard module $\U{n,k}$. But, for the special case $k=n-1$ or $n$, any $n$-link diagrams must have precisely $1$ and $0$ vacancy respectively and $\U{n,n}=\textrm{span\,}\X{n,n}$ and $\U{n,n-1}=\textrm{span\,}\X{n,n-1}$. Therefore any non-zero vector in these modules generates them and the following result follows.
  
\begin{Coro}
$\U{n,n}$ and $\U{n,n-1}$ are irreducible.
\end{Coro}
  
\end{subsection}
  
  
\begin{subsection}{The dimension of $\U{n,k}$}
  
The next step is the computation of the dimensions of the standard modules $\U{n,k}$. This task is made easy by the following ordering of their basis of $n$-link diagrams. (See below for an example.) First start by ordering the $n$-link basis by their number of vacancies $\iota$, where $0\le \iota \le  n-k$ and $\iota \equiv n-k\ \textrm{mod}\ 2$. Second, among those with the same number $\iota$ of vacancies, gather those whose vacancies are at the same positions. The ordering does not need to be specified any further. Now, for a given number of vacancies and their fixed locations, note that the $(n-\iota)$-link diagrams obtained by omitting the vacant positions are in one-to-one correspondence with elements of the link basis of the Temperley-Lieb standard module $\vs{n-\iota,k}$ or, equivalently, $\vs{k+2p,k}$ if the number of arcs $p=(n-\iota-k)/2$ is used. The number of arcs must then be in the range $0\le p\le \lfloor (n-k)/2\rfloor$. For a fixed $\iota$ or $p$, the number of possible positions of the $\iota$ vacancies among the $n$ positions is $\binom{n}{\iota}=\binom{n}{k+2p}$.  Also, recalling the structure of the subalgebra $S_{n}$, the action of this algebra will never change the vacancies of a $n$-link diagrams. We have therefore proved the following proposition and corollary.
\begin{Prop}\label{prop:unk.vs}
As vector spaces,
\begin{equation}
\U{n,k} \simeq \bigoplus_{p=0}^{\lfloor \frac{n-k}{2} \rfloor} \binom{n}{k+2p} \vs{k+2p,k}.
\end{equation}
Furthermore, if we consider $\Res{\U{n,k}}^{\dtl n}_{S_n}$, the restriction of $\U{n,k}$ to the subalgebra $S_{n}$, then this isomorphism is also a $S_{n}$-module isomorphism.
\end{Prop}
\begin{Coro}\label{coro:dimUnk}
The dimension of the standard module $\U{n,k}$ is
\begin{equation}\label{eq:dimUnk}
\dim \U{n,k} = \sum_{p=0}^{\lfloor \frac{n-k}{2} \rfloor} \binom{n}{k+2p}\dim\vs{k+2p,k}
\end{equation}
where $\dim \vs{n,k}=\binom n{(n-k)/2}-\binom n{(n-k)/2-1}$.
\end{Coro}
  
Here is an example, for the module $\U{5,1}$, of the ordering used in the proof. The subset of $5$-links without any vacancy ($p=2$) form a basis of the $\tl 5$-module $\vs{5,1}$:
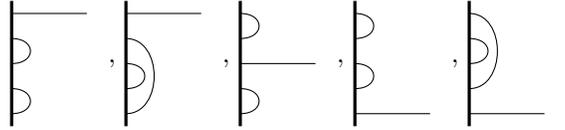
\begin{equation*}
\begin{tikzpicture}[scale=1/3]
	\bord{0}{5};
	\defect{0,4};
	\link{0,0}{1,0}{1,1}{0,1};
	\link{0,2}{1,2}{1,3}{0,3};
	\node at (4,2) {$,$ };
\end{tikzpicture}
\begin{tikzpicture}[scale=1/3]
	\bord{0}{5};
	\defect{0,4};
	\link{0,1}{1,1}{1,2}{0,2};
	\link{0,0}{1.5,0}{1.5,3}{0,3};
	\node at (4,2) {$,$ };
\end{tikzpicture}
\begin{tikzpicture}[scale=1/3]
	\bord{0}{5};
	\defect{0,2};
	\link{0,0}{1,0}{1,1}{0,1};
	\link{0,3}{1,3}{1,4}{0,4};
	\node at (4,2) {$,$ };
\end{tikzpicture}
\begin{tikzpicture}[scale=1/3]
	\bord{0}{5};
	\defect{0,0};
	\link{0,1}{1,1}{1,2}{0,2};
	\link{0,3}{1,3}{1,4}{0,4};
	\node at (4,2) {$,$ };
\end{tikzpicture}\begin{tikzpicture}[scale=1/3]
	\bord{0}{5};
	\defect{0,0};
	\link{0,2}{1,2}{1,3}{0,3};
	\link{0,1}{1.5,1}{1.5,4}{0,4};
	\node at (4,2) {$.$ };
\end{tikzpicture}
\end{equation*}
Now the subset with $\iota=2$ vacancies ($p=1$) contains $20$ link diagrams:
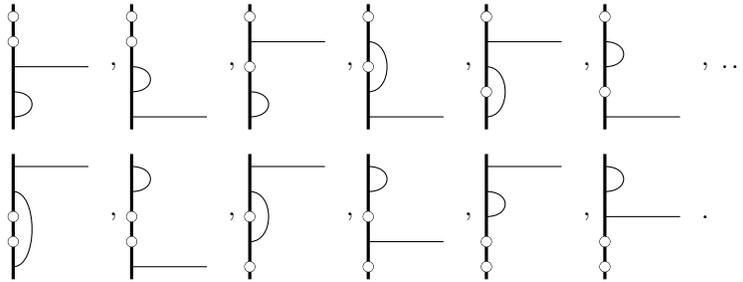
\begin{align*}
\ &\begin{tikzpicture}[scale=1/3]
	\bord{0}{5};
	\vac{0,4};
	\vac{0,3};	
	\defect{0,2};
	\link{0,0}{1,0}{1,1}{0,1};
	\node at (4,2) {$,$ };
\end{tikzpicture}
\begin{tikzpicture}[scale=1/3]
	\bord{0}{5};
	\vac{0,4};
	\vac{0,3};
	\defect{0,0};
	\link{0,1}{1,1}{1,2}{0,2};
	\node at (4,2) {$,$ };
\end{tikzpicture}
\begin{tikzpicture}[scale=1/3]
	\bord{0}{5};
	\vac{0,4};
	\vac{0,2};
	\defect{0,3};
	\link{0,0}{1,0}{1,1}{0,1};
	\node at (4,2) {$,$ };
\end{tikzpicture}
\begin{tikzpicture}[scale=1/3]
	\bord{0}{5};
	\vac{0,4};
	\vac{0,2};
	\defect{0,0};
	\link{0,1}{1,1}{1,3}{0,3};
	\node at (4,2) {$,$ };
\end{tikzpicture}
\begin{tikzpicture}[scale=1/3]
	\bord{0}{5};
	\vac{0,4};
	\vac{0,1};
	\defect{0,3};
	\link{0,0}{1,0}{1,2}{0,2};
	\node at (4,2) {$,$ };
\end{tikzpicture}
\begin{tikzpicture}[scale=1/3]
	\bord{0}{5};
	\vac{0,4};
	\vac{0,1};	
	\defect{0,0};
	\link{0,2}{1,2}{1,3}{0,3};
	\node at (4,2) {$,$ };
	\node at (5,2) {\ \ $\dots $};
\end{tikzpicture}
\\
\ &
\begin{tikzpicture}[scale=1/3]
	\bord{0}{5};
	\vac{0,2};
	\vac{0,1};	
	\defect{0,4};
	\link{0,0}{1,0}{1,3}{0,3};
	\node at (4,2) {$,$ };
\end{tikzpicture}
\begin{tikzpicture}[scale=1/3]
	\bord{0}{5};
	\vac{0,2};
	\vac{0,1};	
	\defect{0,0};
	\link{0,3}{1,3}{1,4}{0,4};
	\node at (4,2) {$,$ };
\end{tikzpicture}
\begin{tikzpicture}[scale=1/3]
	\bord{0}{5};
	\vac{0,2};
	\vac{0,0};	
	\defect{0,4};
	\link{0,1}{1,1}{1,3}{0,3};
	\node at (4,2) {$,$ };
\end{tikzpicture}
\begin{tikzpicture}[scale=1/3]
	\bord{0}{5};
	\vac{0,2};
	\vac{0,0};	
	\defect{0,1};
	\link{0,3}{1,3}{1,4}{0,4};
	\node at (4,2) {$,$ };
\end{tikzpicture}
\begin{tikzpicture}[scale=1/3]
	\bord{0}{5};
	\vac{0,1};
	\vac{0,0};	
	\defect{0,4};
	\link{0,2}{1,2}{1,3}{0,3};
	\node at (4,2) {$,$ };
\end{tikzpicture}
\begin{tikzpicture}[scale=1/3]
	\bord{0}{5};
	\vac{0,1};
	\vac{0,0};
	\defect{0,2};
	\link{0,3}{1,3}{1,4}{0,4};
	\node at (4,2) {$.$ };
\end{tikzpicture}
\end{align*}
Even though some have been omitted, it is clear that, for fixed vacancy positions, the occupied positions are $3$-link diagrams and these form a basis of $\vs{3,1}$. Finally the subset with $\iota=4$ vacancies ($p=0$) is
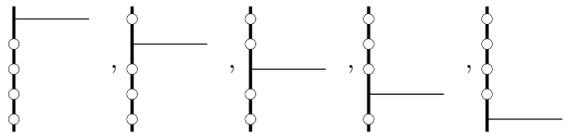
\begin{equation*}
\begin{tikzpicture}[scale=1/3]
	\bord{0}{5};
	\defect{0,4};
	\vac{0,0};
	\vac{0,1};
	\vac{0,2};
	\vac{0,3};
	\node at (4,2) {$,$ };
\end{tikzpicture}
\begin{tikzpicture}[scale=1/3]
	\bord{0}{5};
	\defect{0,3};
	\vac{0,0};
	\vac{0,1};
	\vac{0,2};
	\vac{0,4};
	\node at (4,2) {$,$ };
\end{tikzpicture}\begin{tikzpicture}[scale=1/3]
	\bord{0}{5};
	\defect{0,2};
	\vac{0,0};
	\vac{0,1};
	\vac{0,4};
	\vac{0,3};
	\node at (4,2) {$,$ };
\end{tikzpicture}\begin{tikzpicture}[scale=1/3]
	\bord{0}{5};
	\defect{0,1};
	\vac{0,0};
	\vac{0,4};
	\vac{0,2};
	\vac{0,3};
	\node at (4,2) {$,$ };
\end{tikzpicture}
\begin{tikzpicture}[scale=1/3]
	\bord{0}{5};
	\defect{0,0};
	\vac{0,4};
	\vac{0,1};
	\vac{0,2};
	\vac{0,3};
	\node at (4,2) {$.$ };
\end{tikzpicture}\end{equation*}
This subset contains $\binom{5}{4}=5$ copies of the $1$-link state with a single defect, that is a basis of $\vs{1,0}$. The dimension of $\U{5,1}$ is $30$.
  
Another expression\footnote{We thank A.~Morin-Duchesne for bringing this formula to our attention.} is known for the dimensions of the standard modules, namely
$$\dim\U{n,k}=\begin{pmatrix}n\\k\end{pmatrix}_2-\begin{pmatrix}n\\k+2\end{pmatrix}_2$$
where the trinomial coefficients are defined by the relation $(x+1+x^{-1})^n=\sum_{-n\le k\le n}\left(\begin{smallmatrix}n\\ k\end{smallmatrix}\right)_2x^l$. This can be proved by mapping the link basis of $\U{n,k}$ on leftward walks starting from the origin in $\mathbb Z^2$. (See \cite{RiSY}.) The $i$th step of the walk is determined by the $i$th position of the link: a step $(1,0)$ is taken for a vacancy, a step $(1,1)$ for the opening of the loop or a defect, and $(1,-1)$ for the closing of a loop. All walks corresponding to elements of the basis of $\U{n,k}$ end at $(n,k)$. All walks visiting only points with non-negative vertical coordinates correspond to $n$-links with $k$ defects.\invisible{Il est relativement facile de voir que le nombre des chemins d\'efinis \`a l'aide des vecteurs $(1,0), (1,1)$ et $(1,-1)$ qui commencent en $(0,0)$ et finissent en $(n,k)$ est le coefficient trinomial $\left(\begin{smallmatrix}n\\ k\end{smallmatrix}\right)_2$. Pour le voir il suffit d'associer aux trois vecteurs ci-dessus les poids $1, x$ et $x^{-1}$. Pour se d\'ebarasser des chemins qui visitent les coordonn\'ees verticales n\'egatives, il suffit de les mettre en correspondance biunivoque avec les chemins qui atteignent $(n,-2-k)$. Pour un chemin qui visite les coordonn\'ees verticales n\'egatives, trouver le plus petit $i$ tel que le chemin passe par $(i,-1)$ et faire la r\'eflexion du chemin aprs $i$ par un miroir passant par la coordonn\'ee vertical $-1$. Le chemin r\'efl\'echi finira n\'ecessairement sa course en $(n,-k-2)$. Tout chemin finissant en $(n,-k-2)$ correspond \`a un chemin arrivant $(n,k)$ par le processus inverse et la correspondance est donc biunivoque.}
  
The same method of slicing and unfolding $n$-diagrams used in section \ref{sec:dimension} to obtain the dimensions of $\dtl{n}$ can be used again while keeping track of the number of defects. This leads to another expression for the dimension of the algebra.
\begin{Prop}\label{prop:dimdTLnUnk}The dimension of the dilute Temperley-Lieb algebra $\dtl n$ is also given by
\begin{equation}
\dim \dtl{n} = \sum_{k=0}^{n} \left(\dim \U{n,k}\right)^2.
\end{equation}
\end{Prop} 
  
\end{subsection}
  
  
\begin{subsection}{The restriction of $\U{n,k}$}
  
The next two subsections are devoted to the restriction and induction of the standard modules $\U{n,k}$. The first step, for the study of the restriction, is to decide how the subalgebra $\dtl {n-1}$ is embedded into $\dtl n$. The embedding that we use is realized by adding a pair of points at the bottom of all $(n-1)$-diagrams (the $n$th points) and connecting this pair by a dashed line. As the dashed line is seen to act as the identity on the $n$th points, this is a natural embedding, similar to the one used for the Temperley-Lieb algebra \cite{Westbury, RiSY}. Any $(n-1)$-diagram of $\dtl {n-1}$ is then embedded as the sum of two $n$-diagrams of $\dtl n$.  
The module $\U{n,k}$ seen as a $\dtl{n-1}$-module will be called the \emph{restriction of $\U{n,k}$} and denoted by $\Res{\U{n,k}}$.
\begin{Prop}\label{prop:restr.unk}
With the embedding of $\dtl{n-1}$ in $\dtl{n}$ described above, the short sequence
\begin{equation}\label{eq:restr.unk}
0 \to \U{n-1,k}\oplus \U{n-1,k-1}\to \Res{\U{n,k}} \to \U{n-1,k+1}\to 0
\end{equation}
is exact for all $n\geq 2$ and $k\in\{0,1,\dots,n\}$ and therefore
\begin{equation}\label{eq:restrictionExact}
\Res{\U{n,k}}/{( \U{n-1,k}\oplus \U{n-1,k-1})} \simeq \U{n-1,k+1}.
\end{equation}
Again $\U{m,j}=\{0\}$ if $j\not\in\{0,1,\dots, m\}$.
\end{Prop}
\begin{proof} To show exactness at $\U{n-1,k}\oplus \U{n-1,k-1}$, an injective map $\phi: \U{n-1,k}\oplus \U{n-1,k-1} \to  \Res{\U{n,k}} $ needs to be constructed.
Consider the operation, defined on $(n-1)$-links with $k$ or $k-1$ defects, that consists in adding a point at the bottom of the diagram and putting a defect there if the diagram had $k-1$ defects, and a vacancy otherwise. The result is an $n$-link with precisely $k$ defects. Let $\phi$ be the map that extends linearly this operation to 
$\U{n-1,k}\oplus \U{n-1,k-1}$. Since the elements of $\dtl{n-1}$ do not act on the $n$th point, this is a homomorphism. It should also be clear that it is injective.
  
To define a homomorphism $\psi: \Res{\U{n,k}} \to \U{n-1,k+1}$ such that $\ker\psi=\im\phi$, we again start by defining a diagrammatic operation on $n$-links.
If an $n$-link diagram has a defect or a vacancy at its $n$th position, it is sent to zero in $\U{n-1,k+1}$. Otherwise, its $n$th point is simply removed and the arch which ended at this point is replaced by a defect at its entry (top) point. For example,
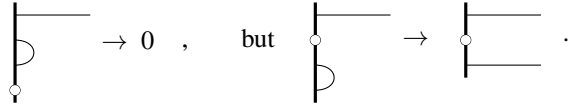
\begin{equation*}
\begin{tikzpicture}[scale=1/3]
	\bord{0}{4};
	\vac{0,0};
	\link{0,1}{1,1}{1,2}{0,2};
	\defect{0,3};
	\node at (4.5,2) {$\rightarrow$\ \ $0$};
	\node at (8.5,2) {,\qquad \textrm{but}\quad};
\end{tikzpicture}
\quad
\begin{tikzpicture}[scale=1/3];
	\bord{0}{4};
	\vac{0,2};
	\link{0,1}{1,1}{1,0}{0,0};
	\defect{0,3};
	\node at (4,2) {$\rightarrow$};
	\draw[\epais] (6,1/2) -- (6,7/2);	
	\vac{6,2};
	\defect{6,1};
	\defect{6,3};
	\node at (10,2) {.};
\end{tikzpicture}
\end{equation*}
The map $\psi$ is defined as the linear extension to $\Res{\U{n,k}}$ of this operation defined on links. To see that this is a homomorphism, suppose that an $n$-diagram in $\dtl{n-1}\subset \dtl n$ transforms the bubble ending at position $n$ of an $n$-link into a defect or a vacancy. This can only be achieved if the opening point of the bubble is linked to a defect by a bubble on the right side of the $n$-diagram. The same diagram applied to the image of the link would then link two of its defects together and would thus correspond to the zero element in $\U{n-1,k+1}$. So $\psi$ is indeed a homomorphism. The map has been constructed so that $ \im \phi \subset\ker \psi$. 
  
To see that $\psi$ is surjective, we construct a pre-image for a general $(n-1)$-link in $\U{n-1,k+1}$. Any such a link has at least one defect since $k+1$ is a positive integer. Then add an $n$th point to the diagram and close the lowest defect in the link onto this new position $n$. This is then an element of $\Res{\U{n,k}}$ whose image by $\psi$ is the original $(n-1)$-link. This construction also shows that there is a one-to-one correspondence between $n$-links in $\Res{\U{n,k}}$ that have a bubble ending at $n$ and $(n-1)$-links in $\U{n-1,k+1}$. Therefore $ \im \phi$ and $\ker \psi$ must coincide.
  
Note finally that the previous constructions for $\phi$ and $\psi$ remain valid when $k=0, n-1$ or $n$ if the modules $\U{n-1,-1}$, $\U{n-1,n}$ and $\U{n-1,n+1}$ are taken to be the trivial ones.
\end{proof}
Note that the exact sequence gives a simple relationship between the dimensions of the $\U{n,k}$s:\begin{equation}\label{eq:simpleRelation}
\dim \U{n,k} =  \dim \U{n-1,k} + \dim \U{n-1,k-1} + \dim \U{n-1,k+1} \text{ .}
\end{equation}
This property could also be proved using the dimension \eqref{eq:dimUnk} of $\U{n,k}$. The module $\U{n-1,k}\oplus \U{n-1,k-1}$ is a direct sum of two submodules of distinct parities. Since $\U{n-1,k+1}$ has the parity of $\U{n-1,k-1}$, the submodule $\U{n-1,k}$ of $\Res{\U{n,k}}$ is the largest of its parity. 
  
\begin{Prop}\label{prop:restr.split}
Let $\beta = q+q^{-1}$ with $q \in \mathbb{C}^\times$. If $q^{2\left(k+1 \right)} \neq 1$, the sequence
\begin{equation}
0 \to \U{n-1,k-1}\to {\Res{\U{n,k}}}/{\U{n-1,k}} \to \U{n-1,k+1}\to 0
\end{equation}
splits and therefore ${\Res{\U{n,k}}}/{\U{n-1,k}}\simeq \U{n-1,k-1} \oplus \U{n-1,k+1}$.
\end{Prop}
\begin{proof}
This proof uses the central element $\DF{n-1}$ defined in appendix \ref{app:Fn}. Since $\DF{n-1}$ is central, its (generalized) eigenspaces in a given $\dtl{n-1}$-module are submodules. The appendix shows that $\DF{n}$ acts on the standard module $\U{n,k}$ as $\df{k}\times \id$ with $\df{k}=q^{k+1}+q^{-(k+1)}$. If $\df{n-1,k-1}$ and $\df{n-1,k+1}$ are different, then ${\Res{\U{n,k}}}/{\U{n-1,k}}$ will contain two eigenspaces of $\DF{n-1}$ of dimensions $\dim \U{n-1,k-1}$ and $\dim \U{n-1,k+1}$ respectively. The exercise consists then in deciding when the two eigenvalues $\df{n-1,k-1}$ and $\df{n-1,k+1}$ are distinct. Their difference is:
\begin{equation*}
\df{n-1,k+1}-\df{n-1,k-1}
= q^{k+2}-q^{k} + q^{-(k+2)}-q^{-k} = (q^2-1)\left(q^{2\left(k+1\right)}-1\right)q^{-k-2}
\end{equation*}
and vanishes if and only if $q^{2\left(k+1 \right)} = 1$.
\end{proof}
\noindent The condition that $q^{2\left(k+1 \right)} \neq 1$ will be fundamental for the rest of the text. An integer $k$ will be called \emph{critical} if $q^{2\left(k+1 \right)}=1$, and \emph{generic} otherwise. We also say that $\U{n,k}$ is critical if $k$ is.
\end{subsection}
  
  
\begin{subsection}{The induction of $\U{n,k}$}\label{sec:induction}
  
After studying the restriction of the $\dtl n$-module $\U{n,k}$, it is natural to ask whether its induction is also part of an exact sequence similar to that satisfied by its restriction. This subsection answers this question.
  
The induction of $\U{n,k}$, denoted by $\Ind{\U{n,k}}$, is defined by the tensor product
$$\Ind{\U{n,k}}=\dtl {n+1}\otimes_{\dtl n}\U{n,k}$$
where the subscript on the tensor product symbol means that the elements of $\dtl n$ (embedded in $\dtl{n+1}$ as in the previous subsection) may pass freely from one of its sides to the other.
  
The first task is to find a finite generating set for $\Ind{\U{n,k}}$ of manageable size. Proposition \ref{prop:cyclic} provides a first simplification. Let $z$ be an $n$-link diagram in $\X{n,k}$. Since $\U{n,k}=\dtl n z$, then
\begin{equation}
\Ind{\U{n,k}} = \dtl{n+1} \otimes_{\dtl{n}} (\dtl n z)=\dtl{n+1}\otimes_{\dtl n} z.
\end{equation}
A further simplification is possible. We introduce for this purpose three ``surgeries'' $\theta_i$, $i\in\{-1,0,1\}$, that transforms an $n$-link diagram $u\in\U{n,k}$ into an $(n+1)$-link one. The first $\theta_1$ adds to the $n$-link $u$ a defect at the bottom, at position $n+1$, and the second $\theta_0$ adds there a vacancy. The last one, $\theta_{-1}$, closes the lowest defect of $u$ into an arc ending at $n+1$, if such a defect exists. If there is none, $\theta_{-1}$ sends the $n$-link to zero. The index on the $\theta_i$ indicates how the number of defects changes. Here are some examples.
\begin{align*}
\begin{tikzpicture}[baseline={(current bounding box.center)},scale=1/3]
	\node at (-3/2,1) {$\theta_1 \bigg($ };
	\bord{0}{3};
	\defect{0,0};
	\link{0,1}{1,1}{1,2}{0,2};
	\node at (5,1) {$\bigg)$ $=$};
\end{tikzpicture}
\begin{tikzpicture}[baseline={(current bounding box.center)},scale=1/3]
	\draw[\epais] (0,-3/2) -- (0,5/2);
	\defect{0,0};
	\link{0,1}{1,1}{1,2}{0,2};
	\defect{0,-1};
\end{tikzpicture}, &\ \qquad
\begin{tikzpicture}[baseline={(current bounding box.center)},scale=1/3]
\node at (-3/2,1) {$\theta_1 \bigg($ };
	\bord{0}{3};
	\vac{0,0};
	\link{0,1}{1,1}{1,2}{0,2};
	\node at (3,1) {$\bigg)$ $=$};
\end{tikzpicture}
\begin{tikzpicture}[baseline={(current bounding box.center)},scale=1/3]
	\draw[\epais] (0,-3/2) -- (0,5/2);
	\vac{0,0};
	\link{0,1}{1,1}{1,2}{0,2};
	\defect{0,-1};
	\node at (3,1) {$,$};
\end{tikzpicture}\\
\begin{tikzpicture}[baseline={(current bounding box.center)},scale=1/3]
	\node at (-3/2,1) {$\theta_0 \bigg($ };
	\bord{0}{3};
	\defect{0,0};
	\link{0,1}{1,1}{1,2}{0,2};
	\node at (5,1) {$\bigg)$ $=$};
\end{tikzpicture}
\begin{tikzpicture}[baseline={(current bounding box.center)},scale=1/3]
	\draw[\epais] (0,-3/2) -- (0,5/2);
	\defect{0,0};
	\link{0,1}{1,1}{1,2}{0,2};
	\vac{0,-1};
\end{tikzpicture}, &\ \qquad
\begin{tikzpicture}[baseline={(current bounding box.center)},scale=1/3]
	\node at (-3/2,1) {$\theta_0 \bigg($ };
	\bord{0}{3};
	\vac{0,0};
	\link{0,1}{1,1}{1,2}{0,2};
	\node at (3,1) {$\bigg)$ $=$};
\end{tikzpicture}
\begin{tikzpicture}[baseline={(current bounding box.center)},scale=1/3]
	\draw[\epais] (0,-3/2) -- (0,5/2);
	\vac{0,0};
	\link{0,1}{1,1}{1,2}{0,2};
	\vac{0,-1};
	\node at (3,1) {$,$};
\end{tikzpicture}\\
\begin{tikzpicture}[baseline={(current bounding box.center)},scale=1/3]
	\node at (-3/2,1) {$\theta_{-1} \bigg($ };
	\bord{0}{3};
	\defect{0,0};
	\link{0,1}{1,1}{1,2}{0,2};
	\node at (5,1) {$\bigg)$ $=$};
\end{tikzpicture}
\begin{tikzpicture}[baseline={(current bounding box.center)},scale=1/3]
	\draw[\epais] (0,-3/2) -- (0,5/2);
	\link{0,0}{1,0}{1,-1}{0,-1};
	\link{0,1}{1,1}{1,2}{0,2};
	\node at (3,1) {$ $};
\end{tikzpicture}, &\ \qquad
\begin{tikzpicture}[baseline={(current bounding box.center)},scale=1/3]
	\node at (-3/2,1) {$\theta_{-1} \bigg($ };
	\bord{0}{3};
	\vac{0,0};
	\link{0,1}{1,1}{1,2}{0,2};
	\node at (3,1) {$\bigg)$ $=\ \ 0$};
\end{tikzpicture}
\begin{tikzpicture}[baseline={(current bounding box.center)},scale=1/3]
	\node at (0,-3/2) {$ $};
	\node at (0,5/2) {$ $};
	\node at (3,1) {$.$};
\end{tikzpicture}
\end{align*}
We now argue that any non-zero element of $\Ind{\U{n,k}}$ can be written as a sum of terms of the form $|u\overline{\theta_i(z)}|\otimes z$ where $i\in\{-1,0,1\}$ and $u\in\U{n+1,k+i}$. It is sufficient to study elements of $\dtl{n+1}$ of the form $|u\bar v|$ with $u$ and $v$ left $(n+1)$-link diagrams.
  
The first case to study is when $v$ is in $\X{n+1,j}$ for some $j$. It is then possible to write $|u\bar v|=|u \bar v||v\bar v|$. Let $v'$ be the $n$-link diagram obtained from $v$ by deleting its position $n+1$ and the vacancy or the defect at this position. Then
\begin{equation*}|v\bar v|=a
\begin{tikzpicture}[baseline={(current bounding box.center)},scale=1/3]
	\bord{0}{2};
	\bord{3}{2};
	\wave{0}{0}
	\node at (0.5,1.3) {$v'$};
	\node at (2.5,1.3) {$\bar v'$};
	\node at (4,0) {$\scriptstyle{n+1}$};
\end{tikzpicture}
\end{equation*}
where $a$ stands for the generator $x_{n+1}$ if $v$ has a vacancy at $n+1$ and for the generator $e_{n+1}$ if instead it has a defect there. (The elements $x_i$ and $e_i$ were defined in subsection \ref{sec:definition}.) Therefore
$$|u\bar v|\otimes z=|u \bar v|a\otimes |v'\bar v'|z$$
with the appropriate $a$. This tensor product is zero unless $v'$ is equal to $z$. That is, when $v$ is an $(n+1)$-link with only defects and vacancies, the vector $|u\bar v|\otimes z$ is non-zero only when $v=\theta_0(z)$ if position $n+1$ of $v$ is vacant and when $v=\theta_1(z)$ if it bears a defect.
  
The second case to study is when $v$ contains an arc between two positions above $n+1$. It is then always possible to find an arc between $i$ and $j$ such that $1\le i< j\le n$ and that all positions $k$ in $v$ with $i<k<j$ are vacant. Then 
\begin{equation}\label{eq:secondCase}|u\bar v|\ \ =\ \ 
\begin{tikzpicture}[baseline={(current bounding box.center)},scale=1/3]
	\bord{0}{7};
	\bord{3}{7};
	\node at (1/2,3) {$u$};
	\node at (5/2,6) {$\bar v_{t}$};
	\node at (5/2,0) {$\bar v_{b}$};
	\node at (7/2,1) {$\scriptstyle{j}$};
	\node at (7/2,5) {$\scriptstyle{i}$};
	\link{3,1}{2,1}{2,5}{3,5};
	\vac{3,2};
	\vac{3,4};
	\draw[dotted] (3,3) circle (1/5); \fill[white] (3,3) circle (1/5);
	\node at (4,3) {$=$};
\end{tikzpicture}
\begin{tikzpicture}[baseline={(current bounding box.center)},scale=1/3]
	\bord{0}{7};
	\bord{3}{7};
	\node at (1/2,3) {$u$};
	\node at (5/2,6) {$\bar v_{t}$};
	\node at (5/2,0) {$\bar v_{b}$};
	\vac{3,1};
	\vac{3,2};
	\vac{3,4};
	\vac{3,5};
	\draw[dotted] (3,3) circle (1/5); \fill[white] (3,3) circle (1/5);
	\bord{4}{7};
	\bord{7}{7};
	\vac{4,1};
	\vac{4,2};
	\vac{4,4};
	\vac{4,5};
	\draw[dotted] (4,3) circle (1/5); \fill[white] (4,3) circle (1/5);
	\link{7,1}{6,1}{6,5}{7,5};
	\vac{7,2};
	\vac{7,4};
	\draw[dotted] (7,3) circle (1/5); \fill[white] (7,3) circle (1/5);
	\wave{4}{0};
	\wave{4}{6};
	\node at (15/2,1) {$\scriptstyle{j}$};
	\node at (15/2,5) {$\scriptstyle{i}$};
\end{tikzpicture}
\end{equation}
where $v_t$ ($v_b$) contains the pattern of positions above $i$ in $v$ (below $j$). There might be arcs going from $v_t$ to $v_b$ as well as arcs between $u$ and the $v_t$ and $v_b$. Consider the rightmost factor of \eqref{eq:secondCase}. All positions corresponding to those of $v_t$ and $v_b$ are occupied by dashed lines. The $n$ top positions of this factor is an element of $\dtl n$ and annihilates $z$, because either the arc joins two defects in $z$ or there is a mismatch between the vacancies and defects of $z$ and those of this factor. Element $|u\bar v|\otimes z$ with $v$ with such an arc are zero and can therefore be ignored.
  
The third and last case is when $v$ contains a single arc whose bottom point is at position $n+1$. If this arc joins position $i$ to $n+1$, then all positions in between must be vacancies. A factorization similar to that used in the first case leads to
\begin{equation*}|u\bar v|=
\begin{tikzpicture}[baseline={(current bounding box.center)},scale=1/3]
	\bord{0}{2};
	\bord{3}{2};
	\bord{4}{2};
	\bord{7}{2};
	\node at (1/2,1/2) {$u$};
	\node at (5/2,1/2) {$\bar v$};
	\node at (9/2+0.1,1) {$v''$};
	\node at (13/2-0.1,1) {$\overline{v''}$};
	\node at (8,0) {$\scriptstyle{n+1}$};
	\wave{4}{0};
\end{tikzpicture}
\end{equation*}
where $v''$ is obtained from $v$ by deleting its position $n+1$ and putting a defect at position $i$. Then $|u\bar v|\otimes z=|u\bar v|\otimes |v'' \overline{v''}|z$ and $|u\bar v|\otimes z$ is non-zero only if ${v''}=z$. Hence, when $v$ has a single arc ending at $n+1$, the element $|u\bar v|\otimes z$ is non-zero only if $\theta_{-1}(z)=v$. The analysis of the above three cases is summed up by saying that $\Ind{\U{n,k}}=\textrm{span\, }B_{n,k}$ where $B_{n,k}$ is the finite set
$$B_{n,k}=\left\{ \ |u\overline{\theta_i(z)}|\otimes z\ \ |\ \ i\in\{-1,0,1\}\textrm{\ and $u$ a link diagram in }\U{n+1,k+i}\right\}.$$
The analysis does not prove that $B_{n,k}$ is a basis however. It does not even rule out some of the elements in $B_{n,k}$ being zero. The main remaining result of the present subsection is that $B_{n,k}$ is indeed a basis.
  
Choose $z\in\X{n,k}$ and let $\phi=\phi_z$ be the linear map $\dtl{n+1}\otimes_{\mathbb C}\U{n,k}\rightarrow \U{n+2,k}$ defined by the following action on elements of the form $|u\bar v|\otimes_{\mathbb C} y$ where $u$ and $v$ are $(n+1)$-links with the same numbers of defects and $y\in\U{n,k}$. (The index on the tensor product sign will be omitted only if it is $\dtl n$.) To compute $\phi(|u\bar v|\otimes_{\mathbb C} y)$, first draw
\begin{equation*}
\begin{tikzpicture}[baseline={(current bounding box.center)},scale=1/3]
	\bord{0}{2};
	\bord{3}{2};
	\bord{6}{2};
	\node at (1/2,1/2) {$u$};
	\node at (5/2,1/2) {$\bar v$};
	\node at (7/2,1) {$y$};
	\node at (11/2,1) {$\bar z$};
	\node at (7,0) {$\scriptstyle{n+1}$};
	\wave{3}{0};
\end{tikzpicture}
\end{equation*}
and then detach the dashed line ending at position $(n+1)$ on the far right to attach it at the bottom of $u$:
\begin{equation}\label{eq:wildPhi}
\begin{tikzpicture}[baseline={(current bounding box.center)},scale=1/3]
	\draw[\epais](0,-3/2) --++ (0,3);
	\bord{3}{2};
	\draw[\epais](6,-1/2) --++ (0,2);
	\node at (1/2,1/2) {$u$};
	\node at (5/2,1/2) {$\bar v$};
	\node at (7/2,1) {$y$};
	\node at (11/2,1) {$\bar z$};
	\node at (-1,-1) {$\scriptstyle{n+2}$};
    \draw[dashed] (3,0) .. controls (4,0) and (4,-1) .. (3,-1);
	\wave{0}{-1};
\end{tikzpicture}
\end{equation}
The object created has $n+2$ positions on its left edge, $n$ on its right one. There are $n+1$ positions on both sides of the central line and one can use the usual rules to multiply diagrams for this new object. If vacancies do not match, then $\phi(|u\bar v|\otimes_{\mathbb C} y)$ is set to zero. If they match, then there exists $w\in \U{n+2,j}$ and $x\in\U{n,j}$ such that the above diagram is $\beta^\# |w\bar x|$ where $\#$ is the number of closed loops in \eqref{eq:wildPhi}. The image $\phi(|u\bar v|\otimes_{\mathbb C} y)$ is non-zero only if $j=k$ and it is then $\beta^\# w$. Note that, if $u', v'$ are some $n$-links with the same numbers of defects, then 
\begin{equation*}\phi \big(
\begin{tikzpicture}[baseline={(current bounding box.center)},scale=1/3]
	\bord{0}{2};
	\bord{3}{2};
	\bord{6}{2};
	\node at (1/2,1/2) {$u$};
	\node at (5/2,1/2) {$\bar v$};
	\node at (7/2,1) {$u'$};
	\node at (11/2,1) {$\bar v'$};
	\node at (15/2,3/4) {$\otimes_{\mathbb C} y$};
	\wave{3}{0};

\end{tikzpicture}\big)
 = \phi(|u\bar v|\otimes_{\mathbb C}|u'\bar v'|y)
\end{equation*}
because the computation of the resulting image is based in both cases on the diagram 
\begin{equation*}
\begin{tikzpicture}[baseline={(current bounding box.center)},scale=1/3]
	\draw[\epais](0,-3/2) --++ (0,3);
	\draw[\epais](3,-1/2) --++ (0,2);
	\draw[\epais](6,-1/2) --++ (0,2);
	\draw[\epais](9,-1/2) --++ (0,2);
	\node at (1/2,1/2) {$u$};
	\node at (5/2,1/2) {$\bar v$};
	\node at (7/2,1) {$u'$};
	\node at (11/2,1) {$\bar v'$};
	\node at (13/2,1) {$y$};
	\node at (17/2,1) {$\bar z$};
	\node at (-1,-1) {$\scriptstyle{n+2}$};
    \draw[dashed] (6,0) .. controls (7,0) and (7,-1) .. (6,-1);
	\wave{3}{0};
	\wave{0}{-1};
	\wave{3}{-1};
\end{tikzpicture}
\end{equation*}
Therefore $\phi$ maps to zero the subspace spanned by $\{ab\otimes_{\mathbb C}y-a\otimes_{\mathbb C}by, a\in\dtl{n+1},b\in\dtl n,y\in\U{n,k}\}$. The linear map $\phi$ thus induces a well-defined linear map 
\begin{equation}\label{eq:mapPhi}
\Phi:\Ind{\U{n,k}}\simeq \frac{\dtl{n+1}\otimes_{\mathbb C}\U{n,k}}{\langle ab\otimes_{\mathbb C}y-a\otimes_{\mathbb C}by\rangle}\rightarrow \Res{\U{n+2,k}}
\end{equation}
  
\begin{Prop}\label{Prop-iso}Let $n\geq 1$ and $k\in\{0,1,\dots, n\}$. Then
  
\noindent(i) the set $B_{n,k}$ is a basis of $\Ind{\U{n,k}}$ and
  
\noindent (ii) $\Ind{\U{n,k}}\simeq \Res{\U{n+2,k}}$ as $\dtl{n+1}$-modules.
\end{Prop}
\begin{proof}The linear map $\Phi$ defined in \eqref{eq:mapPhi} is a $\dtl{n+1}$-homomorphism. This follows by the observation that, if $\Phi(|u\bar v|\otimes y)=\beta^\# w$, then $\Phi(a|u\bar v|\otimes y)$ and $a\Phi(|u\bar v|\otimes y)$ will both give $\beta^\# aw$ for all $a\in\dtl{n+1}$ as can be verified diagrammatically.
  
No elements of the spanning set $B_{n,k}$ is zero. To see this, it is sufficient to note that their images by $\Phi$ are non-zero. Indeed a direct computation shows that, if $u\in\U{n+1,k+i}$ with $i\in\{-1,0,1\}$, then $\phi(|u\overline{\theta_i(z)}|\otimes z)=\theta_{-i}(u)\in\U{n+2,k}$ which is non-zero.
  
To end the proof, it remains to show that the spanning set is linearly independent. Since $|B_{n,k}|=\dim \U{n+2,k}$, it is sufficient to show that any link diagram in $\U{n+2,k}$ has a pre-image in $B_{n,k}$. To find the pre-image of $u$, a $(n+2)$-link in $\U{n+2,k}$, simply construct $|u \bar z|$ and detach the bottom position of $u$ to attach it to $z$. The result is $|u'\overline{\theta_i(z)}|$ for some $i\in\{-1,0,1\}$. Then $\Phi(|u'\overline{\theta_i(z)}|\otimes z)=u$. The spanning set is therefore linearly independent and $\Phi$ is a $\dtl{n+1}$-isomorphism.
\end{proof}
  
The following corollaries are immediate consequences of proposition \ref{Prop-iso} and the properties of the restriction of $\U{n,k}$ obtained in the last subsection.
  
\begin{Coro}\label{coro.ind.ses}
The short sequence
\begin{equation}
0 \to \U{n,k}\oplus \U{n,k-1}\to \Ind{\U{n-1,k}} \to \U{n,k+1}\to 0
\end{equation}
is exact for all $n\geq 2$ and $k\in\{0,1,\dots, n-1\}$.
\end{Coro}
\begin{Coro}\label{coro.ind.split}For all $n\geq 2$ and $k$ generic
in $\{0,1,\dots, n-1\}$
\begin{equation}
\Ind{\U{n-1,k}} \simeq \U{n,k-1} \oplus \U{n,k} \oplus \U{n,k+1} \text{ .}
\end{equation}
\end{Coro}
  
Proposition \ref{Prop-iso} and the analogous result for $\tl n$ differs on one small point. For the latter the isomorphism $\Ind{\vs{n,k}}\simeq\Res{\vs{n+2,k}}$ fails in one particular case, namely when $\beta=0$, then $\Ind{\vs{2,0}}\not\simeq\Res{\vs{4,0}}$. Instead $\dim\Ind{\vs{2,0}}=3>\dim \Res{\vs{4,0}}=2$. The difficulty can easily be seen to occur only at $\beta=0$ because, if $\beta\neq 0$, then $u_1u_2\otimes \cuteBubble=\frac1{\beta} u_1u_2u_1\otimes \cuteBubble=\frac1{\beta}u_1\otimes \cuteBubble=\idtl\otimes \cuteBubble$ where $u_i=b_i^tb_i$. For $\beta=0$ the vectors $u_1u_2\otimes \cuteBubble$ and $\idtl\otimes \cuteBubble$ are linearly independent. The problem does not occur for the dilute modules. For example, the analogous situation is resolved as follows:
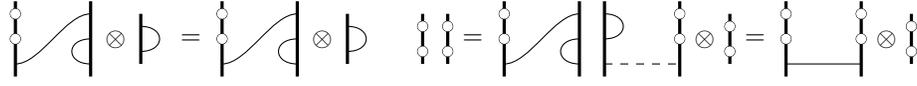
\newline
\begin{center}
\begin{tikzpicture}[scale=1/3]
	\bord{0}{3};
	\bord{3}{3};
	\node at (4,1) {$\otimes$};
	\draw[\epais] (5,0) -- (5,2);
	
	\link{5,1/2}{6,1/2}{6,3/2}{5,3/2};
	
	\link{3,0}{2,0}{2,1}{3,1};
	\link{0,0}{1,0}{2,2}{3,2};
	\vac{0,1};
	\vac{0,2};
	
	\node at (7,1) {$=$};
\end{tikzpicture}
\begin{tikzpicture}[scale=1/3]
	\bord{0}{3};
	\bord{3}{3};
	\node at (4,1) {$\otimes$};
	\draw[\epais] (5,0) -- (5,2);
	\draw[\epais] (8,0) -- (8,2);
	\draw[\epais] (9,0) -- (9,2);
	
	\link{5,1/2}{6,1/2}{6,3/2}{5,3/2};
	\vac{8,1/2};
	\vac{8,3/2};
	\vac{9,1/2};
	\vac{9,3/2};
	
	\link{3,0}{2,0}{2,1}{3,1};
	\link{0,0}{1,0}{2,2}{3,2};
	\vac{0,1};
	\vac{0,2};
	
	\node at (10,1) {$=$};
\end{tikzpicture}
\begin{tikzpicture}[scale=1/3]
	\bord{0}{3};
	\bord{3}{3};
	\node at (8,1) {$\otimes$};
	\draw[\epais] (4,-1/2) -- (4,5/2);
	\draw[\epais] (7,-1/2) -- (7,5/2);
	\draw[\epais] (9,0) -- (9,2);
	
	\link{4,1}{5,1}{5,2}{4,2};
	\vac{7,1};
	\vac{7,2};
	\wave{4}{0}
	\vac{9,1/2};
	\vac{9,3/2};
	
	\link{3,0}{2,0}{2,1}{3,1};
	\link{0,0}{1,0}{2,2}{3,2};
	\vac{0,1};
	\vac{0,2};
	
	\node at (10,1) {$=$};
\end{tikzpicture}
\begin{tikzpicture}[scale=1/3]
	\bord{0}{3};
	\bord{3}{3};
	\node at (4,1) {$\otimes$};
	\draw[\epais] (5,0) -- (5,2);
	
	\vac{5,1/2};
	\vac{5,3/2};
	
	\defect{0,0};
	\vac{0,1};
	\vac{0,2};
	\vac{3,1};
	\vac{3,2};
\end{tikzpicture}.\newline
\end{center}
  
\end{subsection}
\end{section}

%
\begin{section}{The Gram product}\label{sec:Gram}
%
  
This section introduces a bilinear form on standard modules that is invariant under the action of the algebra $\dtl n$ (see lemma \ref{lem.gram.sym}). It is a familiar tool of representation theory since the radical of this bilinear form is a submodule. For $\dtl n$, the radical will be the (unique) maximal submodule. Such a submodule can be non-trivial only if the Gram matrix, representing the bilinear form into some basis, is singular. The Gram determinant and its zeroes can be easily computed. These zeroes occur only when $q$ is a root of unity. The structure of $\dtl n$ is then semisimple when $q$ is generic (not a root of unity) and a complete set of non-isomorphic irreducible modules can be identified (theorem \ref{thm:dtlnGeneric}). The central result of the section concerns non-trivial radicals at $q$ a root of unity. Proposition \ref{prop.rad.iso} shows that they are then irreducible and isomorphic to the irreducible quotients of another standard modules. The section ends with the description of what the irreducible modules $\dl{n,k}$ become under restriction and induction.
  
\begin{subsection}{The bilinear form $\langle *,*\rangle_{n,k}$}\label{sub:gram}
  
The \emph{Gram product} $\langle *,*\rangle_{n,k}:\U{n,k}\times\U{n,k} \to \mathbb{C}$ is a bilinear form defined on $n$-link diagrams and extended linearly. To compute the pairing of two (left) link diagrams, first reflect the first link diagram along its vertical axis and then glue it on the left side of the second one, identifying the corresponding points on both diagrams. If a point containing a string in one of the diagrams is identified with a point containing a vacancy in the other, the result is $0$. Otherwise, the result is non-zero if and only if every defect of the first diagram is linked to a defect of the second. In such cases, the result is $\beta^m$, where $m$ is the number of closed loops formed by the glueing of the two links. For example:
%
  
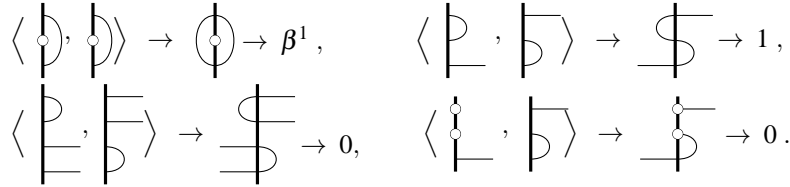
\begin{tikzpicture}[baseline={(current bounding box.center)},scale=1/3]
	\node at (-1,1) {$\Big<$};
	
	\bord{0}{3};
	\link{0,0}{1,0}{1,2}{0,2};
	\vac{0,1};
	
	\node at (1,1) {$,$};
	
	\bord{2}{3};
	\link{2,0}{3,0}{3,2}{2,2};
	\vac{2,1};
	
	\node at (4,1) {$\Big>\ \  \to$};
\end{tikzpicture}
\begin{tikzpicture}[baseline={(current bounding box.center)},scale=1/3]

	\bord{0}{3};
	
	\vac{0,1};
	\link{0,0}{1,0}{1,2}{0,2};
	\link{0,0}{-1,0}{-1,2}{0,2};
	
	\node at (2.7,1) {$\to\ \beta^1$ $,$};
	\node at (7,1) {$ $};
\end{tikzpicture} 
\begin{tikzpicture}[baseline={(current bounding box.center)},scale=1/3]

	\node at (-1,1) {$\Big<$};
	
	\bord{0}{3};
	\defects{0,0};
	\link{0,1}{1,1}{1,2}{0,2};
	
	\node at (2,1) {$,$};
	
	\bord{3}{3};
	\defects{3,2};
	\link{3,0}{4,0}{4,1}{3,1};
	
	\node at (11/2+0.2,1) {$\Big>\ \  \to$};	
	
\end{tikzpicture}
\begin{tikzpicture}[baseline={(current bounding box.center)},scale=1/3]

	\bord{0}{3};
	
	\defects{-3/2,0};
	\defects{0,2};
	\link{0,0}{1,0}{1,1}{0,1};
	\link{0,1}{-1,1}{-1,2}{0,2};
	
	\node at (3.0,1) {$\to\  1$ $,$};
\end{tikzpicture} 

\begin{tikzpicture}[baseline={(current bounding box.center)},scale=1/3]
	\node at (-1,1.5) {$\Big<$};
	
	\bord{0}{4};
	\link{0,2}{1,2}{1,3}{0,3};
	\defects{0,0};
	\defects{0,1};
	
	\node at (1.7,1.5) {$,$};
	
	\bord{2.5}{4};
	\defects{2.5,2};
	\defects{2.5,3};
	\link{2.5,0}{3.5,0}{3.5,1}{2.5,1};
	
	\node at (9/2+0.7,1.5) {$\Big>\ \  \to$};
\end{tikzpicture}
\begin{tikzpicture}[baseline={(current bounding box.center)},scale=1/3]

	\bord{0}{4};
	
	\defects{0,2};
	\defects{0,3};
	\defects{-3/2,0};
	\defects{-3/2,1};
	\link{0,2}{-1,2}{-1,3}{0,3};
	\link{0,0}{1,0}{1,1}{0,1};
	
	\node at (2.9,1) {$\to\  0,$};
	\node at (5.6,1) {$ $};
\end{tikzpicture} 
\begin{tikzpicture}[baseline={(current bounding box.center)},scale=1/3]

	\node at (-1,1) {$\Big<$};
	
	\bord{0}{3};
	\defects{0,0};
	\vac{0,2};
	\vac{0,1};
	
	\node at (2,1) {$,$};
	
	\bord{3}{3};
	\defects{3,2};
	\link{3,0}{4,0}{4,1}{3,1};
	
	\node at (11/2,1) {$\Big>\ \  \to$};	
	
\end{tikzpicture}
\begin{tikzpicture}[baseline={(current bounding box.center)},scale=1/3]

	\bord{0}{3};
	
	\defects{-3/2,0};
	\defects{0,2};
	\link{0,0}{1,0}{1,1}{0,1};
	\vac{0,2};
	\vac{0,1};
	
	\node at (3.2,1) {$\to\  0$ $.$};
\end{tikzpicture} 
\newline
  
\noindent This bilinear form extends that defined on standard modules $\vs{n,k}$ of the Temperley-Lieb algebra $\tl n$ (see appendix \ref{app.tln}). One  difference between the two bilinear forms for $\tl n$ and $\dtl n$ is worth mentioning. It concerns the bilinear form on the standard modules $\U{n,0}$ and $\vs{n,0}$ when $\beta=0$. For $\vs{n,0}$ with $n$ even, the bilinear form is strictly zero, as the pairing of link diagrams always closes at least one loop. A special definition has to be introduced to counter this difficulty \cite{RiSY}. The bilinear form on $\U{n,0}$ as described above is not zero, even when $\beta=0$, as the pairing of the link diagram with $n$ vacancies with itself gives $1$. 
  
The bilinear form is symmetric since exchanging the two arguments amounts to a reflection through a vertical mirror when written in terms of diagrams. We shall say that two elements of $\U{n,k}$ are {\em orthogonal} if their Gram product is zero, even though $\langle *,*\rangle_{n,k}$ can be degenerate.
  
\begin{Lem}\label{lem.gram.sym}
If $x,y\in \U{n,k}$ and $u\in \dtl{n}$ then
\begin{equation}
\langle x,uy\rangle_{n,k} =\langle u^{t}x,y\rangle_{n,k} 
\end{equation}
where $u^{t}$ is the diagram obtained by reflecting $u$ along its vertical axis. If $u$ is a sum of diagrams, the reflection is done on each diagram of the linear combination separately.
\end{Lem}
\begin{proof}
The proof consists in writing the two sides of the equality in terms of diagrams.
\end{proof}
\begin{Lem}\label{lem.gram.diag}
If $x,y,z\in \U{n,k}$, then
\begin{equation}
|x\bar y|z=\langle y,z\rangle_{n,k} x \text{. }\label{eq:gram.action}
\end{equation}
\end{Lem}
\begin{proof}
It is sufficient to verify the relation for link diagrams $x,y,z\in\U{n,k}$, by linearity. Equation \eqref{eq:gram.action} is then non-trivial only if all defects and vacancies of $z$ are respectively linked to defects and vacancies of $y$. In this case, all defects, arcs and vacancies of $x$ will be preserved and remain at their places, so that $|x\bar y|z$ is proportional to $x$. The proportionality constant is the number of closed loops formed which is precisely $\langle y,z\rangle_{n,k}$.
\end{proof}
  
The previous ideas can be extended to the multiplication in $\dtl n$ itself. The ideals $I_k$ that filter $\dtl n$ (see \eqref{eq:filtrationByIdeals}) are spanned by $n$-diagrams with at most $k$ crossing strings and the quotient $I_k/I_{k-1}$ by those with precisely $k$ such strings.
\begin{Lem}\label{lem:axiom3} Let $x,y\in\U{n,k}$ be $n$-links and $u\in\dtl n$. Then there exist $r_u(z,x)\in\mathbb C$ such that
\begin{equation}\label{eq:axiom3}
u|x\bar y|\equiv\sum_{z\in\Y k} r_u(z,x)|z\bar y|\quad\text{\em mod}\ I_{k-1}.
\end{equation}
Moreover, if $x', y'\in\U{n,k}$ are two other $n$-links, then there exists $\phi_u(y,x')\in\mathbb C$ such that
\begin{equation}\label{eq:axiom32}
|x\bar y|u|x'\bar{y'}|=\phi_u(y,x')|x\bar{y'}|\quad\text{\em mod}\ I_{k-1}.
\end{equation}
Clearly $\phi_{\id}(x,y)$ is nothing but $\langle x,y\rangle_{n,k}$.
\end{Lem}
\begin{proof} Since $|x\bar y|$ is an element of the ideal $I_k$, so is $u|x\bar y|$. It can be written as a sum of $n$-diagrams with $k$ crossing strings or less. Those that have precisely $k$ crossing ones have necessarily $\bar y$ as right part by the argument used in the previous proof. Thus $u|x\bar y|\equiv\sum_{z\in\Y k}r_u(z,y,x)|z\bar y|\text{\, mod\,}I_{k-1}$ for some $r_u(z,y,x)\in\mathbb C$. But by the diagrammatic definition of the multiplication, the coefficients $r_u(z,y,x)$ may be computed without even drawing $\bar y$ and may thus depend only on $x$ and $z$. The proof of the second statement repeats the argument for the left part.
\end{proof}
  
The link diagrams in $\X{n,k}$ enjoy a particular property: the Gram product of any pair is $1$ if the two diagrams are the same and $0$ otherwise. Proposition \ref{prop:cyclic} showed that any link diagram in $\X{n,k}$ (or even any non-zero element in its span) is a generator of $\U{n,k}$. The next lemma explains, in terms of the bilinear form $\langle *,*\rangle_{n,k}$, why these link diagrams are generators and identifies a larger set of generators. 
\begin{Lem}\label{lem.gram.gen}
An element $x$ is a generator of $\U{n,k}$ if there exist $y\in\U{n,k}$ such that $\langle x,y\rangle_{n,k}\neq 0$.
\end{Lem}
\begin{proof} 
Let $y\in \U{n,k}$ be such that $\langle y,x\rangle_{n,k} = \alpha \neq 0$. For any $z\in \U{n,k}$, both $z$ and $\bar y$ have the same number of defects and $|z\bar y|$ is thus an element of $\dtl n$. Therefore $\frac{1}{\alpha}|z\bar y|x= z$ and $(\dtl{n})x = \U{n,k}$. 
\end{proof}
Hence any link state that is not orthogonal to all others is a generator. Those that are orthogonal to all others are known to be as important. Their set
$$\dr{n,k}=\{ x\in\U{n,k}\ |\ \langle y,x\rangle_{n,k}=0,\textrm{\ for all }y\in\U{n,k}\}$$
is called 
the {\em (dilute) radical} of $\U{n,k}$. It is easy to see that it is a submodule. Lemma \ref{lem.gram.gen} actually shows that it is its maximal submodule, that is, every proper submodule of $\U{n,k}$ is a submodule of its radical. Moreover the module $\dl{n,k} = \U{n,k} / \dr{n,k}$ is irreducible, since any of its non-zero elements generates it.
  
The Gram product can also be used to restrict morphisms between quotients of standard modules.
\begin{Lem}\label{lem.hom.unk}
Let $N, N'$ be submodules of $\U{n,k}$ and $\U{n,k'}$, respectively, with $k<k'$. Then the only homomorphism $\U{n,k}/N\rightarrow\U{n,k'}/N'$ is the zero homomorphism. 
\end{Lem}
\begin{proof}
Let $\gamma$ be the canonical homomorphism from $\U{n,k}$ to $\U{n,k}/N$ and $\theta$ be a homomorphism from $\U{n,k}/ N$ to $\U{n,k'}/N'$. Choose $y,z\in\U{n,k}$ such that $\langle y,z\rangle_{n,k}=1$. Then for all $x\in \U{n,k}$, 
\begin{equation}
|x\bar y|\theta \left( \gamma \left(z\right)\right) = \theta \left( \gamma \left(|x\bar y|z\right)\right)= \theta \left( \gamma \left(x\right)\right). \label{Lem-Submod}
\end{equation}
Since $\theta \left( \gamma \left(z\right)\right)\in \U{n,k'}/ N'$, the usual representative of this conjugacy class has $k'$ defects. But $|x\bar y|\theta \left( \gamma \left(z\right)\right)$ can have at most $k<k'$ defects and the left side of \eqref{Lem-Submod} must be zero. Therefore $\theta(\gamma(x))$ is zero for all $x$ and, since $\gamma$ is surjective, $\theta$ is zero. 
\end{proof}

\end{subsection}
  
\begin{subsection}{The structure of the radical}
  
Let $\dG{n,k}$ be the matrix representing the bilinear form $\langle *, *\rangle_{n,k}$ in the basis of link diagrams. Similarly denote by $\G{n,k}$ the matrix for the bilinear form for the corresponding standard $\tl n$-module, also in its link basis. These matrices will be called {\em Gram matrices} and, if need be, the adjective {\em dilute} will be added to the first one. The Gram product of two link diagrams in $\U{n,k}$ may be non-zero only if their vacancies coincide. In that case, the product does not depend on their positions and it is equal to the Gram product defined for standard modules of $\tl {n'}$ applied to the two link diagrams obtained from the original ones by deleting their vacancies. (Then $n'$ is $n-\#(\textrm{vacancies})$.) It is then clear that the matrix $\dG{n,k}$ is block-diagonal if the link basis is ordered, first, by gathering links with the same number of vacancies and, second, those with the same positions for these vacancies. The shape of the Gram matrix $\dG{n,k}$ then appears as a consequence of the decomposition of the dilute standard modules into a direct sum of $S_n$-modules (see proposition \ref{prop:unk.vs}). The next result then follows immediately. (The direct sum symbol is used to indicate the block diagonal decomposition of $\dG{n,k}$ and the binomial factors give the multiplicity of each block or vector space.)
\begin{Prop}The dilute Gram matrix for the $\dtl n$-modules $\U{n,k}$ is
\begin{equation}
\dG{n,k} = \bigoplus_{p=0}^{\lfloor \frac{n-k}{2} \rfloor}  \binom{n}{k+2p}\G{k+2p,k}
\end{equation}
where $\G{n,k}$ is the Gram matrix of the $\tl{n}$-module $\vs{n,k}$.
\end{Prop}
  
The following corollaries are immediate consequences.
  
\begin{Coro}\label{coro:DetGram}
The determinant of the Gram matrix is
\begin{equation}
\det\dG{n,k} = \prod_{p=0}^{\lfloor \frac{n-k}{2} \rfloor} \left(\det \G{k+2p,k} \right)^{\binom{n}{k+2p}}
\end{equation}
\end{Coro}
\begin{Coro}The dilute radical $\dr{n,k}$ decomposes as
\begin{equation}\label{eq:decompositiondRnk}
\dr{n,k} \simeq \underset{p=0}{\overset{\lfloor \frac{n-k}{2} \rfloor}{{\bigoplus}'}} \binom{n}{k+2p} \R{k+2p,k}\qquad \textrm{as vector spaces,}
\end{equation}
where $\R{n,k}$ is the radical of the Gram bilinear form on $\vs{n,k}$ and the ``${\oplus}'$'' indicates that the trivial radicals ($=\{0\}$) are omitted of the direct sum. Furthermore this decomposition for $\dr{n,k}\subset \Res{\U{n,k}}^{\dtl n}_{S_n}$ holds as $S_n$-modules.
\end{Coro}
\begin{Coro}\label{Dimens-Rad}
\begin{equation}\label{eq:Dimens-Rad}
\dim \dr{n,k} = \sum_{p=0}^{\lfloor \frac{n-k}{2} \rfloor} \binom{n}{k+2p}\dim \R{k+2p,k}.
\end{equation} 
\end{Coro}
\noindent The last corollary leads to various recurrence relations for the dimensions of the dilute radicals and the irreducible modules. They are simple, though neither compact nor particularly enlightening. They will be presented along with their proofs in appendix \ref{app.dlnk}. 
  
A distinction between the two algebras $\tl n$ and $\dtl n$ at $\beta=0$ follows from the above proposition and corollaries. When $\beta=0$ (and therefore $q=\pm i$), the determinant $G_{n,k}$ vanishes for all even $k$s and is otherwise non-zero. It follows that $\tl n(\beta=0)$ is semisimple if $n$ is odd, because then all its standard modules $\vs{n,k}$ have odd $k$s, and $\tl n(\beta=0)$ is non-semisimple if $n$ is even. It will be shown that the dilute $\dtl n(\beta=0)$ is non-semisimple for all $n>1$.
  
The previous results show that the dilute radical $\dr{n,k}$ is trivial if the radicals $\R{k+2p,k}$, $0\leq p\leq \lfloor (n-k)/2\rfloor$, are all trivial. Since the determinant of $\G{n,k}$ can vanish only at a root of unity distinct from $\pm 1$ (see \eqref{eqn:DetG}), then the following corollaries are straightforward.
\begin{Coro}\label{coro:UnkAreIrreducible}
The dilute standard module $\U{n,k}$ is irreducible if $q$ is not a root of unity.
\end{Coro}
\begin{Coro}\label{coro:irrIfCrit}
The dilute standard module $\U{n,k}$ is irreducible if $k$ is critical.
\end{Coro} 
\begin{proof} We recall that the radical $\R{n,k}$ of the standard $\tl n$-module $\vs{n,k}$ is trivial whenever $k$ is critical, that is when $q^{2(k+1)}=1$. (See proposition \ref{prop:dimIrrTLn}.) This result is independent of $n$ and all vector spaces $\R{k+2p,k}$ appearing in \eqref{eq:decompositiondRnk} are trivial.
\end{proof}
  
\begin{Thm}[Structure of $\dtl n$ for $q$ generic]\label{thm:dtlnGeneric} If $q$ is not a root of unity, then $\dtl n$ is semisimple, the set $\{\U{n,k},0\leq k\leq n\}$ forms a complete set of non-isomorphic irreducible modules and, as a left module, the algebra $\dtl n$ decomposes as
$$\dtl n=\bigoplus_{0\leq k\leq n}(\dim \U{n,k}) \U{n,k}.$$
\end{Thm}
\begin{proof}
Corollary \ref{coro:UnkAreIrreducible} states that the $\U{n,k}$ are irreducible when $q$ is not a root of unity and proposition \ref{prop:nonIsomorphe} that they are non-isomorphic. Weddeburn's theorem \ref{prop:Wedderburn1} and its generalization \ref{prop:Wedderburn1} show that, given a subset $\{I_k, k\in K\}$ of its non-isomorphic irreducible modules, the dimension of an algebra is bounded from below by $\sum_{k\in K}(\dim I_k)^2$. In the present case $\sum_{0\leq k\leq n}(\dim \U{n,k})^2 =\dim\dtl n$ by proposition \ref{prop:dimdTLnUnk}. The three statements then appear as a consequence of Wedderburn's theorem.
\end{proof}
\end{subsection}
  
\begin{subsection}{Symmetric pairs of standard modules} 
  
Let $q$ be a root of unity other than $\pm 1$ and let $\ell$ be the smallest integer such that $q^{2\ell}=1$. Then $\ell\geq 2$. Two non-negative integers $k$ and $k'$ form a symmetric pair if they satisfy
\begin{equation}\label{eq:symm}(k+k')/2+1\equiv 0\textrm{\,mod\,}\ell\qquad\textrm{and} \qquad 0<|k-k'|/2<\ell.
\end{equation}
The Bratteli diagram in Figure \ref{fig:bratteli} explains the meaning of these two conditions. The first equations implies that the average of $k$ and $k'$ falls on a critical line, that is, $k_c=(k+k')/2$ satisfies $q^{2(k_c+1)}=1$. (On the Bratteli diagram with $\ell=4$, the critical lines are through $k=3, 7, \dots$) Consider now the closest critical lines to that going through $k_c$. (If the latter one is the leftmost, the critical line to its left would be one passing through $k=-1$.) The second condition above means that the integers $k$ and $k'$ are strictly between these two closest critical lines. Hence $k$ and $k'$ fall symmetrically on each side of the line $k_c=(k+k')/2$. A pair of standard modules $\U{n,k}$ and $\U{n,k'}$ is also said symmetric if $k$ and $k'$ form a symmetric pair. Note, finally, that there are always a pair of positive integers $a, b$ with $0<b<\ell$ such that $k$ and $k'$ are $k_\pm=a\ell-1\pm b$. When the pair $\U{n,k_+}$ and $\U{n,k_-}$ is symmetric, then the eigenvalues of $\DF n$ on these modules coincide (see proposition \ref{prop:EigenvalueDFn}).
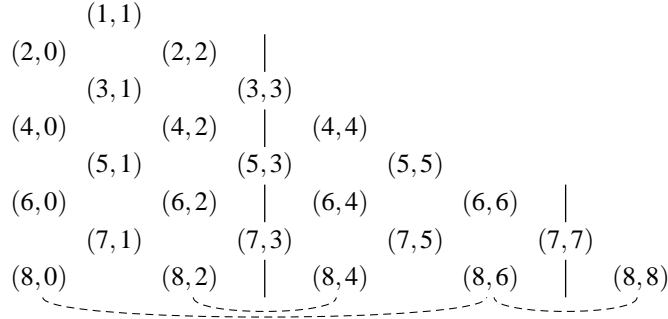
\begin{figure}[h!]\label{fig:bratteli}
\begin{center}
\begin{tikzpicture}[baseline={(current bounding box.center)},scale=1]
	\node at (1,-0.5)  {$(1,1)$};
	\node at (0,-1)  {$(2,0)$}; 
	\node at (2,-1)  {$(2,2)$};
	\node at (1,-1.5)  {$(3,1)$};
	\node at (3,-1.5)  {$(3,3)$};
    \node at (0,-2)  {$(4,0)$};
    \node at (2,-2)  {$(4,2)$};
    \node at (4,-2)  {$(4,4)$};
    \node at (1,-2.5)  {$(5,1)$};
    \node at (3,-2.5)  {$(5,3)$};
    \node at (5,-2.5)  {$(5,5)$};
    \node at (0,-3)  {$(6,0)$};
    \node at (2,-3)  {$(6,2)$};
    \node at (4,-3)  {$(6,4)$};
    \node at (6,-3)  {$(6,6)$};
    \node at (1,-3.5)  {$(7,1)$};
    \node at (3,-3.5)  {$(7,3)$};
    \node at (5,-3.5)  {$(7,5)$};
    \node at (7,-3.5)  {$(7,7)$};
    \node at (0,-4)  {$(8,0)$};
    \node at (2,-4)  {$(8,2)$};
    \node at (4,-4)  {$(8,4)$};
    \node at (6,-4)  {$(8,6)$};
    \node at (8,-4)  {$(8,8)$};
    \draw[line width=0.5pt] (3,-0.75) -- (3,-1.25);
    \draw[line width=0.5pt] (3,-1.75) -- (3,-2.25);
    \draw[line width=0.5pt] (3,-2.75) -- (3,-3.25);
    \draw[line width=0.5pt] (3,-3.75) -- (3,-4.25);
    \draw[line width=0.5pt] (7,-2.75) -- (7,-3.25);
    \draw[line width=0.5pt] (7,-3.75) -- (7,-4.25);
    \draw[densely dashed] (0.05,-4.25) .. controls (0.05,-4.6) and (5.95,-4.6) .. (5.95,-4.25);
    \draw[densely dashed] (2.05,-4.25) .. controls (2.05,-4.5) and (3.95,-4.5) .. (3.95,-4.25);
    \draw[densely dashed] (7.95,-4.25) .. controls (7.95,-4.5) and (6.05,-4.5) .. (6.05,-4.25);
\end{tikzpicture}
\end{center}\caption{The indices $(n,k)$ of even standard modules are presented on a Bratteli diagram. Each line corresponds to a given $n$ and therefore a given $\dtl n$. The vertical lines are the critical lines when $\ell=4$. Symmetric pairs for $n=8$ are joined by dashed lines.}
\end{figure}

For the (original) Temperley-Lieb algebras $\tl n$, it is known that $\R{n,k_-}\simeq\rl{n,k_+}$ for all symmetric pairs with $k_-<k_+\le n$ and, if $k_-\le n<k_+$, then $\R{n,k_-}=\{0\}$. The main result of this section is that these isomorphisms still hold for the dilute family. Even without studying the structure of these modules, one can prove readily the coincidence of their dimensions.
\begin{Lem}\label{lem:coincidenceOfDimensions}If $k_-<k_+$ is a symmetric pair, then $\dim\dr{n,k_{-}} = \dim \dl{n,k_{+}}$.
\end{Lem}
\begin{proof}As above set $b=(k_+-k_-)/2$. Then $\lfloor(n-k_-)/2\rfloor=\lfloor (n-k_+)/2\rfloor+b$. Since $\dim\dl{n,k_+}=\dim\U{n,k_+}-\dim\dr{n,k_+}$, corollaries \ref{coro:dimUnk} and \ref{Dimens-Rad} provide the first equality below. The third one uses the equality of dimensions of radical and irreducible for $\tl n$-modules.
\begin{align*}
\dim\dl{n,k_+}&=
  \sum_{p=0}^{\lfloor\frac{n-k_+}2\rfloor}\binom n{k_++2p}(\dim\vs{k_++2p,k_+}-\dim\R{k_++2p,k_+})\\
  &=\sum_{p=0}^{\lfloor\frac{n-k_+}2\rfloor}\binom n{k_++2p}\dim\rl{k_++2p,k_+}=
  \sum_{p=0}^{\lfloor\frac{n-k_+}2\rfloor}\binom n{k_++2p}\dim\R{k_++2p,k_-}\\
  &=\sum_{q=b}^{\lfloor\frac{n-k_-}2\rfloor}\binom n{k_-+2q}\dim\R{k_-+2q,k_-}\\
  &=\sum_{q=0}^{\lfloor\frac{n-k_-}2\rfloor}\binom n{k_-+2q}\dim\R{k_-+2q,k_-}
  =\dim\dr{n,k_-}.
\end{align*}
In the last line $b$ terms were added. But they are all zero as they are the dimensions of radicals $\R{n',k'}$ indexed by $k'\le n'$ whose symmetric partners fall beyond $n'$.
\end{proof}
  
To probe the structure of the radicals, a few tools will be useful. 
  
Let $z\in\X{n,k}$ be an $n$-link diagram and set $\pi_z=|z\bar z|\in\dtl n$. (Note that $\pi_z$ coincides with the projectors $\pi_A$ introduced in subsection \ref{sec:definition} if $A$ is taken to be the set of positions of the defects of $z$.) Here are some simple observations about $\pi_z$. The set $T_z=\pi_z\dtl n\pi_z$ is spanned by $n$-diagrams that have precisely $(n-k)$ vacancies on their sides, located where the vacancies of $z$ are. The vector space $T_z$ is a subalgebra of $\dtl n$ isomorphic to $\tl k$. This leads to a reformulation of proposition \ref{prop.symmetric.salg}, namely:
$$S_n\simeq \bigoplus_{0\leq k\leq n}\bigoplus_{z\in\X{n,k}}\pi_z\,\dtl n\,\pi_z.$$
Similarly the identity $\idtl\in\dtl n$ can be written as $\idtl=\sum_{0\leq k\leq n}\sum_{z\in\X{n,k}}\pi_z$.
  
Note that $\pi_z\pi_z=\pi_z$ so that $\pi_z$, $z\in\X{n,k}$, acts as a projector. Moreover, for two distinct link diagrams $z\in\X{n,i}$ and $z'\in\X{n,j}$, $0\le i < j\le n$, $\pi_z\pi_{z'}=\pi_{z'}\pi_z=0$ and $\pi_{z'}M\cap\pi_z M=\{0\}$ for any module $M$. 
  
\begin{Lem}\label{lem:OneThatUsedAppendixA}
Let $q$ be a root of unity other than $\pm 1$ and $k$ be critical for this $q$. Let $\psi$ be the endomorphism of $\Ind{\U{n-1,k}}/ \U{n,k}$ defined by left multiplication by the central element $\DF n-\df{k-1}\cdot\idtl$. Then $\psi$ is non-zero.
\end{Lem}
\begin{proof}
The proof builds on that for the Temperley-Lieb algebra. To make contact with this previous result, we need to choose a link diagram $z\in\X{n-1,k}$. The actual one is irrelevant, but the explanations are simpler when $z$ has all its vacancies at the top and its defects at the bottom positions. The vector $v=\pi_{\theta_1(z)}\otimes_{\dtl{n-1}}z$ is then an element of the basis constructed in subsection \ref{sec:induction} for the induced module $\Ind{\U{n-1,k}}$. We claim that $(\DF n-\df{k-1}\cdot\idtl)v$ is non-zero. Note first that
$$(\DF n-\df{k-1}\cdot\idtl)v
=(\DF n-\df{k-1}\cdot\idtl)\pi_{\theta_1(z)}v
=\big(\pi_{\theta_1(z)}(\DF n-\df{k-1}\cdot\idtl)\pi_{\theta_1(z)}\big)v$$
since $\pi_{\theta_1(z)}\pi_{\theta_1(z)}=\pi_{\theta_1(z)}$. Due to proposition \ref{prop:DFnAvecPIz}, $\pi_{\theta_1(z)}\DF n\pi_{\theta_1(z)}$ corresponds to the action of $\F{k+1}$ on the bottom $k+1$ positions, the top ones being forced to be vacancies. The fact that these vacancies do not play any role is useful. Recall that $T_{\pi_{\theta_1(z)}}=\pi_{\theta_1(z)}\dtl n \pi_{\theta_1(z)}$ is a subalgebra isomorphic to $\tl{k+1}$. Similarly $T_z=\pi_z\dtl{n-1}\pi_z\simeq\tl k$ and $\pi_z \U{n-1,k}$ is a $\tl k$-module (with the restricted action) isomorphic to $\vs{k,k}$. With these isomorphisms, the computation of $(\DF n-\df{k-1}\cdot\idtl)v$ amounts to computing the action of $(\F k-\df{k-1}\cdot\idtl_{\tl{k+1}})$ on $\idtl_{\tl{k+1}}\otimes_{\tl k}z_k\in \Ind{\vs{k,k}}$ where $z_k$ is the $k$-link state with $k$ defects. Note that the criterion for criticality does not depend on $n$ and the $\tl k$-module $\vs{k,k}$ also sits on the critical line. Proposition \ref{prop:HomoForTLn} then states readily that $(\F k-\df{k-1}\cdot\idtl_{\tl{k+1}})\idtl_{\tl{k+1}}\otimes_{\tl k}z_k$ is non-zero. One can then conclude that $(\DF n-\df{k-1}\cdot\idtl)\pi_{\theta_1(z)}\otimes_{\dtl{n-1}}z$ is non-zero since $T_{\pi_{\theta_1(z)}}\cdot v$ and $\Ind{\vs{k,k}}$ are isomorphic as modules over the subalgebra $T_{\pi_{\theta_1(z)}}\subset \dtl n$. Clearly the vector $v\in\Ind{\U{n,k}}$ lies in the submodule of $\Ind{\U{n-1,k}}$ that has the parity of $\U{n,k+1}$ and thus projects onto a non-zero vector in $\Ind{\U{n-1,k}}/ \U{n,k}$.
\end{proof}
\begin{Prop}\label{prop.rad.iso}
Let $q$ be a root of unity other than $\pm1$ and let $\U{n,k_-}$ and $\U{n,k_+}$ be two standard $\dtl n$-modules where $k_-$ and $k_+$ form a symmetric pair ($k_-<k_+$). Then 
\begin{equation}
\dr{n,k_-} \simeq \dl{n,k_+}.
\end{equation}
\end{Prop}
\begin{proof}
Let $k=(k_-+k_+)/2$ be the critical $k$ between $k_-$ and $k_+$ and let $b$ such that $k_\pm=k\pm b$. If $b=1$, the short sequence
\begin{equation}
0 \to \U{n,k-1} \overset{\alpha}{\to} \Ind{\U{n-1,k}}/\U{n,k} \overset{\gamma}{\to} \U{n,k+1}\to 0 
\end{equation}
is exact by corollary \ref{coro.ind.ses}. Let $\psi$ be the endomorphism obtained by left multiplying a $\dtl n$-module by $(\DF n-\df{k-1}\cdot\idtl)$. By the previous lemma, this is a non-zero endomorphism on $\Ind{\U{n-1,k}}/\U{n,k}$. But it does act as zero on $\U{n,k-1}$ and therefore $\im \alpha\subset\ker\psi$. It also acts as zero on $\U{n,k+1}$ and $\im\psi\subset\ker\gamma=\im\alpha$. Since $\gamma$ is surjective, for any $w\in\U{n,k+1}$, there is a $v\in \Ind{\U{n-1,k}}/\U{n,k}$ such that $\gamma(v)=w$. If $v'\in \Ind{\U{n-1,k}}/\U{n,k}$ is another vector satisfying $\gamma(v')=w$, then $v-v'\in\ker\gamma\subset \ker\psi$. It thus follows that the map $w\mapsto\psi(v)$ is well-defined. It can be seen to be a module homomorphism $\Psi:\U{n,k+1}\rightarrow \im\alpha\subset \Ind{\U{n-1,k}}/\U{n,k}$. Since $\alpha$ is injective, it has an inverse on im$\psi \subset$ im $\alpha$. Therefore $\alpha^{-1}\circ \Psi:\U{n,k+1}\rightarrow\U{n,k-1}$ is a non-zero homomorphism and $\Hom_{\dtl{n}} \left( \U{n,k+1}, \U{n,k-1}\right)\neq 0$. 
  
Let $b$ be an integer such that $1<b<\ell$ where $\ell$ is the smallest integer such that $q^{2\ell}=1$. Then 
\begin{align*}
&\Hom_{\dtl{n+b}}\left(\U{n+b,k+b}, \U{n+b,k-b}\right) \\
   &\qquad =\Hom_{\dtl{n+b}} \left( \U{n+b,k+b}\oplus\U{n+b,k+b-1}\oplus\U{n+b,k+b-2}, \U{n+b,k-b}\right) \\
   &\qquad =\Hom_{\dtl{n+b}} \left( \Ind{\U{n+b-1,k+b-1}}, \U{n+b,k-b}\right) \\
   &\qquad =\Hom_{\dtl{n+b-1}} \left( \U{n+b-1,k+b-1}, \Res{\U{n+b,k-b}} \right) \\
   &\qquad =\Hom_{\dtl{n+b-1}} \left( \U{n+b-1,k+b-1}, \U{n+b-1,k-b}\oplus \U{n+b-1,k-b+1}\oplus \U{n+b-1,k-b-1} \right) \\
   &\qquad =\Hom_{\dtl{n+b-1}} \left( \U{n+b-1,k+(b-1)}, \U{n+b-1,k-(b-1)}\right). 
\end{align*}
The third equality is due to Frobenius reciprocity theorem and the second and the fourth follow from corollary \ref{coro.ind.split} and the fact that neither $(k+b-1)$ nor $(k-b)$ are critical. The first equality rests upon two slightly different observations. Lemma \ref{lem:deltanPropriete} shows that $\DF{n+b}$ act upon the two modules $\U{n+b,k+b-2}$ and $\U{n+b,k-b}$ with distinct eigenvalues and therefore any homomorphism between them is zero. Similarly, there cannot be a homomorphism between two standard modules of distinct parities and $\Hom_{\dtl{n+b}} (\U{n+b,k+b-1}, \U{n+b,k-b})=0$. The last equality follows from the same two observations. Therefore
\begin{equation}
\Hom_{\dtl{n+b}}\left(\U{n+b,k+b}, \U{n+b,k-b}\right)=\Hom_{\dtl{n}}\left(\U{n,k+1}, \U{n,k-1}\right) \neq \{0\} \text{.}
\end{equation}
  
Let $k_-$ and $k_+$ be a symmetric pair and $f: \U{n,k_+} \to \U{n,k_-}$ a non-zero homomorphism. Its kernel is a proper submodule of $\U{n,k_+}$ and, since the radical of a standard module is a maximal submodule, $\ker f \subset \dr{n,k_+}$. Now, if $f$ is surjective, 
$\U{n,k_{-}} \simeq \U{n,k_{+}}/\ker f$, which contradicts proposition \ref{lem.hom.unk}. We thus conclude that $\im f$ is a proper sub-module of $\U{n,k_{-}}$ and is thus a sub-module of $\dr{n,k_{-}}$ by maximality of the radical. But, we have
$$\dim \dr{n,k_{-}} \geq \dim \im f = \dim \U{n,k_{+}} - \dim \ker f \geq \dim \U{n,k_{+}} - \dim \dr{n,k_{+}} = \dim\dl{n,k_{+}}. $$
But $\dim\dl{n,k_{+}} = \dim \dr{n,k_{-}}$ by lemma \ref{lem:coincidenceOfDimensions}. It then follows that $\dim \im f = \dim  \dr{n,k_{-}}$ and $\dim \ker f = \dim \U{n,k_+} - \dim \im f =\dim \U{n,k_+} - \dim \dl{n,k_+} = \dim \dr{n,k_{+}}$. Thus $\ker f \simeq \dr{n,k_{+}}$ and the first isomorphism theorem then concludes the proof.
\end{proof}
  
Suppose that $(k_-,k_+)$ is a symmetric pair with $k_-\leq n<k_+$. Then the radical $\dr{n,k_-}$ is trivial and $\U{n,k_-}$ irreducible. This can be proved either by extending the previous proof (allowing $\U{n,j}=\{0\}$ whenever $j>n$), or by a careful analysis of the zeroes of $\det\, \dG{n,k}$ (corollary \ref{coro:DetGram}), or by checking with \eqref{eq:dimLnkTL} which radicals of $\tl n$ occuring in corollary \ref{Dimens-Rad} are non-trivial. The last results of this section follow easily from the previous result.
\begin{Coro}The radical $\dr{n,k}$ is either irreducible or trivial.
\end{Coro}
\begin{Coro}\label{cor:exactStan} If $k_-$ and $k_+$ form a symmetric pair ($k_-<k_+$), then the following short sequence is exact:
\begin{equation}\label{eq:exactStan}
0\longrightarrow{\dl{n,k_+}}\longrightarrow{\U{n,k_-}}\longrightarrow{\dl{n,k_-}}\longrightarrow 0.
\end{equation}
\end{Coro}
\begin{proof}The radical is the unique maximal submodule and, by definition, $0\rightarrow\dr{n,k_-}\rightarrow{\U{n,k_-}}\rightarrow\dl{n,k_-}\rightarrow0$. The statement then follows from proposition \ref{prop.rad.iso}.
\end{proof}
  
\begin{Coro}\label{coro.hom.unk}
If $f \in \Hom{\left(\U{n,k},\U{n,k}\right)}$, then $f$ is an isomorphism or zero.
\end{Coro}
\begin{proof}
If $\U{n,k}$ is irreducible, the result is trivial. If $\U{n,k}$ is reducible, then $k$ forms a symmetric pair with some $k_+>k=k_-$. 
Choose a non-zero element $f \in \Hom(\U{n,k},\U{n,k})$. If $\ker f$ is non-zero, then $\ker f=\im f =\dr{n,k}$, since $\dr{n,k}$ is the only non-trivial proper submodule. Then the first isomorphism theorem says $\U{n,k_-}/\dr{n,k_-}\simeq \dr{n,k_-}\simeq \dl{n,k_+}=\U{n,k_+}/\dr{n,k_+}$, contradicting lemma \ref{lem.hom.unk}. So $f$ must be an isomorphism.
\end{proof}
A similar argument gives the following corollary.
  
\begin{Coro}\label{coro.hom.irred}If $\U{n,k}$ is reducible, then
\begin{equation}
\Hom{(\dl{n,k},\U{n,k})} \simeq \Hom{(\U{n,k},\dr{n,k})} \simeq 0 \text{.}
\end{equation}
\end{Coro}
  
\end{subsection}

\begin{subsection}{Restriction and induction of irreducible modules}
  
We complete the analysis of the restriction and induction of the fundamental modules by giving those of the radicals and the irreducible quotients. The results are simple and elegant. Their proofs are straightforward but somewhat long and repetitive. 
  
\begin{Prop}\label{prop:restRadical}
If $\dr{n+1,k} \neq 0$, then
\begin{equation}
\Res{\dr{n+1,k}} \simeq \dr{n,k-1} \oplus \dr{n,k} \oplus \left\{\begin{tabular}{r l}
$\U{n,k+1}$ & if $k+1$ is critical\\
$\dr{n,k+1}$ & otherwise
\end{tabular} \right\} \text{.}
\end{equation}
Some of the direct summands may be trivial.
\end{Prop}
\begin{proof}
If $\dr{n+1,k} \neq 0$, proposition \ref{prop.rad.iso} gives the exactness of the following short sequence of $\dtl{n+1}$-modules:
\begin{equation}
0 \to \dr{n+1,k} \longrightarrow \U{n+1,k} \longrightarrow \dl{n+1,k} \to 0 
\end{equation}
and therefore of its restriction to $\dtl n$:
\begin{equation}\label{eq:exactRestriction}
0 \to \Res{\dr{n+1,k}} \longrightarrow \Res{\U{n+1,k}} \longrightarrow \Res{\dl{n+1,k}} \to 0 \text{.}
\end{equation}
It follows that $\Res{\dr{n+1,k}}$ is isomorphic to a submodule of $\Res{\U{n+1,k}}$ which splits in a direct sum of three modules which are distinct eigenspaces of $\DF n$ of different parity: $\Res{\dr{n+1,k}}\simeq R_{0} \oplus R_{-} \oplus R_{+}$ where $R_{0}$ and the $\R{\pm}$ are submodules of $\U{n,k}$ and $\U{n,k\pm 1}$ respectively. One or more of the $R$s may vanish. (See propositions \ref{prop:restr.split} and \ref{prop:EigenvalueDFn}.)
  
We first study $R_0$. Consider the (restriction of) the injective homomorphism $\phi: \U{n,k} \to \Res{\U{n+1,k}}$ introduced in the proof of proposition \ref{prop:restr.unk} that simply adds a vacancy at the bottom of every link diagram. Let $u\in\Res{\U{n+1,k}}$ and write it as $u' + v'$ where all terms in $u'$ have a vacancy at the position $n+1$ while those in $v'$ do not. Then, if $r$ is in the radical $\dr{n,k}\subset\U{n,k}$
\begin{equation}
\langle\phi(r),u\rangle_{n+1,k} =  \langle\phi(r),u'\rangle_{n+1,k} = \langle r,\phi^{-1}(u')\rangle_{n,k} = 0.
\end{equation} 
The image $\phi(\dr{n,k})$ is thus in $\Res{\dr{n+1,k}}$. Since $R_0$ is the only summand of $\Res{\dr{n+1,k}}$ having the parity of $\dr{n,k}$, it must contain a submodule isomorphic to $\dr{n,k}$.
  
We turn to the other two submodules $R_-$ and $R_+$. Corollary \ref{prop:restr.split} has established the exactness of the short sequence
\begin{equation}
0 \to \U{n+1,k -1}\oplus\U{n+1,k} \longrightarrow \Ind{\U{n,k}} \longrightarrow \U{n+1,k+1} \to 0
\end{equation}
which implies the exactness of (see proposition \ref{prop:rightExactHom})
\begin{multline}
0 \to \Hom{(\U{n+1,k+1},\dr{n+1,k})} \longrightarrow \Hom{(\Ind{\U{n,k}},\dr{n+1,k})}\\ \longrightarrow \Hom{(\U{n+1,k -1}\oplus\U{n+1,k},\dr{n+1,k})} \text{.}
\end{multline}
Corollary \ref{coro.hom.irred}, the linearity of $\Hom$ and the parity of the modules involved lead to
\begin{equation}
\Hom(\U{n+1,k -1}\oplus\U{n+1,k},\dr{n+1,k}) = 0 \qquad\text{and}\qquad
\Hom{(\U{n+1,k+1},\dr{n+1,k})} = 0 \text{.}
\end{equation}
Frobenius theorem then gives
\begin{equation}
\Hom{(\Ind{\U{n,k}},\dr{n+1,k})} \simeq \Hom{(\U{n,k},\Res{\dr{n+1,k}})} \simeq 0.
\end{equation}
Therefore $\Res{\dr{n+1,k}}$ has no (non-trivial) submodule isomorphic to a quotient of $\U{n,k}$. This proves that $R_{0}$ is isomorphic to $\dr{n,k}$.
  
Similarly the short exact sequences
\begin{equation}
0 \to \U{n+1,k \pm 1 -1}\oplus\U{n+1,k \pm 1} \longrightarrow \Ind{\U{n,k \pm 1}} \longrightarrow \U{n+1,k \pm 1 +1} \to 0
\end{equation}
give rise to the exact sequences
\begin{multline}\label{seq.res.r}
0 \to \Hom{(\U{n+1,k \pm 1 +1},\dr{n+1,k})} \longrightarrow \Hom{(\Ind{\U{n,k\pm 1}},\dr{n+1,k})}\\ \longrightarrow \Hom{(\U{n+1,k\pm 1 -1}\oplus\U{n+1,k \pm 1},\dr{n+1,k})} \text{.}
\end{multline}
Note that $\U{n+1,k \pm 1}$ and $\dr{n+1,k}$ always have different parities and
\begin{equation}
\Hom{(\U{n+1,k\pm 1 -1}\oplus\U{n+1,k \pm 1},\dr{n+1,k})} \simeq \Hom{(\U{n+1,k\pm 1 -1},\dr{n+1,k})} \simeq 0 \text{,}
\end{equation}
where the second equality follows from either proposition \ref{lem.hom.unk} or corollary 
\ref{coro.hom.irred}. The argument now splits according to whether $k+1$ is critical or not.
  
If $k+1$ is not critical, the central element $\DF{n+1}$ takes distinct eigenvalues on $\U{n+1,k+2}$ and $\U{n+1,k}$ which forces $\Hom{(\U{n+1,k + 2},\dr{n+1,k})} = 0$. Corollary \ref{coro.hom.irred} also gives $\Hom{(\U{n+1,k},\dr{n+1,k})} = 0$, so Frobenius theorem leads to
\begin{equation}
\Hom{(\Ind{\U{n,k\pm 1}},\dr{n+1,k})} \simeq \Hom{(\U{n,k\pm 1},\Res{\dr{n+1,k}})} \simeq 0 \text{.}
\end{equation}
Therefore, the $\U{n,k\pm 1}$ are not isomorphic to submodules of $\Res{\dr{n+1,k}}$ and in particular $R_{\pm} \neq \U{n,k\pm 1}$. 
  
If $k+1$ is critical, proposition \ref{prop.rad.iso} gives $\dr{n,k} \simeq \dl{n,k+2}$ so that
\begin{equation}
\Hom{(\U{n+1,k +2},\dr{n+1,k})} \simeq \Hom{(\Ind{\U{n,k+1}},\dr{n+1,k})} \simeq \Hom{(\U{n,k+1},\Res{\dr{n+1,k}})} \neq 0
\end{equation}
by the exactness of \eqref{seq.res.r}. Since $\U{n,k+1}$ is irreducible when $k+1$ is critical, the restriction $\Res{\dr{n+1,k}}$ has a submodule isomorphic to $\U{n,k+1}$. But since the parity of $\U{n,k}$ and $\U{n,k+1}$ are different, this submodule cannot be in $R_0$. Again $\DF{n}$ takes distinct eigenvalues on $\U{n,k+1}$ and $\U{n,k-1}$ so that $\U{n,k+1}$ cannot be a submodule of $R_{-}$. (This statement remains true in the special case when $k-1$ is also critical. Then $l=2$, $q=\pm i$ and $\df{k+1}=-\df{k-1}.$) This proves that $\U{n,k+1}$ must be a submodule of $R_{+}$, which is itself a submodule of $\U{n,k+1}$ and thus $\U{n,k+1} \simeq R_{+}$.
  
So far, we have narrowed down the possible submodules of $\Res{\dr{n+1,k}}$ to
\begin{equation}
\Res{\dr{n+1,k}} \simeq \dr{n,k} \oplus \left\{ 0 \text{ or } \dr{n,k-1}\right\} \oplus \left\{ \begin{tabular}{ r r}
$\U{n,k+1} $& if $k+1$ is critical\\
$0$ or $\dr{n,k+1}$& otherwise
\end{tabular} \right\}.
\end{equation}
Equation \eqref{eq:dimdLnk} and proposition \ref{prop.dim.irred} give a formula for the dimension of $\dr{n+1,k}$. The proof ends with a comparison of this dimension with the above possibilities.
\end{proof}
  
Note that equation \eqref{eq:exactRestriction} gives $\Res{\dl{n+1,k}} \simeq \Res{\U{n+1,k}} / \Res{\dr{n+1,k}}$. Combining this observation with the preceding proposition then gives the following corollary.
\begin{Coro}\label{coro.res.dl}
If $\dr{n+1,k} \neq 0$ then
\begin{equation}
\Res{\dl{n+1,k}} \simeq  \dl{n,k-1} \oplus \dl{n,k} \oplus \left\{\begin{tabular}{r l}
$0$ & if $k+1$ is critical\\
$\dl{n,k+1}$ & otherwise
\end{tabular} \right\} \text{.}
\end{equation}
\end{Coro}
Now that we have formulas for the restriction of the irreducible modules, we can use them to prove formulas for their induction.
\begin{Prop}\label{prop:420}
If $\dr{n-1,k} \neq 0$ then
\begin{equation}
\Ind{\dl{n-1,k}} \simeq \dl{n,k-1} \oplus \dl{n,k} \oplus \left\{\begin{tabular}{r l}
$0$ & if $k+1$ is critical\\
$\dl{n,k+1}$ & otherwise
\end{tabular} \right\} \text{.}
\end{equation}
\end{Prop}
\begin{proof}The argument is similar to that of proposition \ref{prop:restRadical} and uses systematically Frobenius theorem, the parity of the modules and the eigenspaces of the central element $\DF{n}$ (or $\DF{n-1}$). If $\dr{n-1,k} \neq 0$, the exactness of
\begin{equation}
0 \to \dr{n-1,k} \longrightarrow \U{n-1,k} \longrightarrow \dl{n-1,k} \to 0 \text{.}
\end{equation}
implies the exactness of the sequence of $\dtl{n+1}$-modules:
\begin{equation}
\Ind{\dr{n-1,k}}\longrightarrow \Ind{\U{n-1,k}} \longrightarrow \Ind{\dl{n-1,k}} \to 0 \text{.}
\end{equation}
Since $\Ind{\U{n-1,k}}$ splits in a direct sum of three modules of distinct parities or on which $\DF{n}$ has different eigenvalues, the module $\Ind{\dl{n-1,k}}$ splits accordingly into $L_- \oplus L_{0} \oplus L_{+}$ where $L_0$  and the $L_{\pm}$ are quotients of $\U{n,k}$ and $\U{n,k\pm 1}$ respectively.
  
We first study $L_0$. Corollary \ref{coro.res.dl} gives
\begin{equation}
\Hom{(\Ind{\dl{n-1,k}},\dl{n,k})} \simeq \Hom{(\dl{n-1,k},\Res{\dl{n,k}})} \simeq \Hom{(\dl{n-1,k},\dl{n-1,k})} \neq 0 \text{.}
\end{equation}
Therefore $L_0$, the only submodule of $\Ind{\dl{n-1,k}}$ of the parity of $\dl{n,k}$, is non-trivial. Moreover proposition \ref{coro.hom.irred} gives
\begin{equation}
\Hom{(\Ind{\dl{n-1,k}},\U{n,k})} \simeq \Hom{(\dl{n-1,k},\U{n-1,k-1} \oplus \U{n-1,k} \oplus \U{n-1,k+1})} \simeq 0.
\end{equation}
Hence $L_0$ is non-trivial, distinct from $\U{n,k}$ and must be isomorphic to $\dl{n,k}$.

We now turn to $L_{-}$. If $k-1$ is not critical, corollary \ref{coro.res.dl} shows again that $\Hom(\Ind{\dl{n-1,k}},\allowbreak \dl{n,k-1})$ is non-trivial and $\dl{n,k-1}$ must be isomorphic to a quotient of $L_-$. The short exact sequence
\begin{equation}
0 \to \U{n-1,k-2}\oplus \U{n-1,k-1} \longrightarrow \Res{\U{n,k-1}} \longrightarrow \U{n-1,k} \to 0 
\end{equation}
gives rise to the exactness of
\begin{equation}
0 \to \Hom{(\dl{n-1,k},\U{n-1,k-2}\oplus \U{n-1,k-1})} \longrightarrow \Hom{(\dl{n-1,k},\Res{\U{n,k-1}})} \longrightarrow \Hom{(\dl{n-1,k},\U{n-1,k})}\text{.}
\end{equation}
Corollary \ref{coro.hom.irred} gives $\Hom{(\dl{n-1,k},\U{n-1,k})} \simeq 0$ and therefore
\begin{equation}
\Hom{(\Ind{\dl{n-1,k}},\U{n,k-1})}\simeq\Hom{(\dl{n-1,k},\U{n-1,k-2} \oplus \U{n-1,k-1})}\simeq 0
\end{equation}
since the three eigenvalues $\df{k-2}$, $\df{k-1}$ and $\df k$ of $\DF{n-1}$ are distinct if both $k-1$ and $k$ are non-critical. The module $\U{n,k-1}$ is not a quotient of $\Ind{\dl{n-1,k}}$ and $L_-$ must therefore be distinct of $\U{n-1,k}$. Hence $L_-$ must be isomorphic to $\dl{n,k-1}$. Finally, if $k-1$ is critical, then $\dl{n-1,k}\simeq \dr{n-1,k-2}$ by proposition \ref{prop.rad.iso} and $\Hom(\Ind{\dl{n-1,k}}, \U{n,k-1})\neq 0$. Since $\U{n,k-1}$ is then irreducible, $L_-\simeq \dl{n,k-1}$.
  
It remains to study $L_{+}$. The exact sequence
\begin{equation}
0 \to \Hom{(\dl{n-1,k},\U{n-1,k}\oplus \U{n-1,k+1})} \longrightarrow \Hom{(\dl{n-1,k},\Res{\U{n,k+1}})} \longrightarrow \Hom{(\dl{n-1,k},\U{n-1,k+2})}
\end{equation}
follows from the exact sequence for $\Res{\U{n,k+1}}$. The two outer $\Hom$ spaces are trivial because of corollary \ref{coro.hom.irred} and lemma \ref{lem.hom.unk}. This proves that $\Hom{(\Ind{\dl{n-1,k}},{\U{n,k+1}})}\simeq 0$ and that
no submodules of $\U{n,k+1}$ are isomorphic to a quotient of $\Ind{\dl{n-1,k}}$ and in particular that $L_{+} \neq \U{n,k+1}$. Now, if $k+1$ is not critical, then $\Hom{(\Ind{\dl{n-1,k}},\dl{n,k+1})}$ is non-trivial by corollary \ref{coro.res.dl} and $L_+$ must therefore be $\dl{n,k+1}$. If $k+1$ is critical, then $\U{n,k+1} \simeq \dl{n,k+1}$ is irreducible and, since $L_{+} \neq \U{n,k+1}$, the submodule $L_{+}$ must  be trivial.
\end{proof}
  
\noindent The use of symmetric pairs and proposition \ref{prop.rad.iso} give the last result of this section.
  
\begin{Coro}
If $\kc$ is critical, $0<i<\ell$ and $\dr{n-1,\kc+i} \neq 0$, then
\begin{equation}
\Ind{\dr{n-1,\kc-i}} \simeq  
\left\{\begin{tabular}{l l}
$0$ & if $i=\ell-1$\\
$\dr{n,\kc-{i-1}}$ & otherwise
\end{tabular} \right\}
\oplus 
\dr{n,\kc-i} \oplus 
\left\{\begin{tabular}{l l}
$\U{n,\kc} $& if $i=1$\\
$\dr{n,\kc-{i+1}}$ & otherwise
\end{tabular} \right\}.
\end{equation} 
Some of these radicals may vanish.
\end{Coro}
\noindent Note that if $\dr{n-1,\kc+i} = 0$ we have $\dr{n-1,\kc-i}\simeq \dl{n-1,\kc+i} \simeq \U{n-1,\kc+i}$. We therefore find simply $\Ind{\dr{n-1,\kc-i}} \simeq \Ind{\U{n-1,\kc+i}}$.
  
\end{subsection}
\end{section}

%
\begin{section}{The structure of $\dtl n$ at a root of unity}\label{sec:AtRootUnity}
%
  
\begin{center}{\em In this section, $q$ is a root of unity and $\ell$ the smallest positive integer such that $q^{2\ell}=1$.}\end{center}
  
\medskip
  
When $q$ is not a root of unity, every standard module of the dilute Temperley-Lieb algebra is irreducible. The algebra is then semisimple and the standard modules form a complete set of irreducible modules (theorem \ref{thm:dtlnGeneric}). However, when $q$ is a root of unity, some of them will be reducible, yet indecomposable. That is, if $q$ is a root of unity, the algebra $\dtl{n}$ is not always semisimple. To probe its structure the first subsection first shows that $\dtl n$ is a {\em cellular algebra}, that is, an example of the associative algebras introduced in 1996. (See also chapter 2 of \cite{Mathas} for a complete overview.) As examples of their cellular algebras Graham and Lehrer \cite{GrahamL} displayed the Hecke algebra, the Brauer's centralizer algebra, the Temperley-Lieb algebra and the Jones algebra. Their results on cellular algebras give a straightforward description of the structure of the principal indecomposable modules; applying these results to $\dtl{n}$ will be the content of the second subsection. The third and last will show how to give a fairly explicit construction of principal modules using induction from $\dtl{n-1}$ to $\dtl{n}$.
%
%
\begin{subsection}{The dilute Temperley-Lieb algebra as a cellular algebra.}\label{sec:cellalgebra}
This section recalls the definition of {\em cellular algebras} over $\mathbb C$ and the results crucial for our task, and shows how the dilute Temperley-Lieb algebra $\dtl n$ satisfies the defining axioms. We refer to \cite{GrahamL, Mathas} for details and proofs.
  
\begin{Def}
A \emph{cellular algebra} over $\mathbb{C}$ is an associative unital algebra $A$, together with \emph{cell datum} $(\Lambda, \Y{},C,\,^t)$ where
\begin{enumerate}
\item $\Lambda$ is a partially ordered set and for each $\lambda \in \Lambda$, $\Y{\lambda}$ is a finite set such that $C: \sqcup_{\lambda \in \Lambda} \Y{\lambda}\times \Y{\lambda} \to A$ is an injective map whose image is a basis of $A$;
\item if $\lambda \in \Lambda$ and $x,y \in \Y{\lambda}$, write $C(x,y)=C^{\lambda}_{x,y} \in A$. The map ``$\ ^t$'' is a linear anti-involution of $A$ such that $(C^{\lambda}_{x,y})^t = C^{\lambda}_{y,x}$;
\item if $\lambda \in \Lambda$ and $x,y \in \Y{\lambda}$, then for any element $u\in A$ we have
\begin{equation*}
u C^{\lambda}_{x,y} \equiv\sum_{z \in \Y{\lambda}} r_{u}(z,x)C^{\lambda}_{z,y} \quad \mod A(<\lambda),
\end{equation*} 
where $ r_{u}(z,x) \in \mathbb{C}$ is independent of $y$ and where $A(<\lambda)$ is the submodule of $A$ generated by  $\big\{ C^{\mu}_{x',y'}| \mu < \lambda ; x',y' \in \Y{\mu}\big\}$. 
  
\end{enumerate}
\end{Def}
The verification that $\dtl n$ is a cellular algebra has essentially been done. The set $\Lambda$ is simply $\{0,1,2,\dots, n\}$ with the usual (total) order $<$ on integers and the sets $\Y k, k\in\Lambda$, are to be identified with the bases $\Y{n,k}$ of the standard modules, that is, the $n$-links with $k$ defects (see subsection \ref{sec:standard}). The injective map $C$ has also been defined in that subsection: The pair $x,y\in\Y{k}$ is mapped by $C$ to the $n$-diagram $|x\bar y|$. Since every $n$-diagram can be cut into its left part $x$ and right one $\bar y$, the image of $C$ contains all $n$-diagrams and is therefore a basis of $\dtl n$. The anti-involution ``$\ ^t$'' is defined on $n$-diagrams by $|x\bar y|\mapsto |y\bar x|$ and was introduced in lemma \ref{lem.gram.sym}. It corresponds to a reflection of any diagram through a vertical mirror. Finally the last axiom is lemma \ref{lem:axiom3}.
  
The following lemma \cite{GrahamL}, a direct consequence of axiom (3), allows for the definition of a bilinear form on some natural modules to be introduced next.
\begin{Lem}Let $A$ be a cellular algebra. For $\lambda\in\Lambda$, $x,y,x',y'\in\Y\lambda$ and $u\in A$, there exists $\phi_u(y,x')\in\mathbb C$ such that
\begin{equation}
C^\lambda_{x,y}uC^\lambda_{x',y'}\equiv \phi_u(y,x')C^\lambda_{x,y'}\quad \mod A(<\lambda).
\end{equation}
The coefficient $\phi_u(y,x')$ is independent of $x$ and $y'$.
\end{Lem}
  
\begin{Def}
For each $\lambda \in \Lambda$, the \emph{cell module} of $A$ corresponding to $\lambda$ is 
$$\U\lambda = \text{span}_{\mathbb{C}}\left\{m_x | x \in \Y\lambda \right\}, $$
where the action of $A$ is defined by
$$u m_x = \sum_{x' \in \Y\lambda} r_{u}(x,x')m_{x'}.$$
For $\lambda \in \Lambda$, define the bilinear form $\langle\,\cdot\, ,\, \cdot\,\rangle_\lambda: \U\lambda\times \U\lambda \to \mathbb{C}$ by $\langle m_x,m_y\rangle_\lambda = \phi_{\id}(x,y)$ for all $x,y \in \Y\lambda$, and extend it linearly.
\end{Def}
  
Since the elements of the basis of $\U\lambda$ are in one-to-one correspondence with elements of $\Y\lambda$, we shall identify $m_x\leftrightarrow x$, $x\in\Y\lambda$. Then, for $A=\dtl n$, the $n$-links with $k$ defects form a basis of its cell module $\Y k,k\in\Lambda$. That the actions on $\U k$ and on $\U{n,k}$ coincide follows from lemma \ref{lem:axiom3}. Finally the bilinear form $\langle\,\cdot\, ,\, \cdot\,\rangle_k$ on $\U k$ and the Gram product $\langle\,\cdot\, ,\, \cdot\,\rangle_{n,k}$ coincide because of lemmas \ref{lem.gram.sym} and \ref{lem:axiom3}. To sum up:
\begin{Prop}The dilute Temperley-Lieb algebra $\dtl n(\beta)$ is cellular. Its cell modules $\mathsf S_k, k\in\{0,1,\dots, n\}$, are isomorphic to the standard modules $\U{n,k}$ and the bilinear form $\langle\,\cdot\, ,\, \cdot\,\rangle_k$ coincides with the Gram product $\langle\,\cdot\, ,\, \cdot\,\rangle_{n,k}$ on $\U{n,k}$.
\end{Prop}
These observations simplify enormously the task of identifying the principal indecomposable modules because of the next theorem. The following notation is needed. The radical $\text{rad}_\lambda$ of the form $\langle\,\cdot\, ,\, \cdot\,\rangle_{\lambda}$ is defined exactly as the radical of the Gram product. The quotient $\U\lambda/\text{rad}_\lambda  = \dl{\lambda}$ can be shown to be (absolutely) irreducible. (Lemma \ref{lem.gram.gen} did it in a direct way for $\dtl n$.) The modules are now assumed to be finite-dimensional.
  
\begin{Thm}[Graham and Lehrer \cite{GrahamL} (see also Mathas \cite{Mathas})]\label{thm:GL}
Let $A$ be cellular algebra and $\Lambda_{0} = \left\{\lambda \in \Lambda\ |\  \langle\,\cdot\, ,\, \cdot\,\rangle_{\lambda} \neq 0 \right\}\subset \Lambda$. Then
\begin{enumerate}
\item The set $\left\{\dl{\lambda}| \lambda \in \Lambda_{0} \right\}$ is a complete set of non-isomorphic irreducible $A$-modules.
\item The set $\left\{\A{}{\lambda}|\lambda \in \Lambda_{0}\right\}$ is a complete set of non-isomorphic projective indecomposable modules, where $\A{}{\lambda}$ is the projective cover of $\U\lambda$.
\item Let $d_{\lambda, \mu}$, $\lambda\in\Lambda,\mu\in\Lambda_0$, be the multiplicity of $\dl{\mu}$ in any composition series of $\U\lambda$ and $c_{\lambda, \mu}$ that of $\dl{\mu}$ in $\A{}{\lambda}$. Then
$$c_{\lambda,\mu} = \sum_{\nu \leq \mu,\lambda} d_{\nu,\mu}d_{\nu,\lambda}.$$
\item For any projective module $\dP{}$, there is a filtration $$0 = M_{0} \subset M_{1} \subset \hdots \subset M_{d-1} \subset M_{d} = \dP{}, $$ such that $d\le |\Lambda|$ and $ M_{i}/M_{i-1}$ is isomorphic to a direct sum of cell 
modules.
\end{enumerate}
\end{Thm} 
\noindent The next section will show how this theorem reveals the structure of the dilute algebra $\dtl n$.
\end{subsection}
%
%
\begin{subsection}{The indecomposable modules $\dP{n,k}$ and the structure of $\dtl n$ for $q$ a root of unity}
  
\begin{center}{\itshape Hereafter the pairs of integers $(k_{-}, k)$ and $(k,k_{+})$ are symmetric pairs with $k_{-} < k<k_{+}$.}\end{center}
  
The first step is to determine the subset $\Lambda_0$ for $\dtl n$. But this was done at the beginning of section \ref{sub:gram} where it was noted that $\langle\,\cdot\, ,\, \cdot\,\rangle_{n,k}$ is never identically zero (contrarily to the Gram product of $\tl n$). So $\Lambda_0=\Lambda=\{0,1,\dots, n\}$. The algebra $\dtl n(\beta=q+q^{-1})$ has thus $n+1$ non-isomorphic irreducible modules and as many principal indecomposable ones. Since this is the number of standard modules $\U{n,k}$, each having a distinct irreducible quotient $\dl{n,k}$ by lemma \ref{lem.hom.unk}, all irreducibles are known. A complete set of projective indecomposable is given by the projective covers of the standard modules $\U{n,k}$. Their structure can be partially revealed by statements (3) and (4) of theorem \ref{thm:GL}.
  
Let $d_{k,j}$ be the number of (irreducible) composition factors $\dl{n,j}$ in $\U{n,k}$ and $c_{k,j}$ their number in $\dP{n,k}$, the projective cover of $\U{n,k}$. By (3)
\begin{equation}
c_{k,j} = \sum_{i\leq k,j} d_{i,j}d_{i,k}.
\end{equation}
Corollary \ref{cor:exactStan} has identified the one or two composition factors of the standard module $\U{n,k}$ when $k$ in non-critical. The matrix $d$ is thus known:
\begin{equation}
d_{k,j} = \delta_{k,j}  + \begin{cases}
\delta_{j,k_{+}}, & \text{ if $k$ is not critical and } k_{+} \leq n \\
0, & \text{ otherwise } 
\end{cases}
\end{equation}
and the composition factors of $\dP{n,k}$ are obtained by matrix multiplication $c=d^td$:
\begin{equation}
c_{k,j} = \delta_{k,j} + \begin{cases}
\delta_{j,k_{+}} + \delta_{j,k} + \delta_{j,k_{-}} & \text{if $k$ is not critical, } k_{-} \geq 0 \\
\delta_{j,k_{+}} & \text{if $k$ is not critical, } k_{-} < 0\\
0 & \text{if $k$ is critical}
\end{cases},
\end{equation}
where it is understood that, if $k_{+} > n$, $\delta_{j,k_{+}}$ is always equal to zero. (Note that the restriction $i\le k,j$ on the sum index is superfluous here as $d$ is an upper triangular matrix.)
  
We now turn to the structure of the principal indecomposable modules $\dP{n,k},0\le k\le n$. If $k$ is critical, $\dP{n,k}$ has a single composition factor and $\dP{n,k}\simeq\dl{n,k}\simeq \U{n,k}$. Thus $\U{n,k}$ is projective when $k$ is critical. When $k < \ell-1$, $\dP{n,k}$ has two composition factors, $\dl{n,k}$ and $\dl{n,k_{+}}$. Since these are precisely the composition factors of $\U{n,k}$ and $\dP{n,k}$ is its projective cover,  $\U{n,k}$ and $\dP{n,k}$ are thus isomorphic for $k <\ell-1$. 
  
When $k > \ell -1$ is not critical, $\dP{n,k}$ has either three or four composition factors among $\dl{n,i_{+}}$,$\dl{n,i_{-}}$ and $\dl{n,i}$ which always appears twice. (If $k_+>n$, then the module $\dl{n,k_+}$ is trivial.) The filtration $0 = M_{0}  \subset \hdots \subset M_{d-1} \subset M_{d} = \dP{n,k}$ of statement (4) shows that $\dP{n,k}$ has a submodule $M_{d-1}$ such that $\dP{n,k}/M_{d-1}$ is isomorphic to a direct sum of standard modules. Since $\dP{n,k}$ is the projective cover of $\U{n,k}$, this direct sum must contain at least a copy of $\U{n,k}$. The composition factors of $\U{n,k}$ are $\dl{n,k}$ and $\dl{n,k_+}$ and the remaining ones are $\dl{n,k}$ and $\dl{n,k_-}$ which are precisely those of $\U{n,k_{-}}$. No standard module contains only the irreducible $\dl{n,k_-}$ and $\U{n,k_{-}}$ is the unique one with the remaining factors. Therefore $\U{n,k_-}$ must be a submodule of $\dP{n,k}$, the quotient $\dP{n,k}/M_{d-1}$ must be isomorphic to $\U{n,k}$ and all composition factors are thus accounted for. The filtration is then simply $0\subset M_1\simeq\U{n,k_-}\subset M_2=\dP{n,k}$. Note finally that $\DF{n}$ takes the same eigenvalue $\df{n,k}$ on both $\U{n,k_{-}}$ and $\U{n,k}$. The only eigenvalue of $\DF{n}$ on $\dP{n,k}$ is thus $\df{n,k}$ as well. This discussion ends the proof of the following statement.
\begin{Prop}\label{prop:ses.pindec}
The set $\{\dP{n,k}\,|\, 0 \leq k\leq n\}$ forms a complete set of non-isomorphic projective indecomposable modules. If $k$ is critical or $k < \ell -1$, $\dP{n,k} \simeq \U{n,k}$. Otherwise, it is the (indecomposable) projective module, unique up to isomorphism, satisfying the short exact sequence
\begin{equation}\label{ses.pindec}
0 \longrightarrow \U{n,k_{-}} \longrightarrow \dP{n,k} \longrightarrow \U{n,k} \longrightarrow 0.
\end{equation}
The central element $\DF{n}$ has a single eigenvalue acting on $\dP{n,k}$, namely $\df{n,k} = q^{k+1} - q^{-(k+1)}$.
\end{Prop}
  
The regular module of $\dtl{n}$ is the algebra itself seen as module for the action given by left multiplication. Wedderburn's theorem states that this module is a direct sum of irreducible modules if the algebra is semisimple, and of principal indecomposable ones if it is not. (See theorems \ref{prop:Wedderburn1} and \ref{prop:Wedderburn2} in appendix \ref{sec:algebre}.) 
Theorem \ref{thm:dtlnGeneric} gave the decomposition of $\dtl n$ in terms of non-isomorphic irreducible modules for $q$ such that $\dtl n$ is semisimple. The following theorem completes the description of the dilute Temperley-Lieb algebras for the case when $q$ is a root of unity. It is a corollary of the previous classification of principal modules and Wedderburn's theorem \ref{prop:Wedderburn2}.
\begin{Thm}[Structure of $\dtl n$ for $q$ a root of unity]\label{prop.struct.dtln}
Let $q$ be a root of unity other than $\pm 1$ and $\ell$ the smallest positive integer such that $q^{2\ell}=1$. Let $K$ be the set of critical integers smaller or equal to $n$. Then
\begin{equation}\label{eq:talam}
\dtl n \simeq \Big(\bigoplus_{0\leq k < \ell-1} \ddl{n,k}\ \U{n,k}\Big)\oplus \Big( \bigoplus_{\kc \in K} \big( \ddl{n,\kc}\,\U{n,\kc} \oplus \big(\bigoplus_{1\leq i<\ell}\ddl{n,\kc+i}\ \dP{n,\kc+i}\big)\big)\Big)
\end{equation}
where $\ddl{n,k}=\dim \dl{n,k}$ if $k\in\{0,1,\dots, n\}$ and $0$ otherwise.
\end{Thm}
\end{subsection}
%
%
\begin{subsection}{Induction of $\dP{n-1,k}$}
We now describe how the projective modules of $\dtl{n-1}$ are related to those of $\dtl{n}$ through induction. When $k<\ell -1$, then $\dP{n,k}\simeq\U{n,k}$ and
corollary \ref{coro.ind.split} gives the result. We concentrate on the other cases. The crucial property here will be that, if $P$ is a projective $\dtl{n-1}$-module, then $\Ind{P}$ is a projective $\dtl n$-module (see Appendix \ref{sec:algebre}).
  
If $k=\kc$ is critical, proposition \ref{coro.ind.ses} gives the short exact sequence
$$0 \longrightarrow \U{n,\kc} \oplus \U{n,\kc-1} \longrightarrow \Ind{\U{n-1,\kc}} \longrightarrow \U{n,\kc+1} \longrightarrow 0.$$ 
Now, since the parity of $\U{n,\kc \pm 1}$ is different from that of $\U{n,\kc}$, the module $\Ind{\U{n-1,\kc}}$ must be isomorphic to $\U{n,\kc} \oplus P$, where $P$ is some projective module satisfying the sequence
$$0 \longrightarrow \U{n,\kc-1} \longrightarrow P \longrightarrow \U{n,\kc+1} \longrightarrow 0.$$ 
Since $\dP{n,\kc+1}$ is the projective cover of $\U{n,\kc +1}$ and satisfies the same sequence, the induced module $\Ind{\U{n-1,\kc}}\simeq \Ind{\dP{n-1,\kc}}$ is
$$\Ind{\U{n-1,\kc}} \simeq \U{n,\kc} \oplus \dP{n,\kc +1} .$$
  
Suppose then that $k$ is non-critical and larger than $\ell-1$. Because induction is right-exact, the sequence \eqref{ses.pindec} yields the exact sequence
$$ \U{n,k_{-}-1} \oplus \U{n,k_{-}} \oplus \U{n,k_{-}+1} \longrightarrow \Ind{\dP{n-1,k}} \longrightarrow  \U{n,k-1} \oplus \U{n,k} \oplus \U{n,k+1} \longrightarrow 0,$$ 
where proposition \ref{coro.ind.split} was used. This already gives an upper bound on the dimension of $\Ind{\dP{n-1,k}}$:
\begin{equation}\label{eq:dimensionofp}
\dim \Ind{\dP{n-1,k}} \leq \dim ( \U{n,k_{-}-1} \oplus \U{n,k_{-}} \oplus \U{n,k_{-}+1} ) + \dim( \U{n,k-1} \oplus \U{n,k} \oplus \U{n,k+1}),
\end{equation}
where the upper bound is reached if the leftmost morphism is injective. Since the projective cover of the standard module $\U{n,k'}$ is $\dP{n,k'}$ for all $k'\in\Lambda$, the induced $\Ind{\dP{n-1,k}}$ must be isomorphic to $P \oplus \dP{n,k-1} \oplus \dP{n, k} \oplus \dP{n,k+1}$ for some projective module $P$. If neither $k+1$ nor $k-1$ is critical, then $\dim \dP{n,i}=\dim\U{n,i}+\dim\U{n,i_-}$ for $i\in\{k-1,k,k+1\}$ and $\dim \Ind{\dP{n-1,k}}=\dim(P\oplus \dP{n,k-1} \oplus \dP{n, k} \oplus \dP{n,k+1})\ge\dim( \dP{n,k-1} \oplus \dP{n, k} \oplus \dP{n,k+1})$ and the upper bound above is also a lower bound. This yields $P \simeq 0$. 
  
A special treatment is required when either $k-1$ or $k+1$ is critical. Suppose $k+1$ is (and then so is $k_--1$). Then $\dP{n,k+1} \simeq \U{n,k+1}$ and the bound only gives $\dim P \leq \dim\U{n,k_{-}-1}$. Since $\dP{n-1,k}$ is projective, the functor $\Hom(\dP{n-1,k},-)$ is exact and, applied to the sequence \eqref{eq:restr.unk}, gives
\begin{align}
0 \longrightarrow &\Hom(\dP{n-1,k}, \U{n-1,k_{-}-1} \oplus \U{n-1,k_{-}-2} )\notag \\
 &\longrightarrow \Hom(\dP{n-1,k},\Res{\U{n,k_{-}-1}} )\longrightarrow \Hom(\dP{n-1,k},\U{n-1,k_{-}}) \longrightarrow 0.
\end{align}
Because of the sequences \eqref{eq:exactStan} and \eqref{ses.pindec}, the third term in the sequence is non-zero ($k$ and $k_{-}$ form a symmetric pair) and thus the middle term cannot vanish. Therefore, by Frobenius reciprocity, $\Hom(\Ind{\dP{n-1,k}}, \U{n,k_{-}-1} ) \simeq \Hom(\dP{n-1,k},\Res{\U{n,k_{-}-1}} )$ is also non-trivial. Since $\U{n,k_{-}-1}$ is irreducible and projective (again, $k_{-}-1$ is critical), there must be a surjective morphism from $\Ind{\dP{n-1,k}}$ onto $\U{n,k_{-} -1}$. But the addition of the composition factors of $\U{n,k_{-} -1}$ saturates the upper bound for the dimension of $\Ind{\dP{n-1,k}}$ and thus $\Ind{\dP{n-1,k}}\simeq\U{n,k_{-} -1}\oplus\dP{n,k-1} \oplus \dP{n, k} \oplus \dP{n,k+1}$. The case when $k-1$ is critical (or when both $k-1$ and $k+1$ are) is treated similarly. We summarize our results in the following proposition. 
  
\begin{Prop}
For $0 \leq k \leq n-1$,
\begin{equation}
\Ind{\dP{n-1,k}} \simeq \dP{n,k} \oplus \dP{n,k+1} \oplus \begin{cases}
0, & \text{ if } k \text{ is critical}\\
\dP{n,k-1} \oplus \dP{n,k-1} \oplus \dP{n,k_{-}-1}, & \text{ if } k \pm 1 \text{ are both critical}\\
\dP{n,k-1} \oplus \dP{n,k_{-}-1}, & \text{ if only } k+1 \text{ is critical}\\
\dP{n,k-1} \oplus \dP{n,k-1}, & \text{ if only } k-1 \text{ is critical}\\
\dP{n,k-1}, & \text{otherwise},
\end{cases}
\end{equation}
where it is understood that $\dP{n,k'} \simeq 0$ if $k' < 0$.
\end{Prop}
This result together with proposition \ref{Prop-iso} give a simple diagrammatic basis for $\dP{n,\kc + 1}$. Similar arguments can then be used to build basis for the other projective indecomposable modules by inducing repeatedly from $\U{n,\kc}$.
\end{subsection}

%
\end{section}
  
%
\begin{section}{Conclusion}
%

The main results of this paper are now reviewed. The dilute Temperley-Lieb algebras  $\dtl n(\beta), n\geq 0,$ form a family of algebras parametrized by a complex (or formal) parameter $\beta$, often written as $\beta=q+q^{-1}$ with $q\in\mathbb C^\times$. The dimension of $\dtl n(\beta)$ is the Motzkin number $M_{2n}$. These algebras decompose into a sum of even and odd parts, and so do their modules. They are examples of cellular algebras \cite{GrahamL}. 
  
Their representation theory is largely based on the study of the standard modules $\U{n,k}$, $0\leq k\leq n$. These are shown to be indecomposable (proposition \ref{prop:UnkIndecomposable}) and cyclic (proposition \ref{prop:cyclic}). Their diagrammatic definition is a technical advantage: it allows for quick computations and observing several of their properties. For example, in the link basis, the matrices representing the generators have at most one non-zero element per column and this element is then a power of $\beta$. It is also easy to observe that any link state with only defects and vacancies is actually a generator. Finally the standard modules are all distinct ($\U{n,k}\simeq\U{n,j}\Leftrightarrow k=j$) and, for neighbouring $n$s, they are related by restriction and induction and $\Ind{\U{n,k}}\simeq \Res{\U{n+2,k}}$ for all $n$ and $0\leq k\leq n$ (proposition \ref{Prop-iso}).
  
The latter property, together with the natural bilinear form $\langle *,*\rangle_{n,k}$ and a particular central element $\DF n$, is sufficient to unravel the structure of the algebra $\dtl n$ when the complex number $q$ is generic, that is not a root of unity. Then $\dtl n(\beta=q+q^{-1})$ is semisimple and the standard modules form a complete set of non-isomorphic irreducible modules (theorem \ref{thm:dtlnGeneric}).
  
If however $q$ is a root of unity, distinct from $\pm 1$, a finer analysis is required. Let $\ell$ be the smallest positive integer such that $q^{2\ell}=1$. An integer $\kc$ is said critical if $\kc+1\equiv 0\text{\ mod\ }\ell$ and a pair $(k_-,k_+)$ of distinct integers form a symmetric pair if their average is critical and $0<(k_+-k_-)/2<\ell$. With this notation the standard module $\U{n,k}$ is reducible, but indecomposable, if $k$ is the smallest element $k_-$ of a symmetric pair with $0\leq k_-<k_+\leq n$. In that case, its maximal proper submodule $\dr{n,k}\subset \U{n,k}$ is the radical of the Gram pairing $\langle *,*\rangle_{n,k}$ and is irreducible. In fact, if $k$ is the $k_-$ of a symmetric pair $(k_-,k_+)$, then $\dr{n,k=k_-}\simeq \dl{n,k_+}$ where $\dl{n,k}$ is the irreducible quotient $\U{n,k}/\dr{n,k}$.
  
The indecomposable projective modules, that is the principal indecomposable ones, can be identified and linked to standard modules thanks to general results that hold for all cellular algebras. Moreover the principal indecomposable modules of $\dtl{n}$ are related to those of $\dtl{n-1}$ through the induction functor described in section \ref{sec:induction}. Alternatively, one could have followed another path, first determining the action of the induction functor on the projective modules of $\dtl{n-1}$ and then using it to build the projective indecomposable modules of $\dtl{n}$. A given principal indecomposable module of $\dtl n$ is then characterized as a direct summand of $(\U{n',\kc})\underbrace{\uparrow\uparrow\hskip-2pt\dots\hskip -2pt\uparrow}_{n-n'}$ (with $n-n'<\ell$) completely determined by its $\DF n$-eigenvalue and parity. Starting with $\dtl{1}\simeq \U{1,1} \oplus \U{1,0}$, this process can be used recursively to construct the principal modules. For example, this approach was used to study the regular Temperley-Lieb algebra $\tl{n}$ in \cite{RiSY}.
  
\begin{center}
\begin{figure}[h!]\label{theLoewy}
\begin{tikzpicture}[thick]
\node (l) at (0,2.5) [] {$\dl{n,k}$};
\node (mt) at (3.5,4) [] {$\dr{n,k_-}=\dl{n,k_+}$};
\node (ml) at (2,2.5) [] {$\dl{n,k_-}$};
\node (mr) at (5,2.5) [] {$\dr{n,k_+}=\dl{n,k_{++}}$};
\node (mb) at (3.5,1) [] {$\dr{n,k_-}=\dl{n,k_+}$};
\node (rt) at (9,4) [] {$\dr{n,k_-}=\dl{n,k_+}$};
\node (rl) at (7.5,2.5) [] {$\dl{n,k_-}$};
\node (rb) at (9,1) [] {$\dr{n,k_-}=\dl{n,k_+}$};
\node (rrt) at (12,3.25) [] {$\dl{n,k}$};
\node (rrb) at (12,1.75) [] {$\dr{n,k}$};
\node at (0,0.) {(a)};
\node at (3.5,0.) {(b)};
\node at (9,0.){(c)};
\node at (12,0.){(d)};
\draw [->] (mt) -- (ml);
\draw [->] (mt) -- (mr);
\draw [->] (ml) -- (mb);
\draw [->] (mr) -- (mb);
\draw [->] (rt) -- (rl);
\draw [->] (rl) -- (rb);
\draw [->] (rrt) -- (rrb);
\end{tikzpicture}
\caption{The Loewy diagrams of the principal indecomposable modules}
\end{figure}
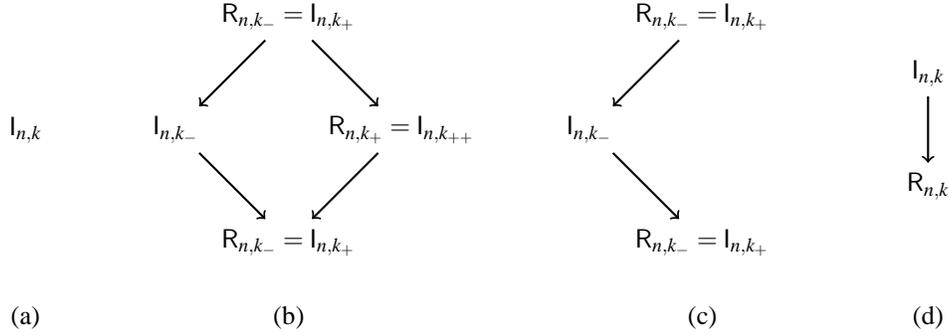
\end{center}
Loewy diagrams are often used in the mathematical physics literature and it is useful to draw them for the principal indecomposable modules. (The construction of the Loewy diagrams for $\dtl n$ is identical to that for $\tl n$ which is described in \cite{RiSY}.) If $k$ is critical, the projective is simply the (irreducible) standard module $\U{n,k}$ and its Loewy diagram contains a single node (figure \ref{theLoewy} (a)). For $k$ non-critical, let $k_-, k_+$ and $k_{++}$ be such that $k_-<k_+=k<k_{++}$ and both $(k_-,k_+)$ and $(k_+,k_{++})$ are symmetric pairs. Then, if $k_-,k_+,k_{++}\in\{0,1,\dots, n\}$, the Loewy diagrams of the principal modules with irreducible quotient $\dl{n,k}$ has the form (b) in the figure. If $k_{++}>n$, then the right node is deleted and the resulting Loewy diagram is of type (c) on the figure. Finally, if $k$ is at the left of the first critical line, then its Loewy diagram is that of the standard $\U{n,k}$ and appears as (d) on the figure.
  
The similarity of these Loewy diagrams with those of the Temperley-Lieb algebra leads to a natural correspondence between their respective irreducible modules:
\begin{equation*}
\dl{n,i} \to \begin{cases}
\irred{n,i}, & \text{ if } n\equiv i \mod 2\\
\irred{n-1,i}, & \text{ otherwise }
\end{cases}.
\end{equation*}
Under this transformation it is clear that the indecomposable projective modules of $\dtl{n} $ are sent to those of $\tl{n}$ and $\tl{n-1}$, except when $\ell =2$, because then $\irred{2n,0} = \{0\} $. This 
suggests that $\dtl{n}$ is Morita equivalent to the direct sum $\tl{n} \oplus \tl{n-1}$ when $\beta \neq 0$.
In fact it can be shown that $\dtl{n}$ and the direct sum $\tl{n} \oplus \tl{n-1}$ are Morita-equivalent when $\beta \neq 0$.\footnote{Though it goes beyond the goals of the paper, here are two paths toward a proof of their Morita-equivalence. In the first, one can put the projective modules of $\tl{n} \oplus \tl{n-1}$ in correspondence with those of the even and odd parts of $\dtl n$, respectively, and show that this one-to-one correspondence preserves all morphisms between them, if $\beta\neq 0$. In the second, suggested by the referee, one may use the (finite) Temperley-Lieb category $\mathbf T=\mathbf T_{\mathbb C, q}$ introduced by Graham and Lehrer (see Definition (2.1) in \cite{GrahamLehrer}). A (full) subcategory is obtained by restraining to diagrams $i\rightarrow j$ for $i, j\leq n$. One the one hand this truncated category can be shown to be Morita-equivalent to $\tl{n} \oplus \tl{n-1}$ when $\beta\neq 0$. And on the other hand, one can establish its Morita-equivalence to $\dtl n$ by noticing that diagrams $i\rightarrow j$ with $i$ or $j$ smaller than $n$ can be understood as diagrams with vacancies.}  
  
What can one learn from these results about limiting structures appearing in physical models like conformal field theories (CFT)? The original Temperley-Lieb algebras, whose representation theory the dilute ones mimic so closely, has been used to understand the representation theory of the Virasoro algebra appearing in the continuum limit of lattice models whose transfer matrix is an element of $\tl n$. The fusion ring, defined formally in \cite{GainutdinovVasseur}, is a natural outcome of the representation theory of these finite-dimensional associative algebras. The result announced there for the $\tl n$ fusion ring is paralleled to the CFT fusion for Virasoro modules, with staggered ones sharing the Loewy structure of the principal indecomposable modules of type (b) in figure \ref{theLoewy}. We hope that the results reported here may help reveal the fusion ring of the dilute Temperley-Lieb algebras.
  
\end{section}

\section*{Acknowledgements}
The authors would like to thank David Ridout for useful discussions 
and the referee whose constructive suggestions brought us to use cellular algebras as the main tool in section \ref{sec:AtRootUnity} and make it easier to understand.
JB holds a scholarship from Fonds de recherche Nature et Technologies (Qu\'ebec).
YSA holds a grant from the Canadian Natural Sciences and Engineering Research Council. This support is gratefully acknowledged.

\appendix
  
%
\begin{section}{The Temperley-Lieb algebra}\label{app.tln}
%
  
The original Temperley-Lieb algebras \cite{TemperleyLieb} were introduced much before the dilute ones. Since the present text studies the latter, the former will be presented starting from the definitions for the dilute objects. Only the results needed here are recalled. They are taken from an article by Ridout and one of the authors \cite{RiSY}. It must be underlined that their paper uses the number $p$ of arcs instead of the number $k$ of defects to characterize link states and modules. The results stated below have been adapted to the labelling in terms of defects.
  
The one-parameter family of Temperley-Lieb algebras $\tl n(\beta)$ is spanned by all $n$-diagrams, defined in subsection \ref{sec:definition}, that contain no vacancies. The product is given by the same rules as for $\dtl n$, the factor $\beta$ also weighting each closed loop generated through concatenation of diagrams. The algebra $\tl n$ has a compact definition in terms of generators $u_i, 1\leq i\leq n-1,$ and the unit $\idtl$. These correspond to the following diagrams:
\begin{equation}
\begin{tikzpicture}[baseline={(current bounding box.center)},scale=1/3]
	\bord{0}{7};
	\bord{3}{7};
	\node at (-3,3) {$\idtl=$};
	\node at (3/2,3/2) {$\vdots$};
	\node at (3/2,11/2) {$\vdots$};
	\node at (-1/2,0) {$\scriptstyle{n}$};
	\node at (-1,2) {$\scriptstyle{i+1}$};
	\node at (-1/2,3) {$\scriptstyle{i}$};
	\node at (-1,4) {$\scriptstyle{i-1}$};
	\node at (-1/2,6) {$\scriptstyle{1}$};
	\defect{0,0};
	\defect{0,2};
	\defect{0,3};
	\defect{0,4};
	\defect{0,6};
\end{tikzpicture}\text{ , }\qquad 
\begin{tikzpicture}[baseline={(current bounding box.center)}, scale=1/3]
	\bord{0}{8};
	\bord{3}{8};
	\node at (-3,7/2) {$u_{i}=$};
	\node at (3/2,3/2) {$\vdots$};
	\node at (3/2,13/2) {$\vdots$};
	\node at (-1/2,0) {$\scriptstyle{n}$};
	\node at (-1,2) {$\scriptstyle{i+2}$};
	\node at (-1,3) {$\scriptstyle{i+1}$};
	\node at (-1/2,4) {$\scriptstyle{i}$};
	\node at (-1,5) {$\scriptstyle{i-1}$};
	\node at (-1/2,7) {$\scriptstyle{1}$};
	\defect{0,0};
	\defect{0,2};
	\link{0,3}{1,3}{1,4}{0,4};
	\link{3,3}{2,3}{2,4}{3,4};
	\defect{0,5};
	\defect{0,7};
\end{tikzpicture}
\end{equation}\label{Set-Diagrams}
and satisfy the relations:
\begin{align*}
u_i^2& =\beta u_i, \\
u_i &= u_iu_{i\pm 1}u_i,\qquad \textrm{if\ }1\leq i, i\pm 1\le n-1,\\
u_iu_j &= u_ju_i,\qquad\quad\ \ |j-i|>1.
\end{align*}
The dimension of $\tl n$ is the Catalan number $C_{n+1}=\frac1{n+1}\left(\begin{smallmatrix}2n\\n\end{smallmatrix}\right)$.
  
The standard module $\vs{n,k}$ is spanned by the basis of $\U{n,k}$ from which all $n$-link states that bear vacancies are discarded. They are defined only when $n$ and $k$ have the same parity. The action of $\tl n$ on $\vs{n,k}$ is defined as that of $\dtl n$ on $\U{n,k}$. Their dimension is given by $\dim \vs{n,k}=\left(\begin{smallmatrix}n\\ (n-k)/2\end{smallmatrix}\right)-\left(\begin{smallmatrix}n\\ (n-k)/2-1\end{smallmatrix}\right)$.
  
The Gram bilinear form $\langle *,*\rangle_{{n,k}}:\vs{n,k}\times \vs{n,k}\rightarrow \mathbb C$ is introduced exactly as the Gram product on $\U{n,k}$. Now the rule stating that $\langle u,v\rangle_{{n,k}}$ is zero whenever unmatched vacancies arise upon glueing of $\bar u$ and $v$ can be ignored safely as no link states with vacancies occur in $\vs{n,k}$. (The same observation holds for the multiplication in $\tl n$ and the action of $\tl n$ on the standard modules $\vs{n,k}$, discussed above.) It is possible to compute the determinant of the matrix $G_{n,k}$ representing the bilinear form $\langle *,*\rangle_{{n,k}}$ in the basis of $n$-link states with $k$ defects. (See for example \cite{Westbury, RiSY}.)
\begin{Prop}\label{prop:GramDetTLn} The Gram determinant for the bilinear form on $\vs{n,k}$ when $\beta=q+q^{-1}$ is given, up to a sign, by
\begin{equation}\label{eqn:DetG}\det G_{n,k}=\prod_{j=1}^{(n-k)/2}\Big(\frac{[k+j+1]_q}{[j]_q}\Big)^{ 
\dim \vs{n,k+2j}}\end{equation}
where $q$-numbers are used: $[m]_q=(q^m-q^{-m})/(q-q^{-1})$.
\end{Prop}

\noindent Note that $\det G_{n,k}$, $n\geq 1, 0\leq k\leq n$, does not vanish at $\beta=\pm 2$, that is, at $q=\pm 1$. 
  
The radical $\R{n,k}=\{ v\in \vs{n,k}\ |\ \langle v,w\rangle_{{n,k}}=0 \textrm{\ for all }w\in\vs{n,k}\}$ is a submodule of the standard module $\vs{n,k}$. It has the following properties.
\begin{Prop}\label{prop.rad} The radical $\R{n,k}$ is the maximal proper submodule of $\vs{n,k}$. It is either trivial ($\simeq \{0\}$) or irreducible.
\end{Prop}
\noindent Like for the dilute ones, the radicals of the Temperley-Lieb standard modules are nontrivial only when $q$ is a root of unity distinct than $\pm 1$. Let $\ell$ be the smallest positive integer such that $q^{2\ell}=1$. An integer $k$ is called critical if $k+1\equiv 0\textrm{\ mod\ }\ell$ and non-critical otherwise. Let $\irred{n,k}$ stand for the irreducible quotient $\vs{n,k}/\R{n,k}$ of the standard module $\vs{n,k}$.
\begin{Prop}\label{prop:dimIrrTLn} With the notation just introduced, the dimensions of the irreducible quotients can be obtained from the following recurrence equations:
\begin{equation}\label{eq:dimLnkTL}
\dim \irred{n,k}=\begin{cases}
\dim \vs{n,k},& \text{if $k$ is critical},\\
\dim \irred{n-1,k-1},& \text{if $k+1$ is critical},\\
\dim \irred{n-1,k-1}+\dim \irred{n-1,k+1}, & \text{otherwise},
\end{cases}
\end{equation}
with initial conditions $\dim \irred{n,n}=1$ for all $n$ and $\dim \irred{n,0}=0$ when $n$ is odd.
\end{Prop}
The algebra $\tl n(\beta)$ has a central element $\F n$ whose eigenvalues can distinguish any pair of standard modules whose labels $k$ and $k'$ fall between two consecutive critical lines. It is defined diagrammatically as the analogous element in $\dtl n$ (see appendix \ref{app:Fn}) by equation \eqref{eq:defDFn}. Here, however, the building tiles are defined by
\begin{align*}
\begin{tikzpicture}[scale=1/3]
	\node[anchor=east] at (0,0) {\dtile{0}{0}};
	\node[anchor=west] at (0,0) {$= \sqrt{q}$\ \ \tilea{-1}{-1}\ $- \frac1{\sqrt{q}}$\ \ \tileb{-1}{-1}\ \ ,};
\end{tikzpicture}\\
\begin{tikzpicture}[scale=1/3]
	\node[anchor=east] at (0,0) {\dtiler{0}{0}};
	\node[anchor=west] at (0,0) {$= \sqrt{q}$\ \  \tileb{-1}{-1}\ $- \frac1{\sqrt{q}}$\ \ \tilea{-1}{-1}\ \ .};
\end{tikzpicture}
\end{align*}
Here are the basic properties of $\F{n}$.
\begin{Prop}\label{prop:HomoForTLn} (i) The element $\F n\in\tl n$ is central, satisfies $F^t=F$ and acts on $\vs{n,k}$ as the identity times $\df k=q^{k+1}+q^{-(k+1)}$.
  
\noindent (ii) Let $q$ be a root of unity distinct from $\pm 1$ and $k\in\{0,1,\dots, n\}$ be critical for this $q$. Let $z_k\in\vs{n,k}$ be the link state with $k$ defects at the lowest positions and arcs between positions $2i-1$ and $2i$ for $1\leq i\leq (n-k)/2$. Then the action of $\F{n+1}$ on $\idtl\otimes z_k\in\Ind{\vs{n,k}}$ has a non-zero component along the vector $y_k=u_1u_2\dots u_n\otimes z_k$ in a basis containing both linearly independent elements $y_k$ and $z_k\in \Ind{\vs{n,k}}$.
\end{Prop}
\noindent A basis $\mathcal S_{n,k}$ of the induced module $\Ind{\vs{n,k}}$ is constructed explicitly in \cite{RiSY}. It is with this basis that the above result {\em (ii)} is stated in that paper. The simpler statement above establishes that the action of $\F{n+1}$ is not a multiple of the identity on $\Ind{\vs{n,k}}$. This is what will be used in the proof of lemma \ref{lem:OneThatUsedAppendixA}.
  
\end{section}

%
\begin{section}{The central element $\DF{n}$}\label{app:Fn}
%
  
One central element of $\dtl n$ plays an important role in the text, starting with the proof of proposition \ref{prop:restr.split}. If $q$ is generic, it has distinct eigenvalues on non-isomorphic standard modules. If $q$ is a root of unity, for certain indecomposable modules, it is not a multiple of the identity, a property that allows one to probe their structure. 
A similar element for $\tl n$ appeared in \cite{KauffmanLins}. A rather different formulation, in terms of diagrams, was used in \cite{MDSA} to probe the Jordan structure of the transfer matrix of loop models. It was also used in \cite{RiSY} to discuss the representation theory of $\tl n$.

The central element $\DF n$ is defined graphically through the following tiles
\begin{align*}
\begin{tikzpicture}[scale=1/3]
	\node[anchor=east] at (0,0) {\dtile{0}{0}};
	\node[anchor=west] at (0,0) {$=$};
	\node[anchor=west] at (1,0) {$\sqrt{q}$\ \ \tilea{-1}{-1}\ $- \frac{1}{\sqrt{q}}$\ \ \tileb{-1}{-1}\ +\ \tilec{-1}{-1}};	
\end{tikzpicture}\\
\begin{tikzpicture}[scale=1/3]
	\node[anchor=east] at (0,0) {\dtiler{0}{0}};
	\node[anchor=west] at (0,0) {$=$};
	\node[anchor=west] at (1,0) {$\sqrt{q}$\ \  \tileb{-1}{-1}\ $- \frac1{\sqrt{q}}$\ \ \tilea{-1}{-1}\ +\  \tilec{-1}{-1}};
\end{tikzpicture}
\end{align*}
which are multiplied according to the rules used for diagrams. It is important to recall that, if a vacancy and a string meet at an edge, the $n$-diagram to which they belong is zero. 
Note that the dashed horizontal segments on these defining tiles underline the fact that, for the three ``states'' of the right-hand sides, the vertical edges are the same: either both receive a link or both are vacancies. A stronger proprety, equation \eqref{eq:DFnProprietes}, will be satisfied by the element $\DF{n}$ whose definition is 
\begin{equation}\label{eq:defDFn}
\DF{n} = 
\begin{tikzpicture}[baseline={(current bounding box.center)},scale=1/3]
	\node at (0,0) {\dtile{0}{0}};
	\node at (2,0) {\dtiler{0}{0}};
	\node at (0,2) {\dtile{0}{0}};
	\node at (2,2) {\dtiler{0}{0}};
	\node at (0,17/4) {$\vdots$};
	\node at (2,17/4) {$\vdots$};
	\node at (0,6) {\dtile{0}{0}};
	\node at (2,6) {\dtiler{0}{0}};
	\node at (0,8) {\dtile{0}{0}};
	\node at (2,8) {\dtiler{0}{0}};
	\draw (0,-1) .. controls (0,-5/2) and (2,-5/2) .. (2,-1);
	\draw (0,9) .. controls (0,21/2) and (2,21/2) .. (2,9);
\end{tikzpicture}\ \ .
\end{equation}
The expansion of the $2n$ tiles leads to $3^{2n}$ different diagrams, most of them being zero. The two first $\DF n$ are
\begin{align*}
\DF{1} & = (q^2 + q^{-2})\ 
\begin{tikzpicture}[baseline={(current bounding box.center)},scale=1/3]
	\bord{0}{1};
	\bord{3}{1};
	\defect{0,0};
\end{tikzpicture}
\ +\ \beta\ 
\begin{tikzpicture}[baseline={(current bounding box.center)},scale=1/3]
	\bord{0}{1};
	\bord{3}{1};
	\vac{0,0};
	\vac{3,0};
\end{tikzpicture}\\
\ &\ \\
\DF{2} & = (q^3 + q^{-3})\ 
\begin{tikzpicture}[baseline={(current bounding box.center)},scale=1/3]
	\bord{0}{2};
	\bord{3}{2};
	\defect{0,0};
	\defect{0,1};
\end{tikzpicture}
\ -\ (q -q^{-1})^2\ 
\begin{tikzpicture}[baseline={(current bounding box.center)},scale=1/3]
	\bord{0}{2};
	\bord{3}{2};
	\link{0,0}{1,0}{1,1}{0,1};
	\link{3,0}{2,0}{2,1}{3,1};
\end{tikzpicture}\\
& \qquad \ +\ (q^2 + q^{-2}) \left(
\begin{tikzpicture}[baseline={(current bounding box.center)},scale=1/3]
	\bord{0}{2};
	\bord{3}{2};
	\defect{0,0};
	\vac{0,1};
	\vac{3,1};
\end{tikzpicture}
\ +\ 
\begin{tikzpicture}[baseline={(current bounding box.center)},scale=1/3]
	\bord{0}{2};
	\bord{3}{2};
	\defect{0,1};
	\vac{0,0};
	\vac{3,0};
\end{tikzpicture}\right)
\ +\ \beta\ 
\begin{tikzpicture}[baseline={(current bounding box.center)},scale=1/3]
	\bord{0}{2};
	\bord{3}{2};
	\vac{0,1};
	\vac{0,0};
	\vac{3,0};
	\vac{3,1};
\end{tikzpicture}\ .
\end{align*}
To verify that it is indeed a central element, we compute its products with the generators of $\dtl{n}$. We start by expanding the tiles of the left column:
\begin{align*}
\begin{tikzpicture}[baseline={(current bounding box.center)},scale=1/3]
	\draw[\epaisdfn] (1,0) -- (1,4);
	\draw[\epaisdfn] (3,0) -- (3,4);
	\vac{1,1};
	\vac{1,3};
	\vac{3,1};
	\vac{3,3};
	\node at (5,1) {\dtiler{0}{0}};
	\node at (5,3) {\dtiler{0}{0}};
\end{tikzpicture}
& =
\begin{tikzpicture}[baseline={(current bounding box.center)},scale=1/3]
	\draw[\epaisdfn] (1,0) -- (1,4);
	\draw[\epaisdfn] (3,0) -- (3,4);
	\vac{1,1};
	\vac{1,3};
	\vac{3,1};
	\vac{3,3};
	\node at (5,1) {\dtile{0}{0}};
	\node at (5,3) {\dtile{0}{0}};
\end{tikzpicture}
=
\begin{tikzpicture}[baseline={(current bounding box.center)},scale=1/3]
	\node at (0,1) {\tilec{0}{0}};
	\node at (0,3) {\tilec{0}{0}};
\end{tikzpicture}
=
\begin{tikzpicture}[baseline={(current bounding box.center)},scale=1/3]
	\draw[\epaisdfn] (3,0) -- (3,4);
	\draw[\epaisdfn] (5,0) -- (5,4);
	\vac{3,1};
	\vac{3,3};
	\vac{5,1};
	\vac{5,3};
	\node at (1,1) {\dtile{0}{0}};
	\node at (1,3) {\dtile{0}{0}};
\end{tikzpicture}
=
\begin{tikzpicture}[baseline={(current bounding box.center)},scale=1/3]
	\draw[\epaisdfn] (3,0) -- (3,4);
	\draw[\epaisdfn] (5,0) -- (5,4);
	\vac{3,1};
	\vac{3,3};
	\vac{5,1};
	\vac{5,3};
	\node at (1,1) {\dtiler{0}{0}};
	\node at (1,3) {\dtiler{0}{0}};
\end{tikzpicture}\\
\begin{tikzpicture}[baseline={(current bounding box.center)},scale=1/3]
	\draw[\epaisdfn] (1,0) -- (1,4);
	\draw[\epaisdfn] (3,0) -- (3,4);
	\vac{1,1};
	\vac{3,3};
	\draw (1,3) -- (3,1);
	\node at (5,1) {\dtile{0}{0}};
	\node at (5,3) {\dtile{0}{0}};
\end{tikzpicture}
& = \sqrt{q}
\begin{tikzpicture}[baseline={(current bounding box.center)},scale=1/3]
	\draw[\epaisdfn] (0,0) -- (0,4);
	\draw[\epaisdfn] (2,0)	-- (2,4);
	\draw (0,3) .. controls (1/2,3) and (1,7/2) .. (1,4);
	\draw (2,1) .. controls (3/2,1) and (1,1/2) .. (1,0);
	\vac{0,1};
	\vac{2,3};
\end{tikzpicture}
-\frac1{\sqrt{q}}
\begin{tikzpicture}[baseline={(current bounding box.center)},scale=1/3]
	\draw[\epaisdfn] (0,0) -- (0,4);
	\draw[\epaisdfn] (2,0)	-- (2,4);
	\draw (0,3) .. controls (1/2,3) and (1,1/2) .. (1,0);
	\draw (2,1) .. controls (3/2,1) and (1,7/2) .. (1,4);
	\vac{0,1};
	\vac{2,3};
\end{tikzpicture}
=
\begin{tikzpicture}[baseline={(current bounding box.center)},scale=1/3]
	\draw[\epaisdfn] (3,0) -- (3,4);
	\draw[\epaisdfn] (5,0) -- (5,4);
	\vac{3,1};
	\vac{5,3};
	\draw (3,3) -- (5,1);
	\node at (1,1) {\dtile{0}{0}};
	\node at (1,3) {\dtile{0}{0}};
\end{tikzpicture}\\
\begin{tikzpicture}[baseline={(current bounding box.center)},scale=1/3]
	\draw[\epaisdfn] (1,0) -- (1,4);
	\draw[\epaisdfn] (3,0) -- (3,4);
	\vac{1,3};
	\vac{3,1};
	\draw (1,1) -- (3,3);
	\node at (5,1) {\dtile{0}{0}};
	\node at (5,3) {\dtile{0}{0}};
\end{tikzpicture}
& = \sqrt{q}
\begin{tikzpicture}[baseline={(current bounding box.center)},scale=1/3]
	\draw[\epaisdfn] (1,0) -- (1,4);
	\draw[\epaisdfn] (3,0) -- (3,4);
	\draw (1,1) .. controls (3/2,1) and (2,7/2) .. (2,4);
	\draw (3,3).. controls (5/2,3) and (2,1/2) .. (2,0);
	\vac{1,3};
	\vac{3,1};
\end{tikzpicture}
-\frac1{\sqrt{q}}
\begin{tikzpicture}[baseline={(current bounding box.center)},scale=1/3]
	\draw[\epaisdfn] (1,0) -- (1,4);
	\draw[\epaisdfn] (3,0) -- (3,4);
	\draw (1,1) .. controls (3/2,1) and (2,1/2) .. (2,0);
	\draw (3,3).. controls (5/2,3) and (2,7/2) .. (2,4);
	\vac{1,3};
	\vac{3,1};
\end{tikzpicture}
=
\begin{tikzpicture}[baseline={(current bounding box.center)},scale=1/3]
	\draw[\epaisdfn] (5,0) -- (5,4);
	\draw[\epaisdfn] (3,0) -- (3,4);
	\vac{3,3};
	\vac{5,1};
	\draw (3,1) -- (5,3);
	\node at (1,1) {\dtile{0}{0}};
	\node at (1,3) {\dtile{0}{0}};
\end{tikzpicture}\\
\begin{tikzpicture}[baseline={(current bounding box.center)},scale=1/3]
	\draw[\epaisdfn] (1,0) -- (1,4);
	\draw[\epaisdfn] (3,0) -- (3,4);
	\vac{1,3};
	\vac{1,1};
	\draw (3,1) .. controls (5/2,1) and (5/2,3) .. (3,3);
	\node at (5,1) {\dtile{0}{0}};
	\node at (5,3) {\dtile{0}{0}};
\end{tikzpicture}
&=
\begin{tikzpicture}[baseline={(current bounding box.center)},scale=1/3]
	\draw[\epaisdfn] (1,0) -- (1,4);
	\draw[\epaisdfn] (3,0) -- (3,4);
	\vac{1,3};
	\vac{1,1};
	\draw (3,1) .. controls (5/2,1) and (5/2,3) .. (3,3);
\end{tikzpicture} 
\left( q
\begin{tikzpicture}[baseline={(current bounding box.center)},scale=1/3]
	\node at (0,0) {\tilea{0}{0}};
	\node at (0,2) {\tilea{0}{0}};
\end{tikzpicture}
-
\begin{tikzpicture}[baseline={(current bounding box.center)},scale=1/3]
	\node at (0,0) {\tilea{0}{0}};
	\node at (0,2) {\tileb{0}{0}};
\end{tikzpicture}
-
\begin{tikzpicture}[baseline={(current bounding box.center)},scale=1/3]
	\node at (0,0) {\tileb{0}{0}};
	\node at (0,2) {\tilea{0}{0}};
\end{tikzpicture}
+ \frac1q
\begin{tikzpicture}[baseline={(current bounding box.center)},scale=1/3]
	\node at (0,0) {\tileb{0}{0}};
	\node at (0,2) {\tileb{0}{0}};
\end{tikzpicture}
\right)\\
& = 
-
\begin{tikzpicture}[baseline={(current bounding box.center)},scale=1/3]
	\draw[\epaisdfn] (1,0) -- (1,4);
	\draw[\epaisdfn] (3,0) -- (3,4);
	\vac{1,1};
	\vac{1,3};
	\draw (3,1) .. controls (5/2,1) and (5/2,3) .. (3,3);
	\draw (2,0) -- (2,4);
\end{tikzpicture}
=-
\begin{tikzpicture}[baseline={(current bounding box.center)},scale=1/3]
	\draw[\epaisdfn] (5,0) -- (5,4);
	\draw[\epaisdfn] (3,0) -- (3,4);
	\vac{3,3};
	\vac{3,1};
	\draw (5,1) .. controls (9/2,1) and (9/2,3) .. (5,3);
	\node at (1,1) {\dtile{0}{0}};
	\node at (1,3) {\dtile{0}{0}};
\end{tikzpicture}\\
\begin{tikzpicture}[baseline={(current bounding box.center)},scale=1/3]
	\draw[\epaisdfn] (1,0) -- (1,4);
	\draw[\epaisdfn] (3,0) -- (3,4);
	\vac{3,3};
	\vac{3,1};
	\draw (1,1) .. controls (3/2,1) and (3/2,3) .. (1,3);
	\node at (5,1) {\dtile{0}{0}};
	\node at (5,3) {\dtile{0}{0}};
\end{tikzpicture}
& =
\begin{tikzpicture}[baseline={(current bounding box.center)},scale=1/3]
	\draw[\epaisdfn] (1,0) -- (1,4);
	\draw[\epaisdfn] (3,0) -- (3,4);
	\vac{3,1};
	\vac{3,3};
	\draw (1,1) .. controls (3/2,1) and (3/2,3) .. (1,3);
	\draw (2,0) -- (2,4);
\end{tikzpicture} 
=
-\left( q
\begin{tikzpicture}[baseline={(current bounding box.center)},scale=1/3]
	\node at (0,0) {\tilea{0}{0}};
	\node at (0,2) {\tilea{0}{0}};
\end{tikzpicture}
-
\begin{tikzpicture}[baseline={(current bounding box.center)},scale=1/3]
	\node at (0,0) {\tilea{0}{0}};
	\node at (0,2) {\tileb{0}{0}};
\end{tikzpicture}
-
\begin{tikzpicture}[baseline={(current bounding box.center)},scale=1/3]
	\node at (0,0) {\tileb{0}{0}};
	\node at (0,2) {\tilea{0}{0}};
\end{tikzpicture}
+ \frac1{q}
\begin{tikzpicture}[baseline={(current bounding box.center)},scale=1/3]
	\node at (0,0) {\tileb{0}{0}};
	\node at (0,2) {\tileb{0}{0}};
\end{tikzpicture}
\right)
\begin{tikzpicture}[baseline={(current bounding box.center)},scale=1/3]
	\draw[\epaisdfn] (1,0) -- (1,4);
	\draw[\epaisdfn] (3,0) -- (3,4);
	\vac{3,3};
	\vac{3,1};
	\draw (1,1) .. controls (3/2,1) and (3/2,3) .. (1,3);
\end{tikzpicture}\\
& = -
\begin{tikzpicture}[baseline={(current bounding box.center)},scale=1/3]
	\draw[\epaisdfn] (5,0) -- (5,4);
	\draw[\epaisdfn] (3,0) -- (3,4);
	\vac{5,3};
	\vac{5,1};
	\draw (3,1) .. controls (7/2,1) and (7/2,3) .. (3,3);
	\node at (1,1) {\dtile{0}{0}};
	\node at (1,3) {\dtile{0}{0}};
\end{tikzpicture}
\end{align*}
Note that a sign appears during the commutation of the two last generators. The computation for the right column is obtained from that for the left by the exchange 
$\sqrt{q}\leftrightarrow -1/\sqrt{q}$. The following result is thus proved.
\begin{Prop}
$\DF{n}$ is a central element of $\dtl{n}$.
\end{Prop}
  
\noindent The eigenvalues of $\DF n$ on standard modules $\U{n,k}$ are easily computed. Before doing so, it is useful to note that rows of the defining tiles, acting on vacancies or arcs of a link state, have the following properties
\begin{equation}\label{eq:DFnProprietes}
\begin{tikzpicture}[baseline={(current bounding box.center)},scale=1/3]
	\node at (0,0) {\dtile{0}{0}};
	\node at (2,0) {\dtiler{0}{0}};
	\vac{3,0};
\end{tikzpicture}\ = \ 
\begin{tikzpicture}[baseline={(current bounding box.center)},scale=1/3]
	\draw (0,0) --++ (4,0) --++ (0,2) --++ (-4,0) --++ (0,-2);
	\vac{0,1};
	\draw (1,0) -- (1,2);
	\draw (3,0) -- (3,2);
\end{tikzpicture}\qquad\textrm{and}\qquad
\begin{tikzpicture}[baseline={(current bounding box.center)},scale=1/3]
	\node at (0,0) {\dtile{0}{0}};
	\node at (2,0) {\dtiler{0}{0}};
	\node at (0,2) {\dtile{0}{0}};
	\node at (2,2) {\dtiler{0}{0}};
	\link{3,0}{4,0}{4,2}{3,2};
\end{tikzpicture}\ = \ 
\begin{tikzpicture}[baseline={(current bounding box.center)},scale=1/3]
	\draw (0,0) --++ (4,0) --++ (0,4) --++ (-4,0) --++ (0,-4);
	\link{0,1}{0.75,1}{0.75,3}{0,3};
	\draw (1,0) -- (1,4);
	\draw (3,0) -- (3,4);
\end{tikzpicture}
\end{equation}
as a direct expansion of the tiles shows. The first property above indicates that $\DF n$ is actually an element of the subalgebra $S_n$. To see this, let $u\in\dtl n$ be an $n$-diagram. If $I=\{i_1,i_2,\dots, i_{n-k}\}$ is the set of positions of its vacancies on its left side, then $u=\pi_zu$ where $z\in\X{n,k}$ is the link diagram with vacancies at these same positions. (The element $\pi_z=|z\bar z|\in\dtl n$ is introduced and discussed in Section \ref{sec:Gram}.) The product $\DF nu=\DF n \pi_z u$ is simplified by the observation that all vacancies of $\pi_z$ go through $\DF n$ due to \eqref{eq:DFnProprietes} and $\DF nu=(\pi_z\DF n \pi_z) u$. The sums in the remaining tiles may omit the tile $\tilecMini{0}{0}$, as a link in $\pi_z$ is connected on either side of each tile to be summed. These sums are then precisely those intervening in the definition of $\F k$ of $\tl k$. 
\begin{Prop}\label{prop:DFnAvecPIz}
The central element $\DF n\in S_n\subset\dtl n$ can be written as
\begin{equation}\label{eq:DFnAvecPIz}
\DF n=\sum_{0\le k\le n}\sum_{z\in\X{n,k}}\pi_z\DF n\pi_z
\end{equation}
where each summand $\pi_z\DF n\pi_z$ is constructed by insertion in $\F k$ of $(n-k)$ lines of vacancies to match those in $z$.
\end{Prop}

\begin{Prop}\label{prop:EigenvalueDFn}
On $\U{n,k}$, the element $\DF{n}$ acts as $\df{k}\cdot \id$ where $\df{k} = q^{k+1} + q^{-(k+1)}$.
\end{Prop}
\begin{proof}
Since $\DF{n}$ is central and the modules $\U{n,k}$ are indecomposable (proposition \ref{prop:UnkIndecomposable}), the endomorphism defined by left multiplication by $\DF n$ can only have one eigenvalue. By the previous proposition and the properties \eqref{eq:DFnProprietes}, the tiles on lines of $\DF n$ acting on vacancies or on arcs of a link state are therefore completely determined, they contribute an overall factor of $1$ and can be left out of the computation. For $z$ a link diagram in $\U{n,k}$, the computation of $\DF nz$ thus reduces to that of $\DF kz_0$ where $z_0$ is the unique $k$-link diagram with $k$ defects. Moreover the computation of $\DF kz_0$ does not involve anymore the tile $\tilecMini{0}{0}$ and it becomes identical to that for the action of the central element $\F k$ on $\vs{k,k}$. This computation was done in \cite{RiSY}: $\DF kz_0=\F kz_0=(q^{k+1} + q^{-(k+1)})z_0$. (See proposition \ref{prop:HomoForTLn}.) Since the link diagrams form a basis of $\U{n,k}$, the element $\DF n$ acts as a multiple of the identity and the result follows.
\end{proof}
  
\noindent These eigenvalues of the central element $\DF n$ provide a good way to distinguish between standard modules. More precisely:
\begin{Lem}\label{lem:deltanPropriete} Let $n$ be a positive integer. 
  
\noindent (i) If $q$ is not a root of unity, then $\df j\neq\df k$ if $j\neq k$.
  
\noindent Let $q$ be a root of unity other than $\pm 1$ and $\ell$ be the smallest positive integer such that $q^{2\ell}=1$. Let $k_c$ be critical or $-1$ and $K_{\mathsf e}$ (resp.~$K_{\mathsf o}$) denote the set of $k$s such that $k_c< k\leq k_c+\ell$ and $k$ has the parity of $n-k_c$ (resp.~of $n-k_c-1$).
  
\noindent (ii)  If $j$ and $k$ are distinct and both in $K_{\mathsf e}$ (or both in $K_{\mathsf o}$), then $\df j\neq\df k$. 
  
\noindent (iii) The intersection $K_{\mathsf e}\cap K_{\mathsf o}$ is non-empty if and only if $q$ is of the form $e^{2i\pi m/\ell}$ with $\gcd(m,l)=1$ and $l$ odd.
  
\noindent (iv) The function $\df k$ is even with respect to a mirror reflection through a critical line.
\end{Lem}
\begin{proof}
If $\theta\in\mathbb C$ is chosen such that $q=e^{i\theta}$, then $\df j=\df k$ is equivalent to $\cos ((j+1)\theta)=\cos ((k+1)\theta)$ which in turn amounts to either {\em (a)} $(k+1)\theta=(j+1)\theta+2\pi p$ or {\em (b)} $(k+1)\theta=-(j+1)\theta+2\pi p$ for some integer $p$. If $j\neq k$, then $\theta$ must be a (real) rational multiple of $\pi$ and {\em (i)} follows. If $q^{2\ell}=1$ with $\ell$ the smallest possible, then either {\em (c)} $q=e^{2\pi im/\ell}$ with $\textrm{gcd}(m,\ell)=1$ and $\ell$ odd or {\em (d)} $q=e^{\pi i(2m+1)/\ell}$ with $\textrm{gcd}(2m+1,\ell)=1$. The equation {\em (a)} requires that $(k-j)\theta$ be an integer multiple of $2\pi$. But $(k-j)\theta$ is either $2\pi(k-j)m/\ell$ or $\pi(k-j)(2m+1)/\ell$. Both forms require that the difference $k-j$ be a multiple of $\ell$ which is impossible since $k_c<k,j\leq k_c+\ell$.
  
To study the case {\em (b)}, write $k=k_c+\bar k$ and $j=k_c+\bar j$ with $0<\bar k, \bar j\leq \ell$. Since $k_c+1\equiv 0\ \textrm{mod}\ \ell$, the equation {\em (b)} forces $(k+j+2)\theta$ (or equivalently $(\bar k+\bar j)\theta$) to be an integer multiple of $2\pi$. For the case (d), this is impossible since $(\bar k+\bar j)(2m+1)/\ell = 2 p$ would mean that $(\bar k+\bar j)$ is an even multiple of $\ell$. However, in the case {\em (c)}, $\ell$ is always odd (and $\geq 3$) and the equation $(\bar k+\bar j)m/\ell = p$ has always the solution $\bar k=1$ and $\bar j=\ell-1(\neq \bar k)$. Note that this solution and all others ($\bar k=i$ and $\bar j=\ell-i$) have $k$ and $\ell$ of distinct parity. This proves both {\em (ii)} and {\em (iii)}. 
  
If $k_\pm=k_c\pm m$, then $q^{k_++1}=q^{k_c+m+1}=q^{-k_c+m-1}=q^{-(k_-+1)}$ where the criticality of $k_c$ was used. The last statement follows.
\end{proof}
  
\end{section}
  
%
\begin{section}{The dimensions of the irreducible modules $\dl{n,k}$}\label{app.dlnk}
%
The dimensions of the irreducible quotients $\dl{n,k}$ satisfy recurrence relations that can be used efficiently to compute them. 
We gather here these relations and their proofs. Tables containing the dimensions of standard modules and of irreducible quotients for $\ell=3$ and $\ell=4$ are also given. 
  
As usual $q$ is a root of unity other than $\pm 1$ and $\ell$ is the smallest positive integer such that $q^{2\ell}=1$ (and $\ell\geq 2$). The notation
$$\ddl{n,k} = \dim \dl{n,k},\quad \ddr{n,k} = \dim \dr{n,k}, \quad \dU{n,k} = \dim \U{n,k}\ \ \textrm{and}\ \ 
\drl{n,k}=\dim \rl{n,k}.$$
is used throughout. We recall that the symbol $\dl{n,k}$ is used for the irreducible module over $\dtl n$ and $\rl{n,k}$ for that over $\tl n$. 
  
The module $\dl{n,k}$ is defined to be the irreducible quotient $\U{n,k}/\dr{n,k}$. Corollaries \ref{coro:dimUnk} and \ref{Dimens-Rad} then give a simple formula for its dimension in terms of those for the irreducibles $\rl{n,k}$ of the (original) Temperley-Lieb algebra:
\begin{equation}\label{eq:dimdLnk}
\ddl{n,k}=\dim \U{n,k}-\dim \dr{n,k}=\sum_{p=0}^{\lfloor(n-k)/2\rfloor} \binom{n}{k+2p}\drl{n,k}.\end{equation}

\begin{Prop} \label{prop.dim.irred}
Let $n\geq 1$ and $\kc$ be an integer critical for $q$. Then the three following recurrence relations hold:
\begin{align}
\ddl{n+1,\kc-1} &= \ddl{n,\kc-2} + \ddl{n,\kc-1},\label{eq:lnk1}\\
\ddl{n+1,\kc+i} &= \ddl{n,\kc+i+1}+\ddl{n,\kc+i}+\ddl{n,\kc+i-1},\qquad 1\leq i\leq \ell-2,\label{eq:lnk2}\\
\ddl{n+1,\kc} &= \ddl{n,\kc-1}+\ddl{n,\kc}+2\ddl{n,\kc+1}+\ddl{n,\kc+2\ell-1},\label{eq:lnk3}
\end{align}
where any $\ddl{n,j}$ with a $j$ outside the set $\{0,1,\dots, n\}$ is zero. With this convention, the second equation also holds for $\kc=-1$. The boundary conditions are $\ddl{n,-1}=0$ and $\ddl{n,n}=1$. 
\end{Prop}
\begin{proof} For the first of these recurrences, use \eqref{eq:dimdLnk} to write $\ddl{n+1,\kc-1}$ in terms of the $\drl{}$s and split the sum into two using the binomial identity $\big(\begin{smallmatrix}n+1\\ j\end{smallmatrix}\big)=\left(\begin{smallmatrix}n\\ j\end{smallmatrix}\right)+\left(\begin{smallmatrix}n\\ j-1\end{smallmatrix}\right)$. The summation index of the second sum, that containing the binomial $\left(\begin{smallmatrix}n\\ j-1\end{smallmatrix}\right)$, is then shifted using \eqref{eq:dimLnkTL} of proposition \ref{prop.rad}. Terms that are missing at either end of the sums can be added as they are weighted by a binomial that vanishes. The proof of the second recurrence follows the same lines.
  
The last recurrence is proved as follows:
\begin{align*}\ddl{n+1,\kc} &= \dU{n+1,\kc}\\
	&= \dU{n,\kc-1} + \ddl{n,\kc} + \dU{n,\kc+1}\\
	&= \ddl{n,\kc-1}+\ddr{n,\kc-1}+\ddl{n,\kc}+\ddl{n,\kc+1}+\ddr{n,\kc+1}\\
	&= \ddl{n,\kc-1}+\ddl{n,\kc+1}+\ddl{n,\kc}+\ddl{n,\kc+1}+\ddl{n,\kc+2\ell-1}.
\end{align*}
The first equation is simply the irreducibility of $\U{n+1,\kc}$, the second line follows from the restriction of $\U{n+1,\kc}$ (see \eqref{eq:simpleRelation}), the last line is a consequence of proposition (\ref{prop.rad.iso}).
\end{proof}

The dimensions of $\U{n,k}$ are showed in table \ref{dim.unk} for $n\leq 10$, and the dimension of $\dl{n,k}$ are showed in tables \ref{dim.lnk3} and \ref{dim.lnk4} for $\ell=3$ and $\ell=4$, respectively.
\begin{table}
\begin{tabular}{c |c c c c c c c c c c c}
$n$ / $k$ & $0$ & $1$ & $2$ & $3$ & $4$ & $5$ & $6$ & $7$ & $8$ & $9$ & $10$\\ \hline
 $1$ & $1$ & $1$ & $ $ & $ $ & $ $ & $ $ & $ $ & $ $ & $ $ & $ $ & $$  \\
 $2$ & $2$ & $2$ & $1$ & $ $ & $ $ & $ $ & $ $ & $ $ & $ $ & $ $ & $$  \\
 $3$ & $4$ & $5$ & $3$ & $1$ & $ $ & $ $ & $ $ & $ $ & $ $ & $ $ & $$  \\
 $4$ & $9$ & $12$ & $9$ & $4$ & $1$ & $ $ & $ $ & $ $ & $ $ & $ $ & $$  \\
 $5$ & $21$ & $30$ & $25$ & $14$ & $5$ & $1$ & $ $ & $ $ & $ $ & $ $ & $$  \\
 $6$ & $51$ & $76$ & $69$ & $44$ & $20$ & $6$ & $1$ & $ $ & $ $ & $ $ & $$  \\
 $7$ & $127$ & $196$ & $189$ & $133$ & $70$ & $27$ & $7$ & $1$ & $ $ & $ $ & $ $  \\
 $8$ & $323$ & $512$ & $518$ & $392$ & $230$ & $104$ & $35$ & $8$ & $1$ & $ $ & $ $  \\
 $9$ & $835$ & $1353$ & $1422$ & $1140$ & $726$ & $369$ & $147$ & $44$ & $9$ & $1$ & $ $  \\
 $10$ & $2188$ & $3610$ & $3915$ & $3288$ & $2235$ & $1242$ & $560$ & $200$ & $54$ & $10$ & $1$
\end{tabular}
\caption{Dimensions $\dU{n,k}=\dim\U{n,k}$}
\label{dim.unk}
\end{table}
\begin{table}
\begin{tabular}{c| c c c c c c c c c c c}
$n$ / $k$ & $0$ & $1$ & $2$ & $3$ & $4$ & $5$ & $6$ & $7$ & $8$ & $9$ & $10$\\ \hline
 $1$ & $1$ & $1$ & $ $ & $ $ & $ $ & $ $ & $ $ & $ $ & $ $ & $ $ & $ $  \\
 $2$ & $2$ & $2$ & $1$ & $ $ & $ $ & $ $ & $ $ & $ $ & $ $ & $ $ & $ $  \\
 $3$ & $4$ & $4$ & $3$ & $1$ & $ $ & $ $ & $ $ & $ $ & $ $ & $ $ & $ $  \\
 $4$ & $8$ & $8$ & $9$ & $4$ & $1$ & $ $ & $ $ & $ $ & $ $ & $ $ & $ $  \\
 $5$ & $16$ & $16$ & $25$ & $14$ & $5$ & $1$ & $ $ & $ $ & $ $ & $ $ & $ $  \\
 $6$ & $32$ & $32$ & $69$ & $44$ & $19$ & $6$ & $1$ & $ $ & $ $ & $ $ & $ $  \\
 $7$ & $64$ & $64$ & $189$ & $132$ & $63$ & $27$ & $7$ & $1$ & $ $ & $ $ & $ $  \\
 $8$ & $128$ & $128$ & $518$ & $384$ & $195$ & $104$ & $35$ & $8$ & $1$ & $ $ & $ $  \\
 $9$ & $256$ & $256$ & $1422$ & $1097$ & $579$ & $369$ & $147$ & $43$ & $9$ & $1$ & $ $  \\
 $10$ & $512$ & $512$ & $3915$ & $3098$ & $1676$ & $1242$ & $559$ & $190$ & $54$ & $10$ & $1$
\end{tabular}
\caption{Dimensions $\ddl{n,k}= \dim \dl{n,k}$ for $\ell=3$}
\label{dim.lnk3}
\end{table}
  
\begin{table}
\begin{tabular}{c|ccccccccccc}
$n$ / $k$ & $0$ & $1$ & $2$ & $3$ & $4$ & $5$ & $6$ & $7$ & $8$ & $9$ & $10$ \\ \hline
 $1$ & $1$ & $1$ & $ $ & $ $ & $ $ & $ $ & $ $ & $ $ & $ $ & $ $ & $ $ \\
 $2$ & $2$ & $2$ & $1$ & $ $ & $ $ & $ $ & $ $ & $ $ & $ $ & $ $ & $ $\\
 $3$ & $4$ & $5$ & $3$ & $1$ & $ $ & $ $ & $ $ & $ $ & $ $ & $ $ &  $ $\\
 $4$ & $9$ & $12$ & $8$ & $4$ & $1$ & $ $ & $ $ & $ $ & $ $ & $ $ &  $ $\\
 $5$ & $21$ & $29$ & $20$ & $14$ & $5$ & $1$ & $ $ & $ $ & $ $ & $ $ & $ $ \\
 $6$ & $50$ & $70$ & $49$ & $44$ & $20$ & $6$ & $1$ & $ $ & $ $ & $ $ & $ $ \\
 $7$ & $120$ & $169$ & $119$ & $133$ & $70$ & $27$ & $7$ & $1$ & $ $ & $ $ & $ $ \\
 $8$ & $289$ & $408$ & $288$ & $392$ & $230$ & $104$ & $34$ & $8$ & $1$ & $ $ & $ $ \\
 $9$ & $697$ & $985$ & $696$ & $1140$ & $726$ & $368$ & $138$ & $44$ & $9$ & $1$ & $ $ \\
 $10$ & $1682$ & $2378$ & $1681$ & $3288$ & $2234$ & $1232$ & $506$ & $200$ & $54$ & $10$ & $1$
\end{tabular}
\caption{Dimensions $\ddl{n,k}= \dim \dl{n,k}$ for $\ell=4$.}
\label{dim.lnk4}
\end{table}

\end{section}
  
%
\begin{section}{Tools from algebra}\label{sec:algebre}
%
  
We review here concepts and results in algebra that are used in the article and might not be familiar to some readers. We start by presenting short exact sequences and proceed to projective modules. The interplay between induction and the tensor product is then recalled. We finally recall Wedderburn's theorem, and its generalization, and Frobenius reciprocity theorem.
  
Throughout the appendix, $A$ is a unital associative algebra over $\mathbb C$, $B$ a subalgebra of $A$. Unless otherwise stated, $L, M, N$ and $P$ are $A$-modules.
   
%
\begin{subsection}{Short exact sequences}
%
  
Let $L\overset{f}{\to} M$ and $M \overset{g}{\to} N$ be two module homomorphisms. The sequence $L\overset{f}{\to} M \overset{g}{\to} N$ is said to be \emph{exact} (or {\em exact at $M$}) if the kernel of $g$ is equal to the image of $f$. A \emph{short exact sequence} is a sequence of homomorphisms $$0 \longrightarrow L\overset{f}{\longrightarrow} M \overset{g}{\longrightarrow} N \longrightarrow 0$$
that is exact at $L$, $M$ and $N$. This is equivalent to saying that the sequence is exact at $M$ with $f$ and $g$ being injective and surjective, respectively.
\begin{Prop}\label{prop.short}
A sequence 
$$0 \longrightarrow L\overset{f}{\longrightarrow} M \overset{g}{\longrightarrow} N \longrightarrow 0$$ is exact if and only if it verifies the three following conditions:
  
(i) $gf =0$
  
(ii) if there is a module $U$ and a homomorphism $u:U \to M$ such that $gu =0$, then there is a unique homomorphism $\bar{u}: U \to L$ such that $f\bar{u} = u$;
  
(iii) if there is a module $V$ and a homomorphism $v:M \to V$ such that $vf =0$, then there is a unique homomorphism $\bar{v}:N \to V$ such that $\bar{v}g = v$.
\end{Prop}
  
The short exact sequence of proposition \ref{prop.short} is called \emph{split} if $M\simeq L \oplus N$.
  
\begin{Prop}\label{prop.split}
If the short sequence $$0 \longrightarrow L\overset{f}{\longrightarrow} M \overset{g}{\longrightarrow} N \longrightarrow 0$$ is exact, the three following statements are equivalent:
  
(i) the sequence splits;
  
(ii) there is a homomorphism $\bar{f}: M \to L$ such that $\bar{f}f = \id_{L}$;
  
(iii) there is a homomorphism $\bar{g}: N \to M$ such that $g\bar{g} = \id_{N}$.
\end{Prop}
  
\end{subsection}

%
\begin{subsection}{Projective modules}\label{sec:projectiveAndFlat}
%
A module $P$ is said to be \emph{projective} if for all modules $M$ and $N$ and all homomorphisms $f:M \to N$ and $g:P \to N$ with $f$ surjective, there is a homomorphism $h:P \to M$ such that $f\circ h = g$. In other words, given homomorphisms $f$ and $g$ as in the diagram below with an exact horizontal row, then there exist $h$ that makes the diagram commute.
\begin{center}
\begin{tikzpicture}[node distance=1.2cm, auto]
  \node (P) {$P$};
  \node (N) [below of=P] {$ N$};
  \node (M) [left of=N] {$ M$};
  \node (O) [right of=N] {$0$.};
  \draw[->] (P) to node {$g$} (N);
  \draw[->, dashed] (P) to node [swap] {$h$} (M);
  \draw[->] (M) to node [swap] {$f$} (N);
  \draw[->] (N) to node [swap] {$\ $} (O);
\end{tikzpicture}
\end{center}
Because of proposition (\ref{prop.split}), the above definition (with $P=N$ and $g=\id_P$) gives:
\begin{Prop}
A module $P$ is projective if and only if all short exact sequences 
\begin{equation*}
0 \longrightarrow L \longrightarrow M \longrightarrow P \longrightarrow 0
\end{equation*}
split.
\end{Prop}
\noindent Direct sums and direct summands of projective modules are also projective. Note also that an algebra seen as a module over itself is always projective. 
  
A \emph{projective cover} of a module $M$ is a pair $(P,f)$ with $P$ a projective module and $P\overset{f}{\rightarrow}M$ a surjective morphism having the following property. If $(P',g)$ is another pair where $P'$ is projective and $g$ is a surjective morphism from $P'$ to $M$, then $P' \simeq P \oplus Q$ for some module $Q$. Therefore projective covers are unique up to isomorphism. 
  
Write $\Hom(M,N)$ for the vector space of $A$-homomorphisms of $M$ into $N$.
  
\begin{Prop}\label{prop:rightExactHom} If $0 \to M\to N\to P\to 0$ is exact, then, for any other module $L$, so are
$$0\longrightarrow 
\Hom(L,M)\longrightarrow
\Hom(L,N)\longrightarrow
\Hom(L,P), $$ and $$0\longrightarrow 
\Hom(P,L)\longrightarrow
\Hom(N,L)\longrightarrow
\Hom(M,L).$$
Moreover, if $L$ is projective, then the last homomorphism in the first sequence above is surjective.
\end{Prop}
  
\end{subsection}
  
%
\begin{subsection}{Restriction and induction}
%
If an algebra $A$ has a subalgebra $B$, it is natural to ask how a given $A$-module $M$ would behave as a $B$-module. Since $B$ is a subalgebra of $A$, the space $M$ can be seen as a $B$-module for the same action and the $B$-module thus obtained is called the \emph{restriction} of $M$ to $B$, and is noted $\Res{M}$ (or $\Res{M}_B^A$). It can be shown that restriction preserves short exact sequences, that is:
\begin{Prop}
Let $A$ be an associative algebra and $B$ a sub-algebra of $A$, the sequence
$$0 \longrightarrow L \longrightarrow M \longrightarrow P \longrightarrow 0$$ of $A$-modules is exact if and only if the sequence $$0 \longrightarrow \Res{L}_B^A \longrightarrow \Res{M}_B^A  \longrightarrow \Res{P}_B^A  \longrightarrow 0$$ of $B$-modules is exact.
\end{Prop}
 It is also natural to do the ``reverse process'', that is, to transform a $B$-module into an $A$-module. This process is slightly more complex, and the resulting module is called the \emph{induction} of $M$ to $A$, noted $\Ind{M}$ (or $\Ind{M}_B^A$). It is defined as the tensor product of $A\otimes_B M$. The regular module structure then carries over to the tensor product: $a'(a\otimes m)=(a'a)\otimes m$ for all $a, a'\in A$ and $m\in M$. It can be shown that the induction preserves parts of exact short sequences. More precisely, one has:
\begin{Prop}
Let $A$ be an associative algebra and $B$ a subalgebra of $A$, the sequence 
$$A\otimes L \longrightarrow A \otimes M \longrightarrow A \otimes P \longrightarrow 0 $$ 
of $A$-modules is exact if the sequence 
$$ L \longrightarrow M \longrightarrow P \longrightarrow 0 $$
 of $B$-modules is exact.
\end{Prop}
  
There are cases when the homomorphism $A\otimes L\rightarrow A\otimes M$ fails to be injective even when the $B$-module homomorphism $L\rightarrow M$ is. 
  
\end{subsection}
  
\begin{subsection}{Frobenius reciprocity, Wedderburn and Jordan-H\"older theorem}
  
The operations of restriction and induction were presented as ``reverse processes". This is particularly meaningful in view of the next result.
\begin{Prop}[Frobenius reciprocity theorem]
Let $A$ be a finite-dimensional associative algebra over $\mathbb{C}$ and $B$ a subalgebra of $A$. Let $M$ be an $A$-module and $N$ be a $B$-module. Then, as vector spaces
\begin{equation}
\Hom_B \left(N, \Res{M} \right) \simeq \Hom_A \left(\Ind{N}, M \right).
\end{equation}
\end{Prop}
  
The algebra $A$ can be seen as a left $A$-module where the action is simply left multiplication. This module is called the \emph{regular} module and one may write ${}_{A}A$ to emphasize the left module structure. The algebra is called {\em semisimple} if its regular module is completely reducible, that is, it is isomorphic to a direct sum of irreducible modules. A key property of semisimple algebras is the following.
\begin{Thm}[Wedderburn's theorem]\label{prop:Wedderburn1}
Let $A$ be a finite-dimensional associative algebra over $\mathbb{C}$. $A$ is semisimple if and only if the regular module decomposes as $${}_{A}A \simeq \bigoplus_i \left( \dim L_{i}\right) L_{i}$$ where the set $\{ L_{i}\}$ forms a complete set of non-isomorphic irreducible A-modules.
\end{Thm}
It can also be shown that $A$ is semisimple if and only if every $A$-module is projective. If an algebra is not semisimple, there will be indecomposable yet reducible modules. When $A$ is not semisimple, Wedderburn's theorem no longer holds, and it is replaced by the following generalisation.
\begin{Thm}\label{prop:Wedderburn2}
Let $A$ be a finite-dimensional associative algebra over $\mathbb{C}$. The regular module decomposes as $${}_{A}A \simeq \bigoplus_i  \left( \dim L_{i}\right) P_{i}$$ where the set $\{ P_{i}\}$ forms a complete set of non-isomorphic projective indecomposable A-modules, and $L_{i}$ is the unique irreducible quotient of $P_{i}$.
\end{Thm}
The projective indecomposables in this last proposition are called \emph{principal indecomposable modules}. It can be shown that any projective module is a direct sum of principal indecomposable ones.
  
Note that induction of the regular module ${}_BB$ is simply $\Ind{B}^A_B={}_AA\otimes_B{}_BB\simeq {}_AA$. If ${}_B B=\oplus_i B_i$ is the decomposition of $B$ into its principal indecomposable modules, then ${}_AA\simeq\oplus_i \Ind{B_i}$ and, since they appear as direct summands of the free module ${}_AA$, the $A$-modules $\Ind{B_i}$ are projective. (They might not be indecomposable.) Therefore the induction of a projective module is projective.
  
A {\em composition series} of the module $M$ is a filtration $0=M_0\subset M_1\subset \dots \subset M_{k-1}\subset M_k=M$ such that all quotients $M_i/M_{i-1}$, $0<i\le k$, are irreducible. The quotients $M_i/M_{i-1}$ are called the {\em composition factors} of $M$.
\begin{Thm}[Jordan-H\"older's theorem]\label{prop:Jordan}Any finite-dimensional module $M$ has a composition series. Moreover, if $0=M_0\subset M_1\subset \dots \subset M_{k-1}\subset M_k=M$ and $0=N_0\subset N_1\subset \dots \subset N_{l-1}\subset N_l=M$ are two composition series of $M$, then $k=l$ and the two sets of composition factors coincide up to permutation.
\end{Thm}
  
\end{subsection}
  
\end{section}

\end{document}